\documentclass[fleqn,usenatbib]{mnras}

\usepackage{newtxtext,newtxmath}

\usepackage[T1]{fontenc}
\usepackage{xfrac}

\DeclareRobustCommand{\VAN}[3]{#2}
\let\VANthebibliography\thebibliography
\def\thebibliography{\DeclareRobustCommand{\VAN}[3]{##3}\VANthebibliography}

\usepackage{graphicx}	
\usepackage{amsmath}	
\usepackage{amssymb}	
\usepackage{makecell}
\usepackage{multirow}
\usepackage{ulem}
\usepackage{xcolor}

\newcommand{\kms}{km s$^{-1}$}
\newcommand{\dego}{$^\circ$}
\newcommand{\msun}{M$_\odot$}

\title[R Dor: a global view]{Morpho-kinematics of the circumstellar envelope of the AGB star R Dor: a global view}

\author[P.T. Nhung et al.]{{P.T. Nhung$^1$\thanks{E-mail: pttnhung@vnsc.org.vn},
D.T. Hoai$^1$\thanks{E-mail: dthoai@vnsc.org.vn},
P. Tuan-Anh$^1$,
P. Darriulat$^1$,
P.N. Diep$^1$,
N.B. Ngoc$^1$,
  N.T. Phuong$^{1,2}$}
\newauthor{and T.T. Thai$^1$}
\\
$^1$Department of Astrophysics, Vietnam National Space Center, Vietnam Academy of Science and Technology, \\
18, Hoang Quoc Viet, Nghia Do, Cau Giay, Ha Noi, Vietnam\\
$^2$Korea Astronomy and Space Science Institute, 776 Daedeokdae-ro, Yuseong-gu, Daejon 34055, Republic of Korea\\}

\date{Accepted XXX. Received YYY; in original form ZZZ}

\pubyear{2021}

\begin{document}
\label{firstpage}
\pagerange{\pageref{firstpage}--\pageref{lastpage}}
\maketitle

\begin{abstract}
  We analyse new ALMA observations of the $^{29}$SiO ($\nu$=0, $J$=8$-$7) and SO$_2$($\nu$=0, $34_{3,31}$$-$$34_{2,32}$) line emissions of the circumstellar envelope (CSE) of the oxygen-rich AGB star R Dor. With a spatial resolution of $\sim$2.3 au, they cover distances below $\sim$30 au from the star providing a link between earlier observations and clarifying some open issues. The main conclusions are: 1) Rotation is confined below $\sim$15 au from the star, with velocity reaching a maximum below 10 au and morphology showing no significant disc-like flattening. 2) In the south-eastern quadrant, a large Doppler velocity gas stream is studied in more detail than previously possible and its possible association with an evaporating planetary companion is questioned. 3) A crude evaluation of the respective contributions of rotation, expansion and turbulence to the morpho-kinematics is presented. Significant line broadening occurs below $\sim$12 au from the star and causes the presence of high Doppler velocity components near the line of sight pointing to the centre of the star. 4) Strong absorption of the continuum emission of the stellar disc and its immediate dusty environment is observed to extend beyond the disc in the form of self-absorption. The presence of a cold SiO layer extending up to some 60 au from the star is shown to be the cause. 5) Line emissions from SO, $^{28}$SiO, CO and HCN molecules are used to probe the CSE up to some 100 au from the star and reveal the presence of two broad back-to-back outflows, the morphology of which is studied in finer detail than in earlier work.

\end{abstract}




\section{Introduction}

In the recent years, the availability of high sensitivity and high angular resolution observations, in particular from the Atacama Large Millimeter/submillimeter Array (ALMA), has shed new light on the mechanisms that govern the mass loss of evolved stars. At the same time it has revealed their complexity and many open questions are still awaiting an answer. Such is particularly the case of oxygen-rich AGB stars, for which the details of the formation of a dust-grain-driven wind are not fully understood. Recently, several such stars have been the object of detailed observations and studies; among these is R Dor, which is the subject of the present article.

R Dor, an oxygen-rich AGB star, is a semi-regular variable of the SRb type, belonging to spectral class M8IIIe, having an initial mass estimated between 1 and 1.5 \msun\ from the value of the $^{17}$O/$^{18}$O ratio \citep{Danilovich2017}, a mass-loss rate of $\sim$1.6 10$^{-7}$ \msun yr$^{-1}$ \citep{Ramstedt2014, Maercker2016} and a temperature of $\sim$3000 K \citep{Dumm1998}; it is close to the Sun, at a distance of only $\sim$59 pc \citep{Knapp2003}; it displays no technetium in its spectrum \citep{Lebzelter1999} and has a $^{12}$CO/$^{13}$CO abundance ratio of $\sim$10 \citep{Ramstedt2014}. It has a dual period of 175 and 332 days \citep{Bedding1998} and its infrared emission above black body between 1 and 40 $\mu$m wavelength shows an enhancement of aluminium oxide and melilite in the dust \citep{Heras2005}.

\begin{figure*}
  \centering
  \includegraphics[width=15cm,trim=.05cm 0.cm 0.cm 0.cm,clip]{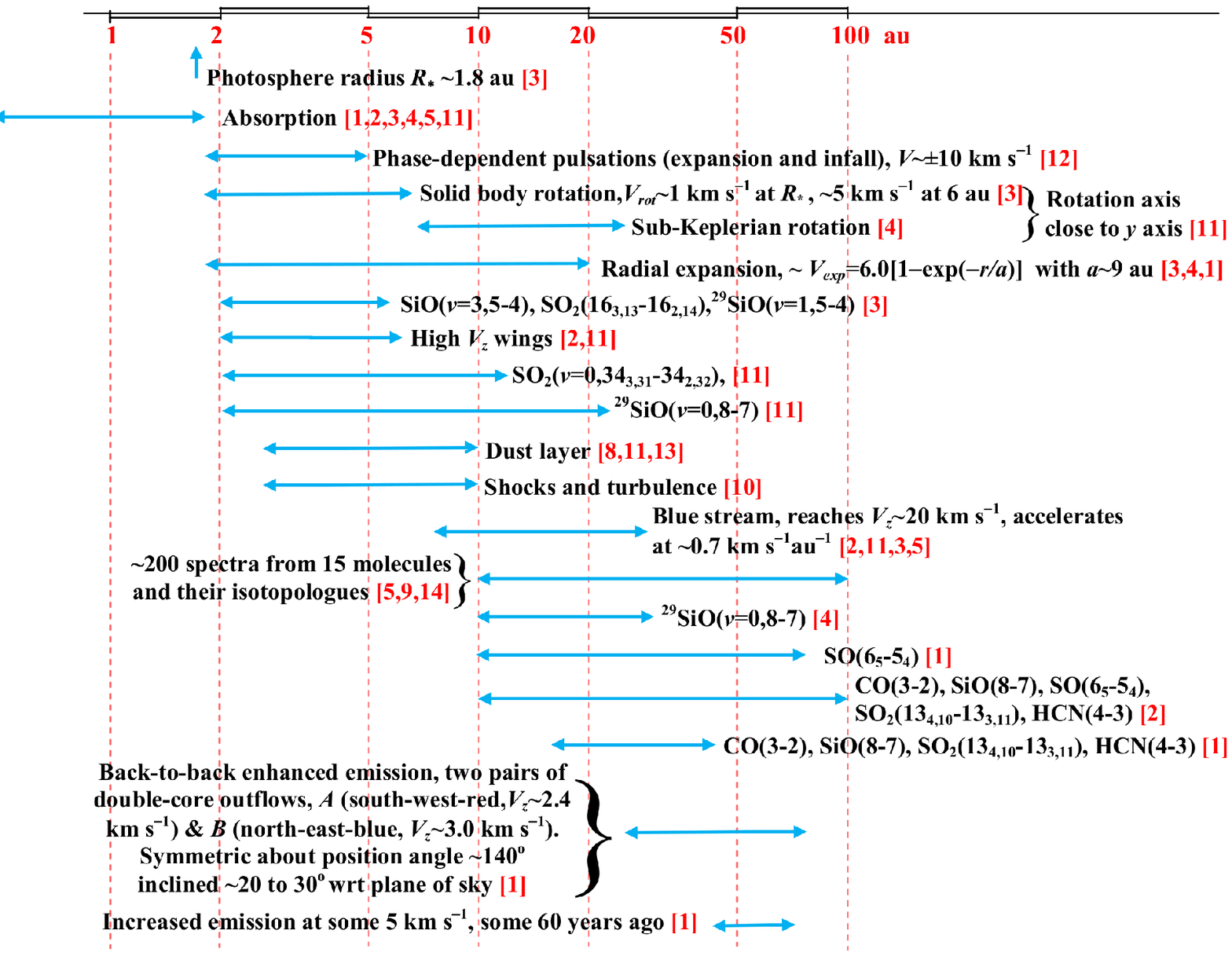}
   \caption{A summary of recently published articles on R Dor. Blue arrows indicate the radial range covered by each study. Red numbers in square brackets refer to the following articles: 1. \citet{Nhung2019} 2. \citet{Hoai2020} 3. \citet{Vlemmings2018} 4. \citet{Homan2018} 5. \citet{Decin2018} 6. \citet{VandeSande2018} 7. \citet{DeBeck2018} 8. \citet{Khouri2016} 9. \citet{Danilovich2016} 10. \citet{Vlemmings2019} 11. This work. 12. \citet{Nowotny2010} 13. \citet{Decin2017} 14. \citet{Danilovich2020}}
 \label{fig1}
\end{figure*}

Recent studies of the morpho-kinematics of the CSE of R Dor are summarized in Figure \ref{fig1}, which lists relevant references. The stellar disc and the stellar atmosphere that extends a very few au above it have been probed over a large range of wavelengths. Observations in the millimetre and sub-millimetre range are mostly from the Atacama Large Millimeter-sub-millimeter Array (ALMA) and use continuum or molecular lines emissions that probe preferentially the close neighbourhood of the star, such as the $\Lambda$ doublet OH line \citep{Vlemmings2019, Khouri2019}. Evidence is found for short time variability (at month scale),  hot spots covering a small fraction of the stellar disc, radial velocities at the 10 \kms\ scale, all revealing the impact of shocks associated with pulsations and convective cells. Similar results are obtained from observations in the optical and near-IR ranges, using the Very Large Telescope (VLT) equipped with SPHERE \citep{Khouri2016}. Large Doppler velocity wings are observed near the line of sight crossing the centre of the stellar disc, revealing the broad velocity dispersion in the inner layer \citep{Decin2018,Hoai2020}. In addition to the evidence for strong inhomogeneity and time variability, the stellar disc radius is measured at a 1.5 to 2 au scale dependent on wavelength; temperatures are estimated in the 2700$\pm$300 K range.
  
Evidence for the presence of large dust grains, at 0.1 micron scale, within a very few au from the radiosphere is obtained from the study of continuum emission as well as from the observation of polarized light using SPHERE/ZIMPOL on the VLT \citep{Khouri2016}. Observations at infrared wavelengths using NACO on VLT \citep{Norris2012} give similar results. This suggests a scenario where the wind is driven by photon scattering on large transparent grains very close to the star, rather than by radiation pressure on silicate grains, which is too small to drive the winds of O-rich AGB stars \citep{Woitke2006}. Numerous multi-wavelength studies, using in particular Herschel and ALMA observations, have addressed the question of the composition of the dust and of the nature of its main precursors, generally concluding that aluminium bearing molecules play a dominant role \citep{Heras2005, Decin2018}. \citet{Decin2017} study the line profiles of AlO, AlOH and AlCl and suggest the presence of large gas-phase (Al$_2$O$_3$)$_{\rm{n}}$ clusters ($n$$>$34) close to the star. Moreover, \citet{Takigawa2019} have shown that transition alumina containing $\sim$10\% of Si rather than amorphous alumina with pure Al$_2$O$_3$ composition is the most plausible source of the dust emission from alumina-rich AGB stars \citep[see also][]{Karovicova2013}. Analyses of molecular line profiles from single dish observations provide a host of useful information on the gas and dust composition of the CSE up to a few 100 au from the star, an important result being the confinement of SiO to shorter distances than CO, interpreted in part as the result of a progressive condensation of SiO gas molecules onto dust grains \citep{Schoier2004}. \citet{DeBeck2018} and \citet{Decin2018} perform extensive spectral scans of the envelope, the latter detecting gaseous precursors of dust grains such as SiO, AlO, AlOH, TiO, and TiO$_2$. \citet{Danilovich2016, Danilovich2017, Danilovich2019,Danilovich2020} study sulphur-bearing and water molecules and HCN abundances are measured by \citet{Schoier2013} and by \citet{VandeSande2018}.

Significant rotation of the inner CSE has been detected by \citet{Homan2018} and \citet{Vlemmings2018}. The former authors use ALMA observations of the $^{28}$SiO($\nu$=1,$J$=8-7) line emission to probe the $\sim$6 to $\sim$25 au radial range with an angular resolution of $\sim$9 au. Their data are well described by a nearly edge-on disc in sub-Keplerian rotation, with a projected rotation velocity of $\sim$6 \kms\ at a distance of 12 au. The latter authors use ALMA observations of the SiO($\nu$=3,$J$=5$-$4) and SO$_2$($J_{Ka,Kc}$=16$_{3,13}$$-$16$_{2,14}$) line emissions with an angular resolution of $\sim$2 au to probe the inner layer from the radiosphere up to some 15 au. Their data give no evidence for disc morphology and are well described by solid-body rotation, again with a projection rotation velocity of $\sim$6 \kms\ at a distance of 12 au. The difference between the two models is largely the result of the different radial ranges probed by the observations. Both authors argue that the presence of significant rotation is most easily explained by the transfer of angular momentum from a companion to the star. The presence of a blue-shifted feature in the south-eastern quadrant, extending to some 25 au in projected distance from the star and reaching near 20 \kms\ in Doppler velocity \citep{Decin2018, Vlemmings2018, Hoai2020} is invoked as suggesting its association with such companion.

Finally, a detailed study of the outer region of the CSE, beyond some 25 au, shows the presence of outflows and of cavities suggesting a complex history of mass ejection \citep[][hereafter referred to as Paper I]{Nhung2019}.

The present work studies the emissions of two molecular lines that have been observed by ALMA: $^{29}$SiO ($\nu$=0, $J$=8$-$7) and SO$_2$($\nu$=0, $34_{3,31}$$-$$34_{2,32}$), thereafter simply referred to as $^{29}$SiO and SO$_2$. Parameters of relevance are listed in Table \ref{tab1}. 
      
\begin{table}
  \centering
  \caption{Parameters of the $^{29}$SiO and SO$_2$ emission lines. $E_{\rm{u}}$ is the energy of the upper level of the transition. The frequency $f$ corresponds to a wavelength $\lambda=0.875$ mm and a transition temperature of 16.5 K.}
  \label{tab1}
  \begin{tabular}{cccc}
    \hline
    Molecule&Transition&$f$ (GHz)&$E_{\rm{u}}$ (K)\\
    \hline
    $^{29}$SiO&  $\nu$=0, $J$=8$-$7 &  342.98084800&    74.08\\
    SO$_2$&    $\nu$=0, $34_{3,31}$$-$$34_{2,32}$ &    342.76162320 &    581.92 \\
    \hline
    
  \end{tabular}
\end{table}

The paper is organized as follows: after a description of observations and data reduction, we present in Section 3 a general picture of the morpho-kinematics of the CSE as obtained from the new results discussed in the present work and taking into account analyses published earlier. Section 4 discusses some properties of the inner layer, accounting for absorption and including rotation, expansion and contributions to the line width. Section 5 addresses questions related to the observed properties of the south-eastern blue-shifted feature. Section 6 explores the CSE to larger distances from the star using other molecular line emissions as probes with the aim of explaining what causes the strong absorption of the SiO lines. A last section summarizes the main results and concludes.

\section{Observations and data reduction}

\begin{figure*}
  \centering
  \includegraphics[width=4.4cm,trim=0.0cm .5cm 1.cm 1.5cm,clip]{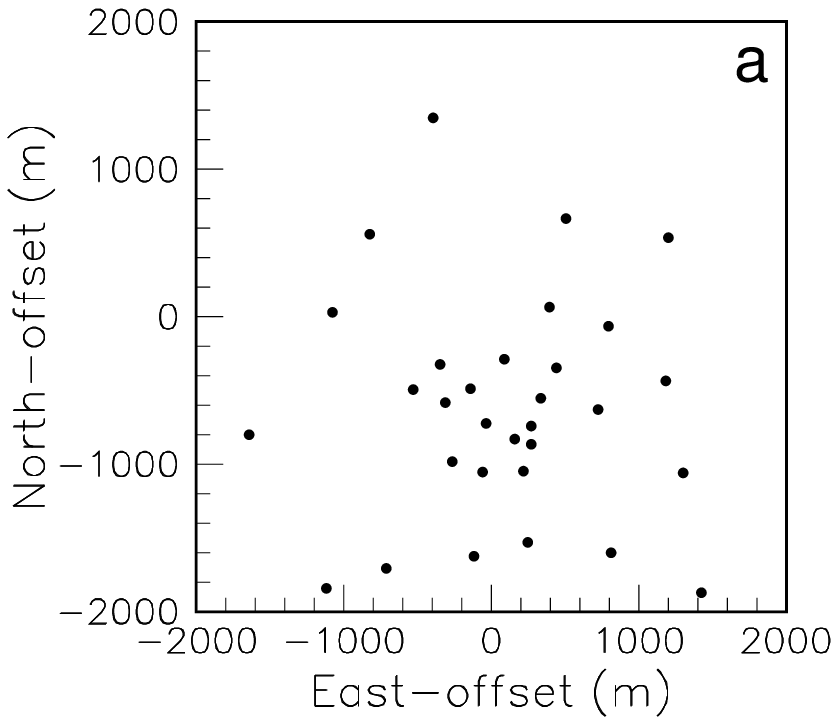}
  \includegraphics[width=4.4cm,trim=0.3cm .5cm .7cm 1.5cm,clip]{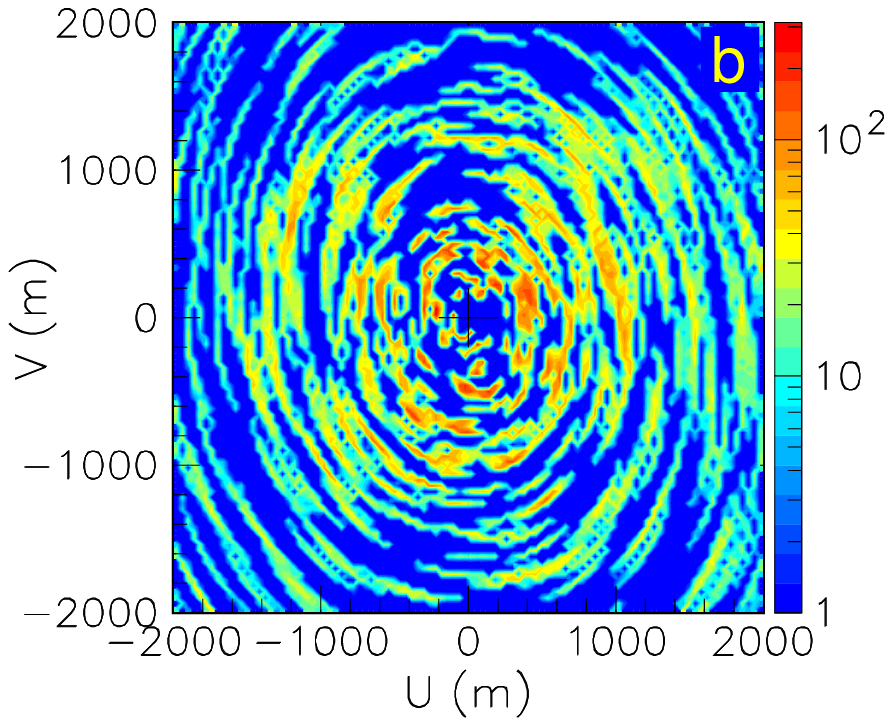}
  \includegraphics[width=4.4cm,trim=0.0cm .5cm 1.cm 1.5cm,clip]{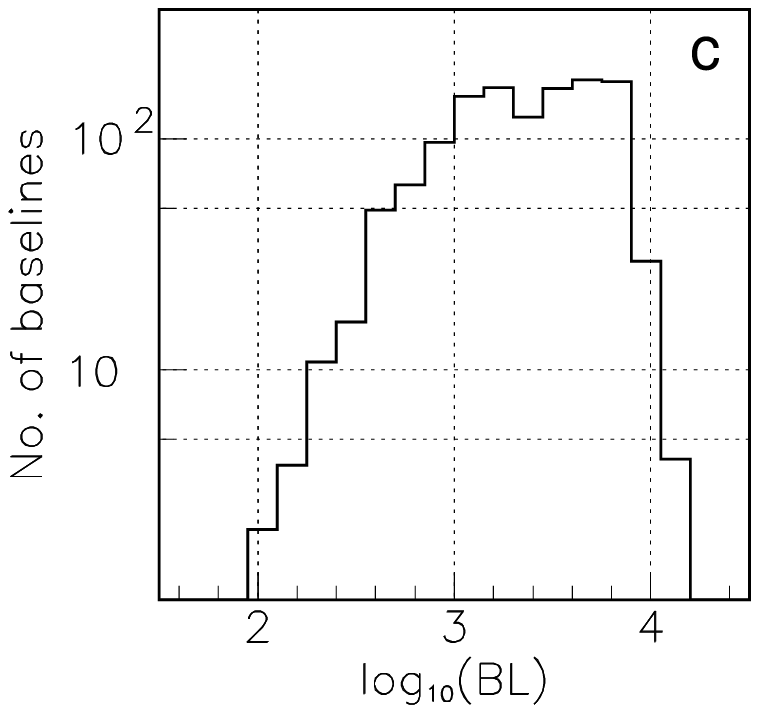}
  \includegraphics[width=4.4cm,trim=0.0cm .5cm 1.cm 1.5cm,clip]{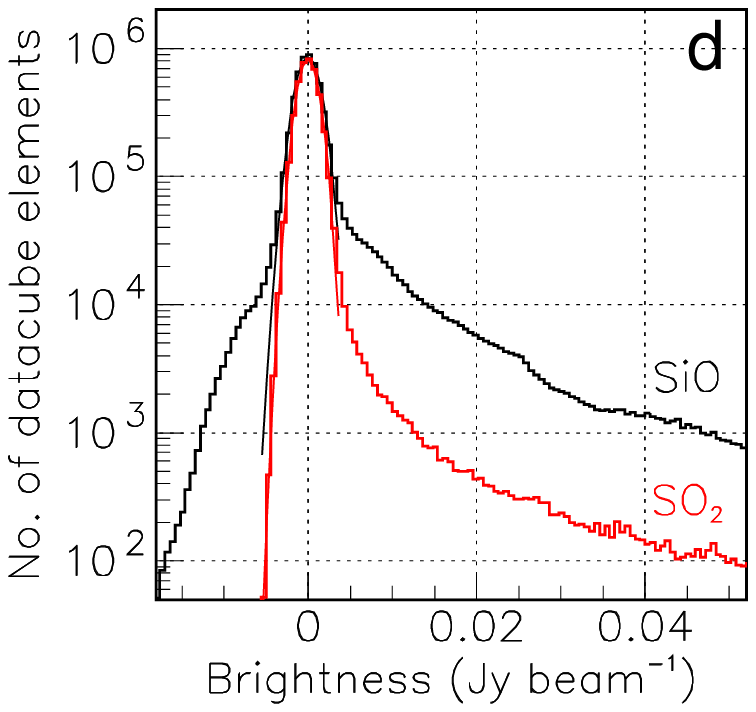}
   \caption{The antenna configuration (a) and $uv$ coverage (b) are shown for the central region (|$u$| and |$v$|< 2000 m). The colour scale in panel b shows the density of visibility measurements (in 40$\times$40 m$^2$ bin) (c): distribution of the baselines; the abscissa is log$_{10}$(BL), the decimal logarithm of the baseline length measured in meters. (d): brightness distributions for $^{29}$SiO (black) and SO$_2$ (red) in a $R<1$ arcsec circle. }
 \label{fig2}
\end{figure*}

The data are retrieved from ALMA archives, project 2017.1.00191.S, observed in November 2017 in band 7 with antennas in the C43-9 configuration. The observations were made in two blocks using 44 and respectively 45 antennas.  The total time on source was 90 min, 45 min for each block. Visibilities are averaged over 20 s wide time intervals. Baselines span between 100 m and 12 km (Figure \ref{fig2}) and the maximal recoverable scale, evaluated for a wavelength of 0.87 mm and an effective minimal baseline of 300 m, is 0.36 arcsec. We select a frequency interval of 81 channels corresponding to a Doppler velocity interval of 69 \kms, each Doppler velocity bin being 0.854 \kms\ wide. The systemic velocity of R Dor (7 \kms) is taken as origin. The data were reduced using standard scripts for continuum emission and two line emissions: $^{29}$SiO($\nu$=0, $J$=8$-$7) and SO$_2$($34_{3,31}$$-$$34_{2,32}$).

We calibrated the data using the scripts provided by the ALMA staff. Deconvolution and cleaning were done with natural weighting using two different codes: CASA\footnote{http//casa.nrao.edu} and GILDAS\footnote{https://www.iram.fr/IRAMFR/GILDAS}. The line data were processed without continuum subtraction. The beams obtained with CASA and GILDAS (FWHM) were 34$\times$25 mas$^2$ and 41$\times$35 mas$^2$ respectively. In both cases we use pixels of 5$\times$5 mas$^2$. We found that the differences between the results obtained with the two codes are simply explained by the differences in beam size and have negligible impact on the results of our analyses. In what follows we use the GILDAS results as default but we checked their consistency with the CASA results whenever possibly relevant. Brightness distributions (Figure \ref{fig2}) show a noise level (1$\sigma$) of 1.4 mJy\,beam$^{-1}$ for $^{29}$SiO and 1.1 mJy\,beam$^{-1}$ for SO$_2$. Additional information is provided in Section 6.1. Channel maps are shown in the Appendix (Figures \ref{figA1} and \ref{figA2}).

\begin{table*}
  \centering
  \caption{$^{28}$SiO, CO, HCN and SO observations. $f$ is the frequency, MRS the maximal recoverable scale, $N$ the number of antennas.}
  \label{tab3}
  \begin{tabular}{cccccccc}
    \hline
    Line &$f$& Beam& MRS& $N$&Noise&Date&Time on source\\
    &(GHz)&(mas$^2$)&(au)&&(mJy\,beam$^{-1}$)&&\\
    \hline
    $^{28}$SiO II&    347.3&    220$\times$170&    80&    38&    18&    Aug 2015&    21 mn\\
    CO&    345.8&    180$\times$140&    80&32&14&Sep 2015&21 mn\\
    HCN&354.5&157$\times$145&78&34&8&Aug 2015&26 mn\\
    SO&251.8&154$\times$147&136&45&1.2&Dec 2017&2.7 hr\\
    \hline
  \end{tabular}
\end{table*}

\begin{figure*}
  \centering
  \includegraphics[width=5cm,trim=0.cm 0.5cm 1.2cm 1.8cm,clip]{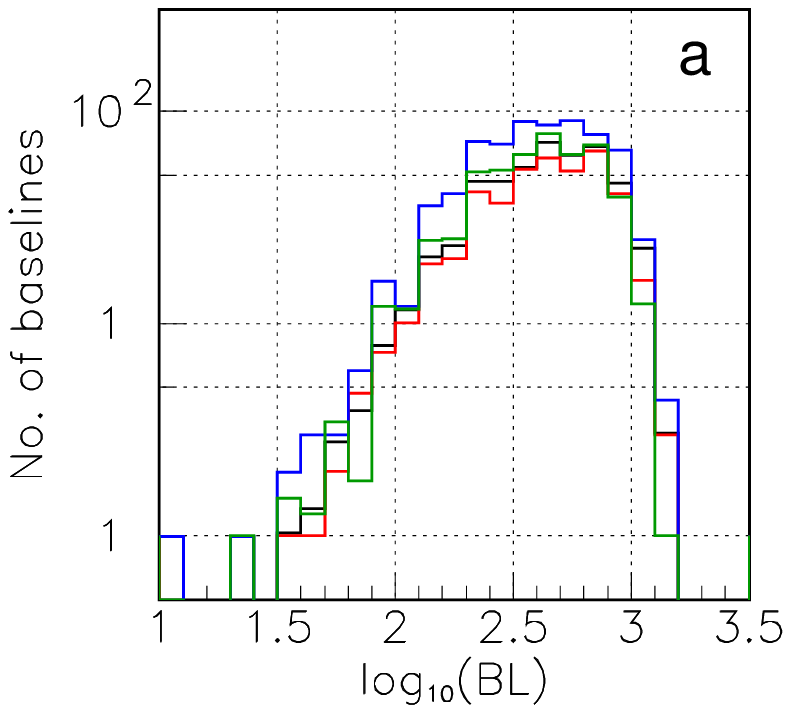}
  \includegraphics[width=5cm,trim=0.cm 0.5cm 1.2cm 1.8cm,clip]{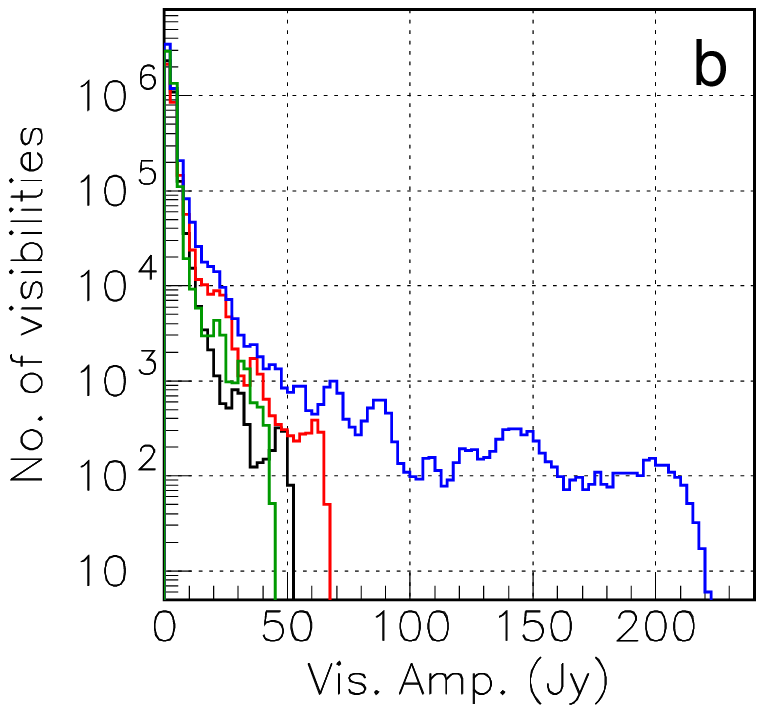}
  \includegraphics[width=5cm,trim=0.cm 0.5cm 1.2cm 1.8cm,clip]{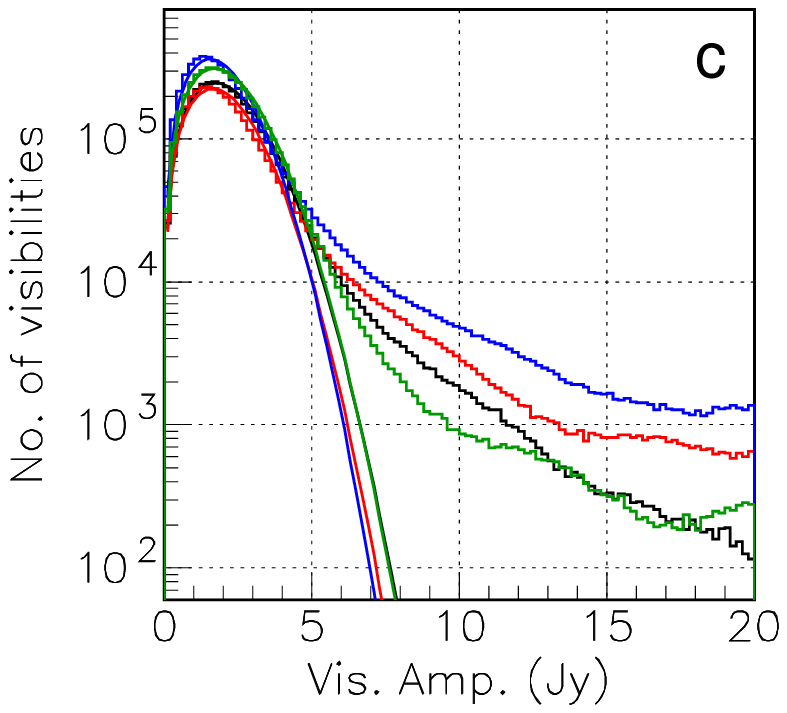}
   \caption{Observation and reduction of the $^{28}$SiO I (red), $^{28}$SiO II (blue), CO (black) and HCN (green) data. (a): dependence of the number of baselines on the decimal logarithm of the baseline lengths measured in meters, log$_{10}$(BL). (b): distribution of the visibility amplitude |$U$| in the interval |$V_z$|<7 \kms. (c): zoom of panel b showing noise fits of the form |$U$|$\exp(-\sfrac{1}{2}|U|^2/\sigma^2)$. They give $\sigma$=1.6, 1.5, 1.7 and 1.7 Jy for $^{28}$SiO I, $^{28}$SiO II, CO and HCN respectively.}
 \label{fig21}
\end{figure*}

Moreover, in Section 6, we use observations of other molecular lines that have been briefly considered in Paper I: $^{28}$SiO(8$-$7), CO(3$-$2) and HCN(4$-$3). Their emission was detected with a similar beam size but in less favourable observational conditions than for the SO($J_K$=6$_5$$-$5$_4$) line, which was the central subject of Paper I (see Section 6.2). They are from project 2013.1.00166.S observed in summer 2015 in band 7 with an average of 35 antennas in the C34-7 configuration. They include two independent sets of $^{28}$SiO data, hereafter referred to as $^{28}$SiO I and $^{28}$SiO II. Paper I used data reduced by the ALMA staff with continuum subtraction in the $uv$ plane. As the $uv$ coverage for short baselines is poor, we reduced these data anew, using GILDAS with natural weighting and without continuum subtraction. A summary of the main parameters and results is listed in Table \ref{tab3} and illustrated in Figure \ref{fig21}, which shows the baseline and brightness distributions obtained in each case. We paid much attention to the impact of the short spacing problem on the quality and reliability of the reduced data. This is illustrated in Figure \ref{fig21}b: the $^{28}$SiO I and $^{28}$SiO II sets, although having similar baseline distributions, display very different visibility distributions. The simple addition of a few short baselines is the cause. Accordingly, we restrict the analysis to distances from the star not exceeding 60 au (1 arcsec). Moreover, we disregard the $^{28}$SiO I set of data as not adding significant information to that obtained from the $^{28}$SiO II set.
  
We use orthonormal coordinates with the $x$ axis pointing east, the $y$ axis pointing north and the $z$ axis pointing away from Earth. Position angles, $\omega$, are measured counter-clockwise from north. 

\section{Overall picture of the CSE}

\begin{figure*}
  \centering
  \includegraphics[width=4.4cm,trim=0.cm 0.5cm 1.5cm 0.cm,clip]{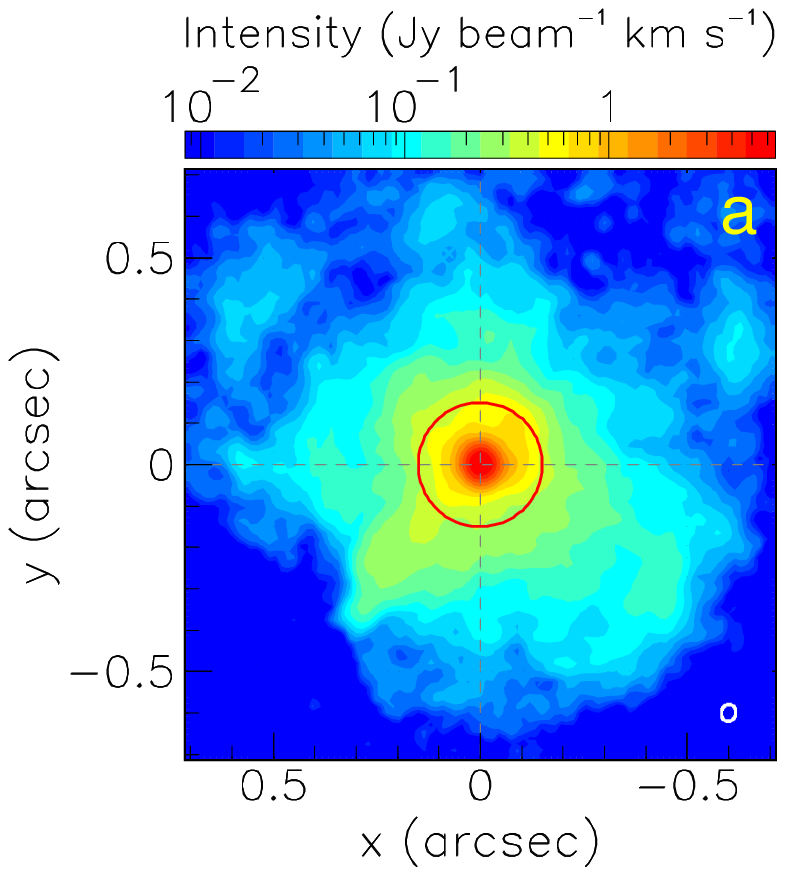}
  \includegraphics[width=4.4cm,trim=0.cm 0.5cm 1.5cm 0.cm,clip]{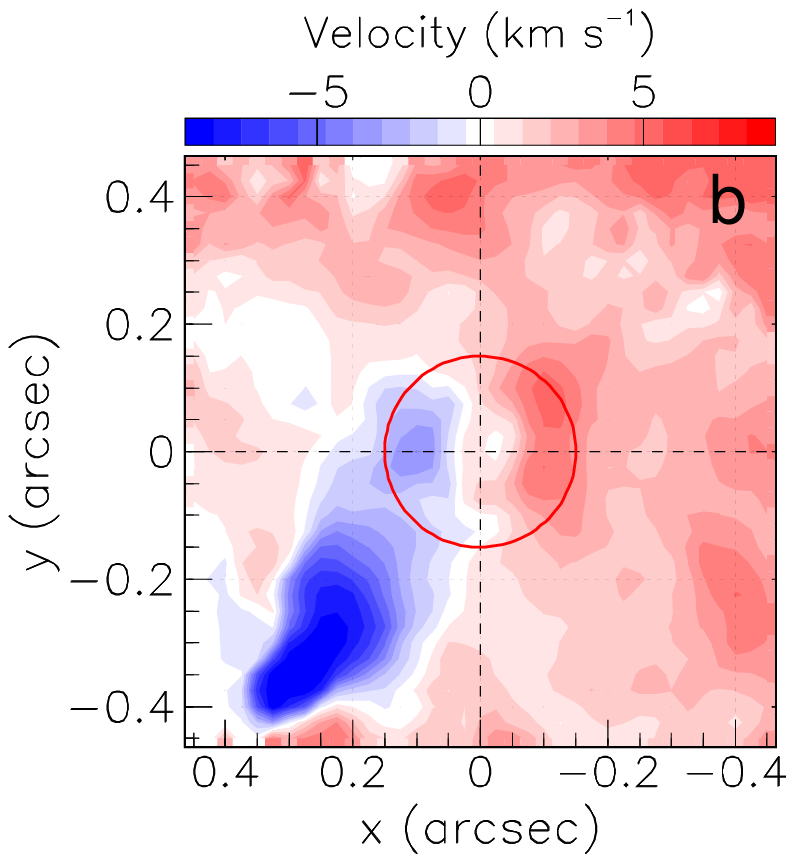}
  \includegraphics[width=4.4cm,trim=0.cm 0.5cm 1.5cm 0.cm,clip]{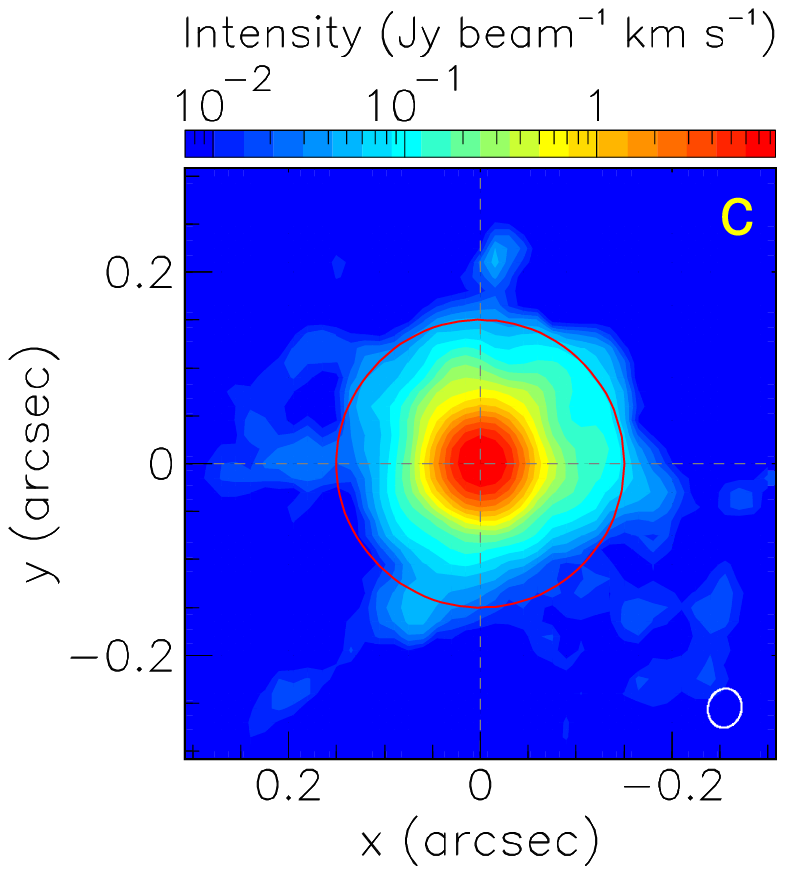}
  \includegraphics[width=4.4cm,trim=0.cm 0.5cm 1.5cm 0.cm,clip]{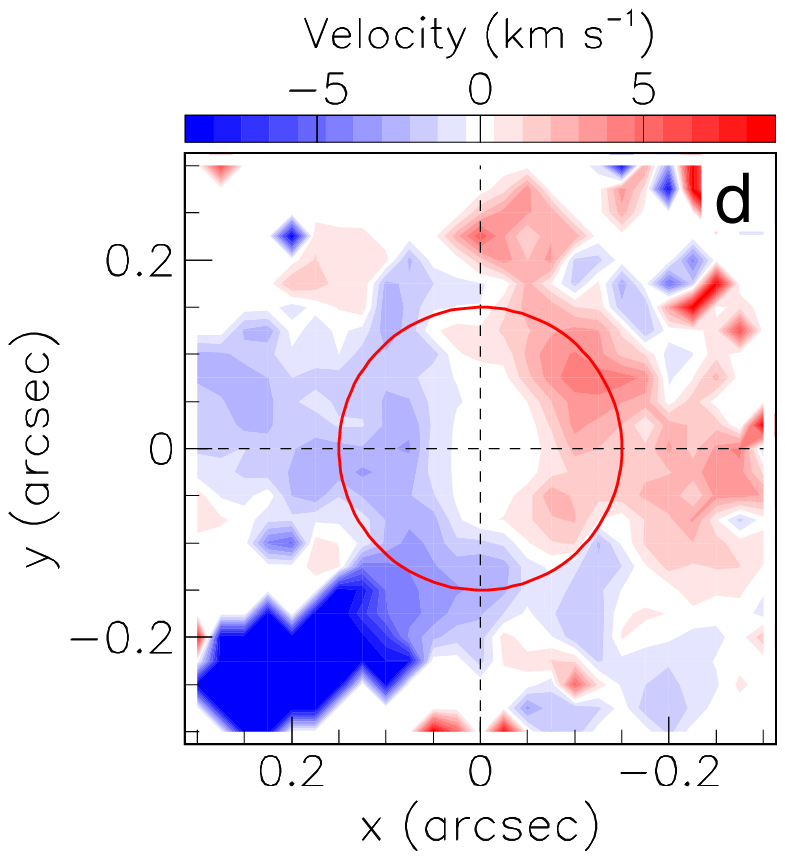}
   \caption{Maps of the intensity (a,c) and of <$V_{z}$> (b,d), respectively integrated and averaged over $|V_{z}|$<24 \kms\ for $^{29}$SiO (a,b) and SO$_2$ (c,d) emissions. Note the different scales. In both cases a 3$\sigma$ cut has been applied. In the SO$_2$<$V_{z}$> map the colour scale is saturated over the south eastern blue stream where the intensity is near noise level. The red circle ($R$<9 au $=$ 0.15 arcsec) shows the inner region dominated by rotation. }
 \label{fig3}
\end{figure*}

Figure \ref{fig3} displays the sky maps of the integrated intensity (moment 0) and mean Doppler velocity<$V_{z}$> (moment 1) for both the $^{29}$SiO and SO$_2$ lines. While $^{29}$SiO emission is seen to extend up to some 40 au projected distance from the star, SO$_2$ emission is limited to some 12 au. The high excitation energy of the SO$_2$ line prevents it from populating the outer CSE as the temperature can be expected to take a value of $\sim$1100 K at 6 au from the star and to decrease as  $r^{-0.65}$ \citep{Decin2018, Homan2018,VandeSande2018}. The south-eastern blue stream stands out in the $^{29}$SiO map, more clearly than in previous publications, both because of the better match with the radial range being probed and the better sensitivity and angular resolution. For both lines the north-western half of the <$V_{z}$> map is red-shifted, the south-eastern half blue-shifted and the inner layer of the CSE gives evidence for rotation about an axis projecting close to north.

\subsection{The inner layer}

\begin{figure}
  \centering
   \includegraphics[width=8.5cm,trim=3.4cm 3.cm 3.4cm 3.cm,clip]{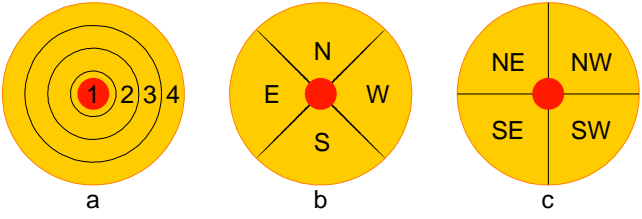}
   \caption{Numbering of the rings (a) and quadrants (b,c). The red circle shows the star with a radius of 0.035 arcsec $=$ 2.1 au; the yellow circle has a radius of 12 au $=$ 0.2 arcsec; each ring is 3 au $=$ 0.05 arcsec wide.}
 \label{fig4}
\end{figure}

\begin{figure}
  \centering
   \includegraphics[width=9cm,trim=0.cm 0.5cm 1.cm 1.5cm,clip]{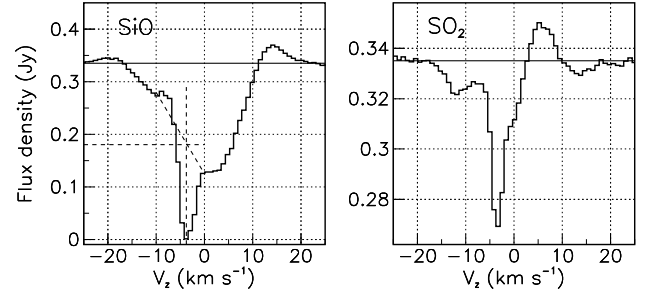}
   \caption{Doppler velocity spectra integrated over the stellar disc ($R$<0.035 arcsec, region 1 of Figure \ref{fig4}) for $^{29}$SiO (left) and SO$_2$ (right, note the zero-suppressed scale of ordinate). The dashed lines in the left panel illustrate how we estimate $I_{\rm{mid}}$ in Section 3.2.}
 \label{fig5}
\end{figure}

\begin{figure*}
  \centering
  \includegraphics[width=8.7cm,trim=0.cm 0.5cm 0.cm 1.cm,clip]{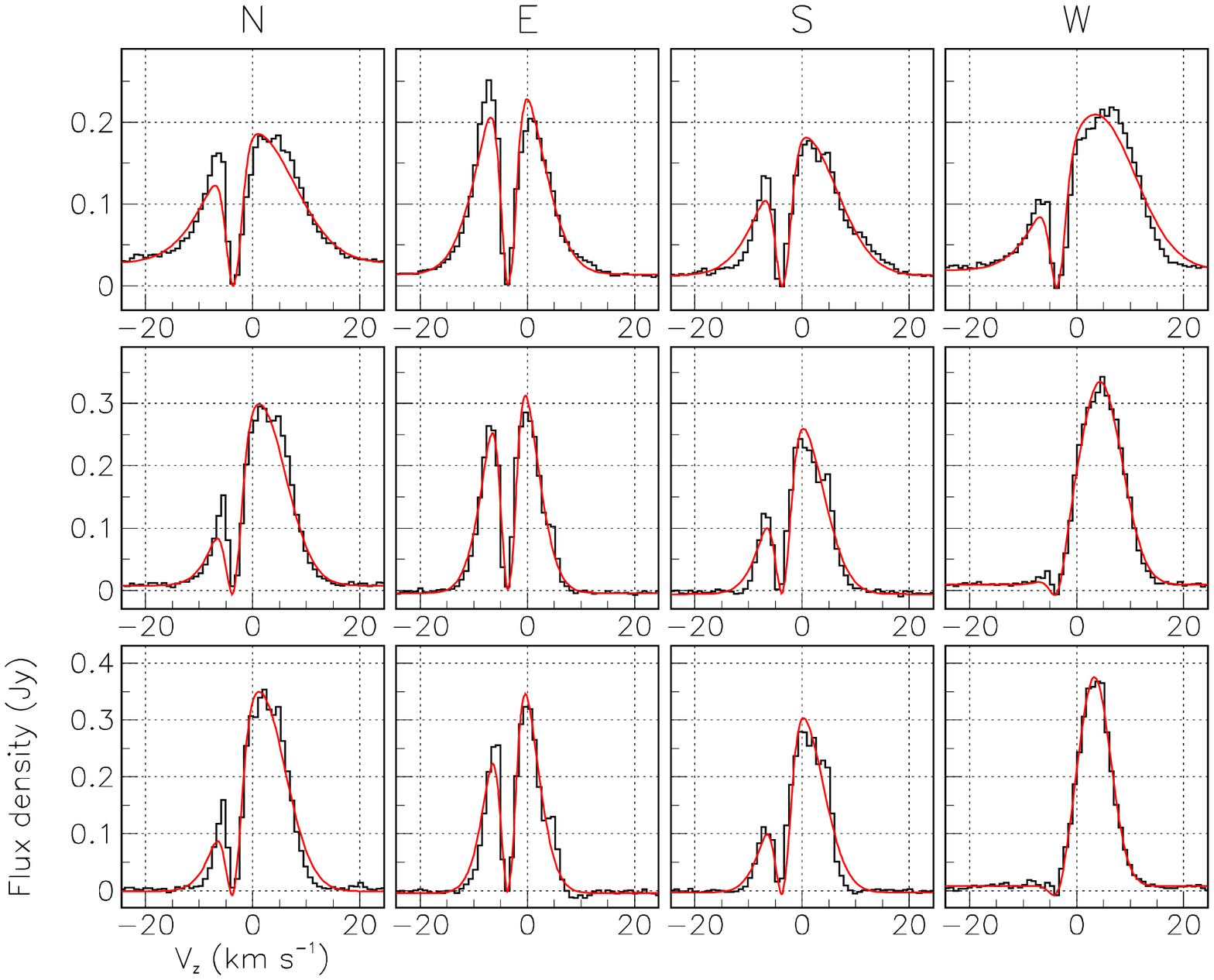}
  \includegraphics[width=8.7cm,trim=0.cm 0.5cm 0.cm 1.cm,clip]{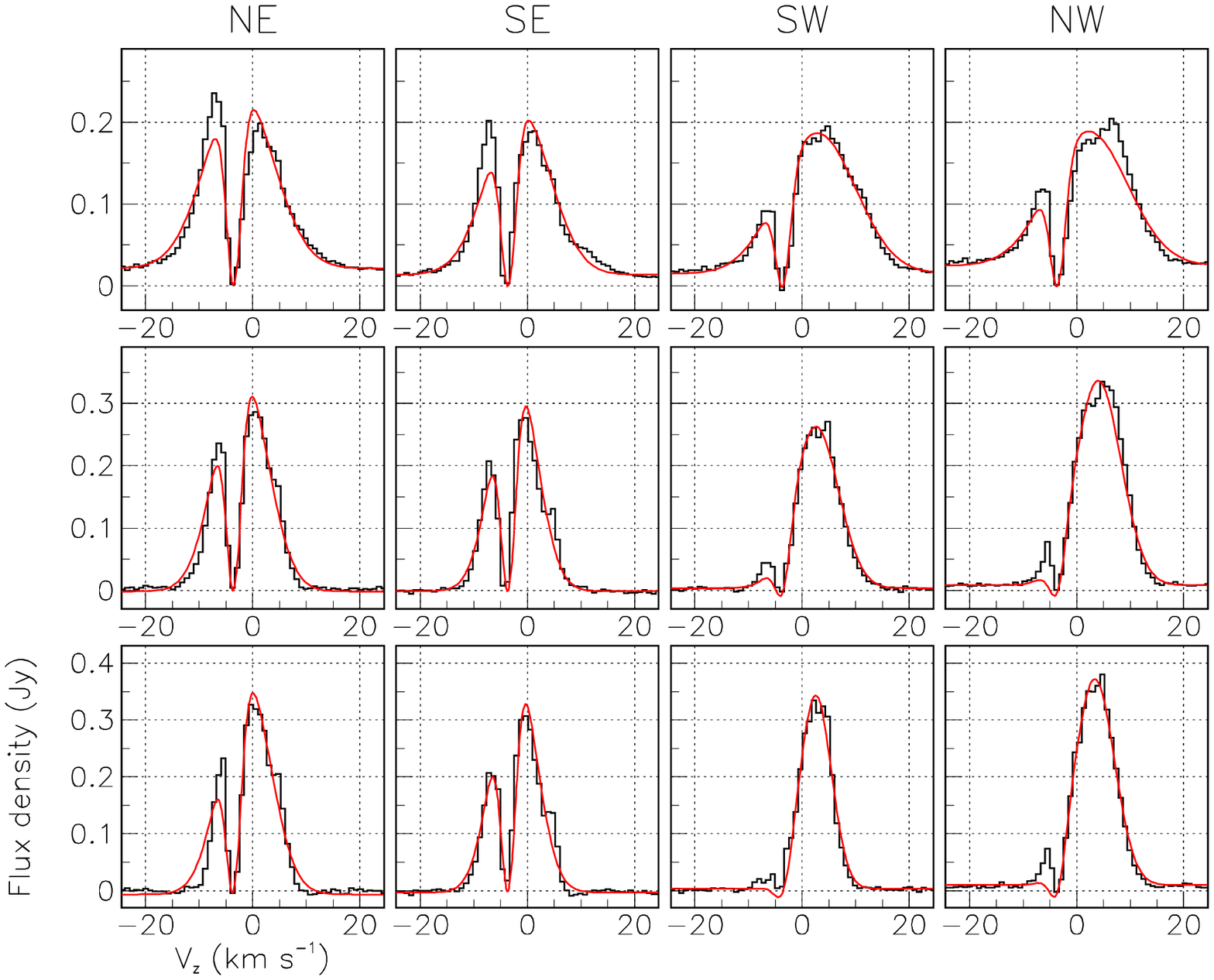}
   \caption{Doppler velocity spectra (black) of $^{29}$SiO emission. From left to right: quadrants N, E, S,  W, NE, SE, SW and NW (see Figure \ref{fig4}). From up down: rings 2, 3 and 4. The fits (red) are described in the text and their parameters listed in Table \ref{tabA1}.}
 \label{fig6}
\end{figure*}

\begin{figure*}
  \centering
   \includegraphics[width=8.7cm,trim=0.cm 0.5cm 0.cm 1.cm,clip]{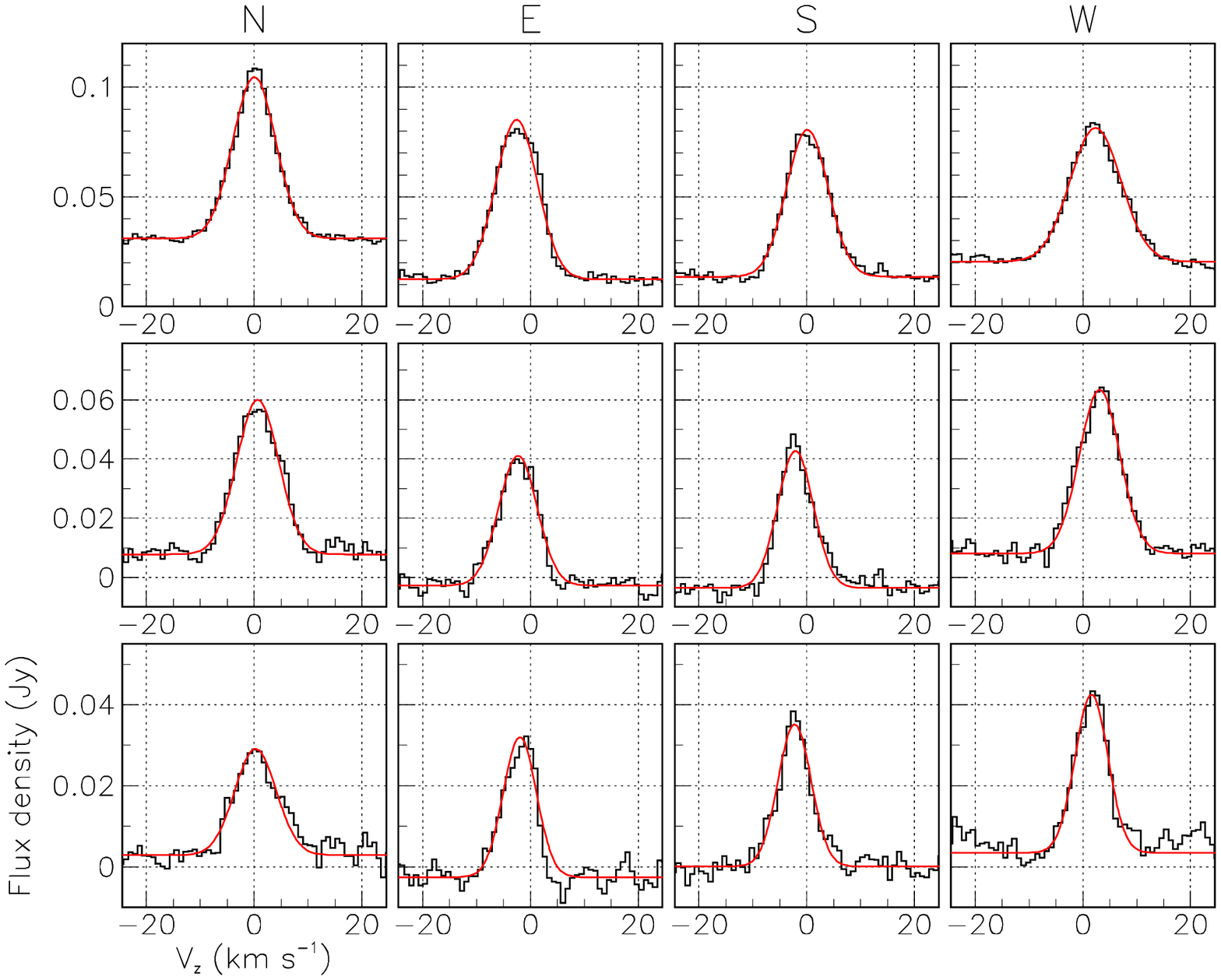}
   \includegraphics[width=8.7cm,trim=0.cm 0.5cm 0.cm 1.cm,clip]{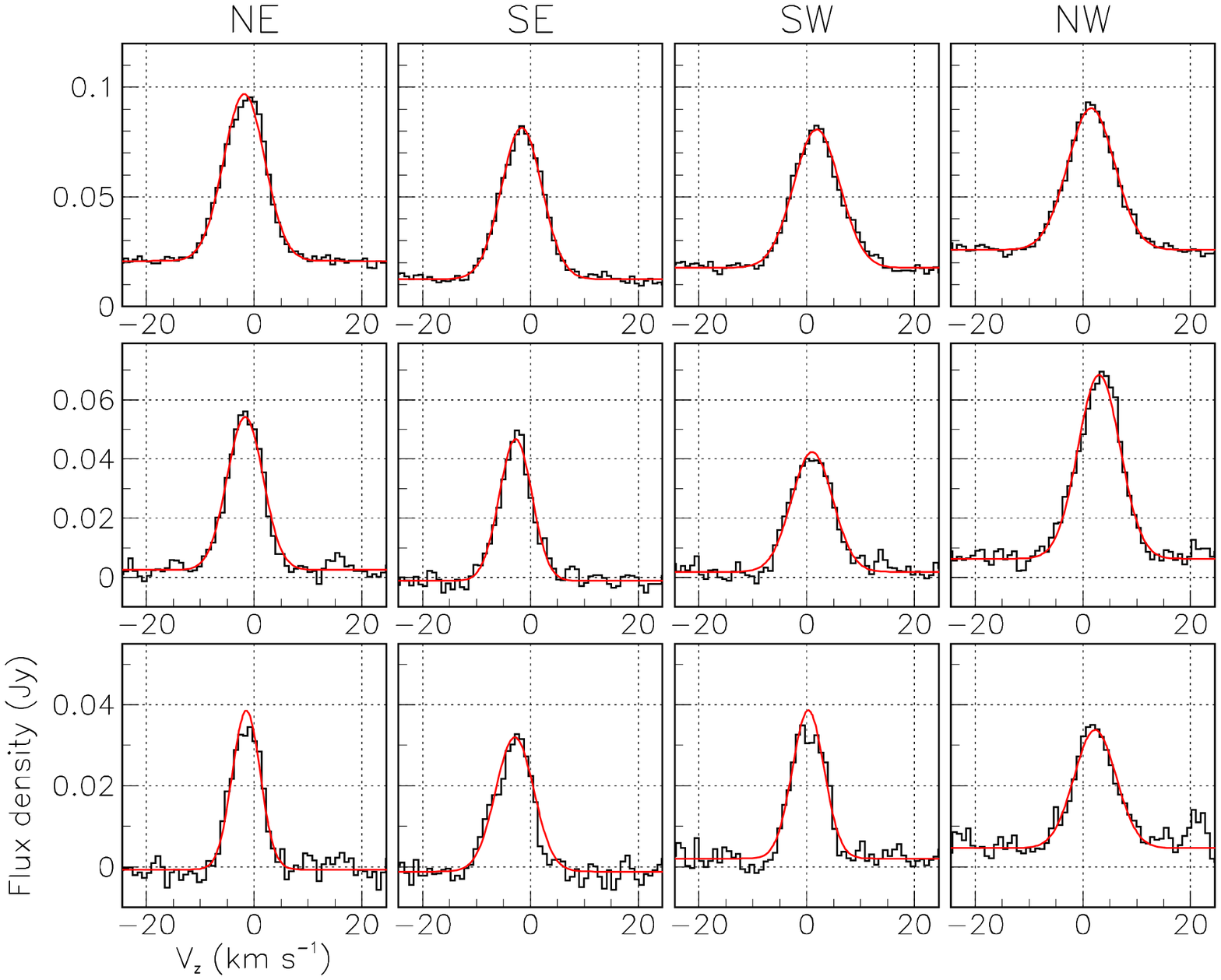}
   \caption{Doppler velocity spectra (black) of SO$_2$ emission. From left to right: quadrants N, E, S, W, NE, SE, SW and NW (see Figure \ref{fig4}). From up down: rings 2, 3 and 4. The fits (red) are described in the text and their parameters listed in Table \ref{tabA2}.}
 \label{fig7}
\end{figure*}

Figures \ref{fig4} to \ref{fig7} illustrate the main properties of the inner CSE layer, which we present in a set of rings surrounding the star, each 0.05 arcsec (3 au) wide and covering 0.05<$R$<0.20 arcsec as illustrated in Figure \ref{fig4}. The stellar disc, known to have a radius of $\sim$1.8 au (0.03 arcsec) at mm wavelength \citep{Vlemmings2018}, is here represented with a radius of 0.035 arcsec.  Figure \ref{fig5} displays the Doppler velocity spectra integrated over it for both $^{29}$SiO and SO$_2$ emissions. They are mostly seen in absorption. The $^{29}$SiO spectrum shows a clear absorption peak occurring in the outer layer where the slow wind has reached a velocity of $\sim$4 \kms. In addition, it also displays absorption by the inner layer, revealing a Doppler velocity distribution extending up to $|V_{z}|$$\sim$20 \kms. Its precise shape is not expected to be significantly affected by rotation but to tell about the radial increase of the expansion velocity as well as about line width; the latter are expected to receive important contributions from shocks associated with pulsations and convective cell ejections \citep[][and references therein]{Freytag2017}. The red side of the inner layer is seen in emission, the line width of the outer layer being too small to significantly populate Doppler velocities exceeding $\sim$10 \kms, $\sim$15 \kms\ above the mean Doppler velocity at such a distance from the star.

The SO$_2$ spectrum is also mostly seen in absorption, displaying a shape qualitatively similar to that of the $^{29}$SiO spectrum, with a narrow absorption peak near $-$4 \kms. However, the scale is very different, both in flux density and in Doppler velocity. The amplitude of the absorption peak is only 18\% of the continuum level instead of being total for $^{29}$SiO and the part of the spectrum seen in emission is centred at $\sim$6 \kms\ instead of $\sim$13 \kms. A reliable interpretation of the features displayed in Figure \ref{fig5} requires accounting, at least approximately, for optical thickness, which we do in Section 4.

Figures \ref{fig6} and \ref{fig7} display the Doppler velocity spectra averaged in the rings and quadrants defined in Figure \ref{fig4} and covering projected distances from the star between 0.05 and 0.20 arcsec (3 and 12 au). While the $^{29}$SiO spectra show strong absorption by the outer CSE layer, seen as a peak at $\sim$$-$4 \kms, the SO$_2$ spectra are approximately Gaussian and display no absorption feature. Good fits to both $^{29}$SiO and SO$_2$ spectra are accordingly obtained using the simple form
\begin{align}
  &I=I^*+I_0G(V_{z},V_{z0},\sigma_0)-A G(V_{z},-4.0,1.2) \nonumber \\ 
    \mbox{where\,}& G(V_{z},a,b)=\exp(-\sfrac{1}{2}[V_{z}-a]^2/b^2).\nonumber 
\end{align}

The first term, $I^*$, accounts for a small continuum contribution, particularly significant in the inner ring; $I_0$, $V_{z0}$ and $\sigma_0$ define the Gaussian emission and the last term accounts for the absorption by the outer layer. $A$ is set to zero in the SO$_2$ case and is such that the xabsorption be total at $V_{z}=-4.0$ \kms\ in the $^{29}$SiO case, namely $A=I^*+I_0G(V_{z},-4.0,1.2)$; the width of the absorption term is fixed with a $\sigma$ of 1.2 \kms.

Tables \ref{tabA1} and \ref{tabA2} of the Appendix list the values of the best-fit parameters.  For both lines the values of $I^*$ show a same excess of N and NW over S and SE by a factor 2 in the innermost ring, reflecting a small asymmetry of continuum emission.

\begin{figure*}
  \centering
  \includegraphics[width=4.3cm,trim=0.cm 0.5cm 1.9cm 1.5cm,clip]{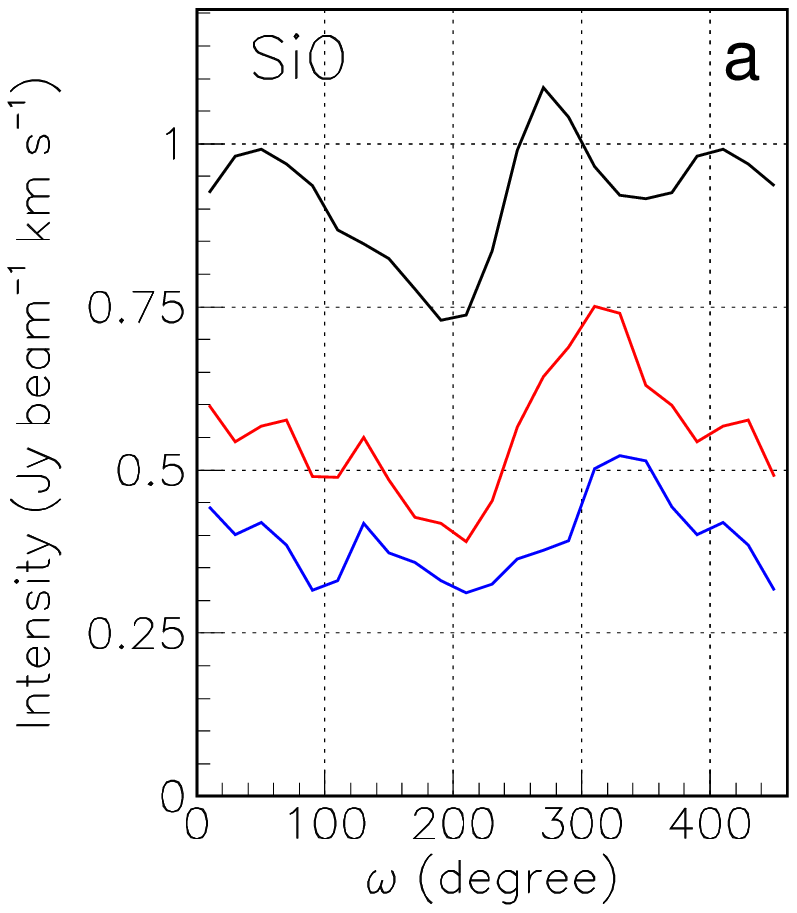}
  \includegraphics[width=4.3cm,trim=0.cm 0.5cm 1.9cm 1.5cm,clip]{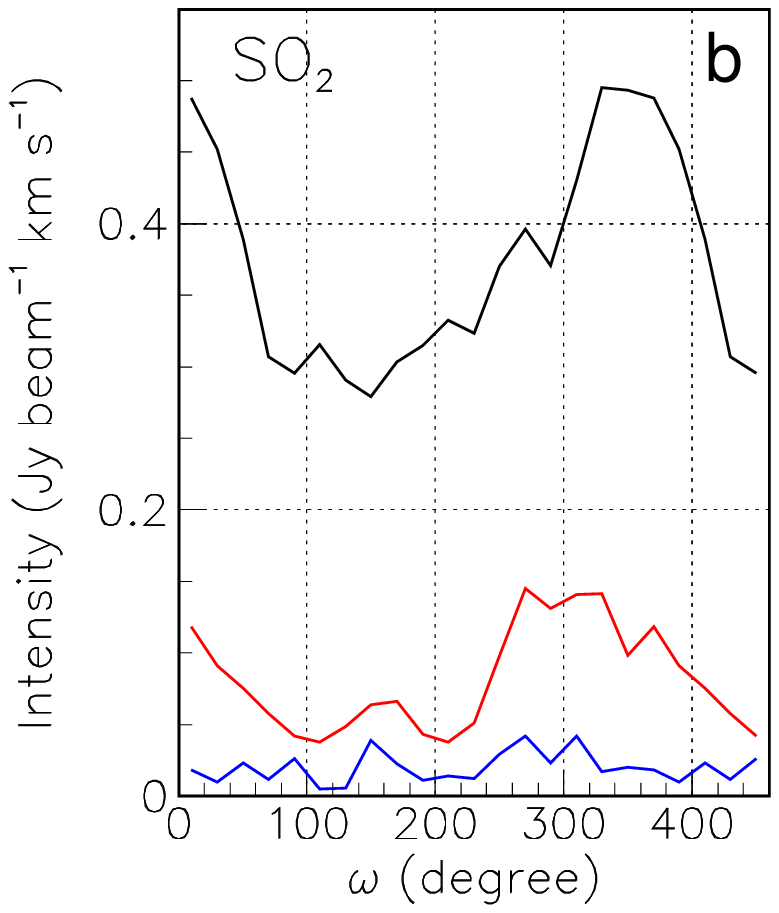}
  \includegraphics[width=4.3cm,trim=0.cm 0.5cm 1.9cm 1.5cm,clip]{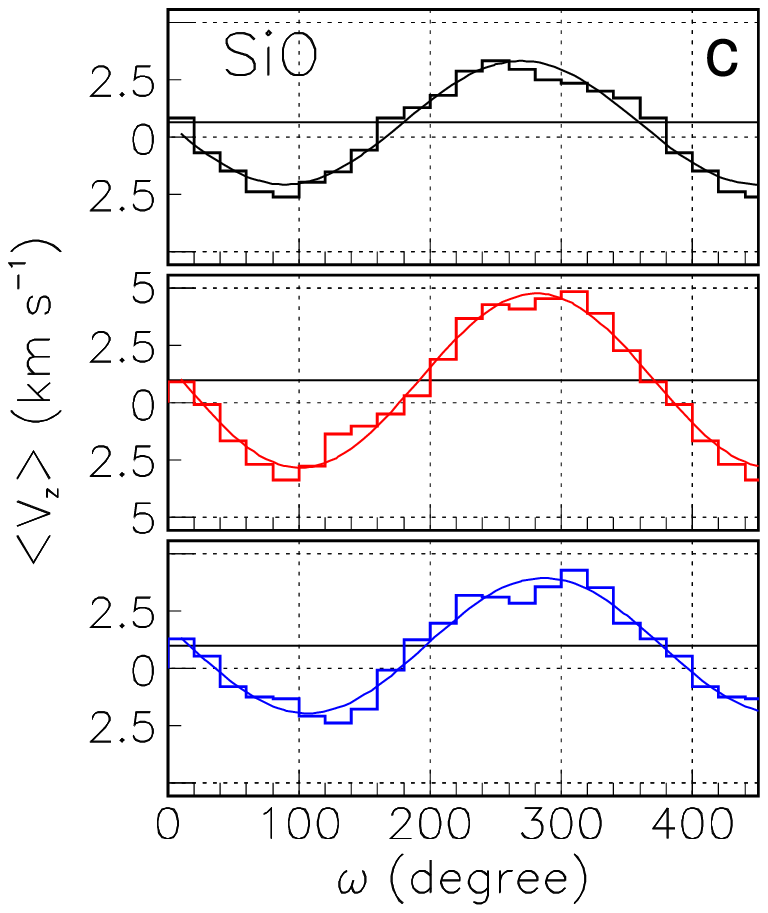}
  \includegraphics[width=4.3cm,trim=0.cm 0.5cm 1.9cm 1.5cm,clip]{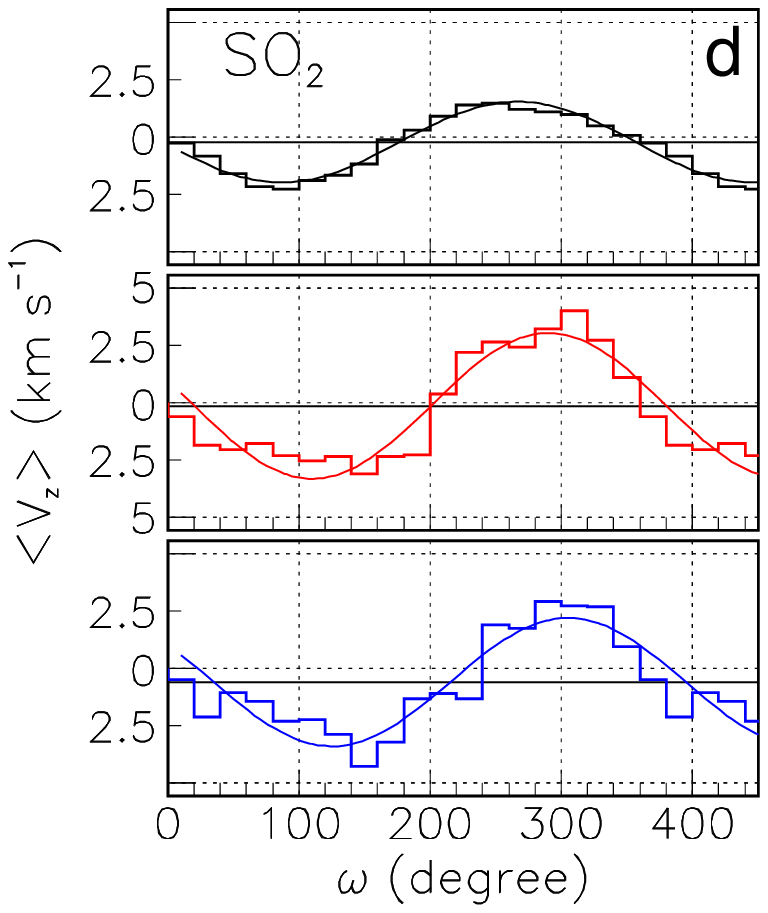}
   \caption{Dependence on position angle of the intensity (a,b) and mean Doppler velocity (c,d) averaged over rings 2 (black), 3 (red) and 4 (blue) for $^{29}$SiO (a,c) and SO$_2$ (b,d). The intensity is integrated and the velocity averaged over the interval $|V_z|$<12 \kms.}
 \label{fig8}
\end{figure*}

If the density of the gas in rotation had a significantly flattened distribution along the equator, in the form of a nearly edge-on disc as suggested by \citet{Homan2018}, the intensity projected on the sky plane should display significant ellipticity: its distribution in each ring as a function of position angle $\omega$ should take the form of a $1-\zeta\cos(2\omega)$ term with $\zeta$ providing a measure of the flattening and the distribution reaching its two maxima near the E and W directions and its two minima near the S and N directions. Figure \ref{fig8}a,b  displays these distributions; there is no evidence for such a quadripolar term but instead dipole sine waves, of the form $A+B\sin(\omega-\omega_0)$ are seen to give reasonable fits. They reach their maxima in the NW direction with $\omega_0$$\sim$330\dego\ and the amplitudes of the oscillation are $\sim$20\% of the average emission level. The absence of a quadripolar term sheds doubts on the previously suggested evidence for the presence of an edge-on disc. Figure \ref{fig8}c,d displays the distributions of the mean Doppler velocity; sine waves, as expected for rotation, are seen to give good fits (Table \ref{tab2}). They reach their maxima in the W direction with $\omega_0$$\sim$190\dego\ and the amplitude of the oscillations reaches $\sim$4 \kms. From the east-west asymmetry displayed by the values of $V_{z0}$, one can see that the rotation reaches a maximal velocity within the displayed range of $R$, namely below 0.15 arcsec (9 au); beyond the outer ring the blue stream starts contributing significantly to the mean Doppler velocity. A detailed discussion of these observations is presented in Section 4. A clear enhancement of the line width is seen in the inner ring for both $^{29}$SiO and SO$_2$.

\begin{table}
  \caption{Parameters of the sine wave fits of the form $A+B\sin(\omega-\omega_0)$ to the mean Doppler velocity distributions displayed in Figure \ref{fig8}. A cut of 3$\sigma$ has been applied to the data; the intensity is integrated and the Doppler velocity averaged over the |$V_z$|<12 \kms\ interval. }
  \centering
  \label{tab2}
  \begin{tabular}{ccccc}
  \hline
  
  Line& $R$ range (mas)&\multicolumn{3}{c}{Mean Doppler velocity}\\
  &&$A$&$B$&$\omega_0$ (\dego)\\
  \hline
\multirow{3}{*}{$^{29}$SiO}&50-100&0.6&2.7&179\\
 
&100-150&
1.0&
3.8&
191\\

&150-200&
1.0&
3.0&
196\\
\hline
\multirow{3}{*}{SO$_2$}&
50-100&
$-$0.2&
1.8&
177\\

&100-150&
$-$0.1&
3.2&
200\\

&150-200&
$-$0.6&
2.8&
215\\
\hline
\end{tabular}
\end{table}

\begin{figure*}
   \includegraphics[height=5.4cm,trim=.2cm 1.4cm 0.4cm 0.3cm,clip]{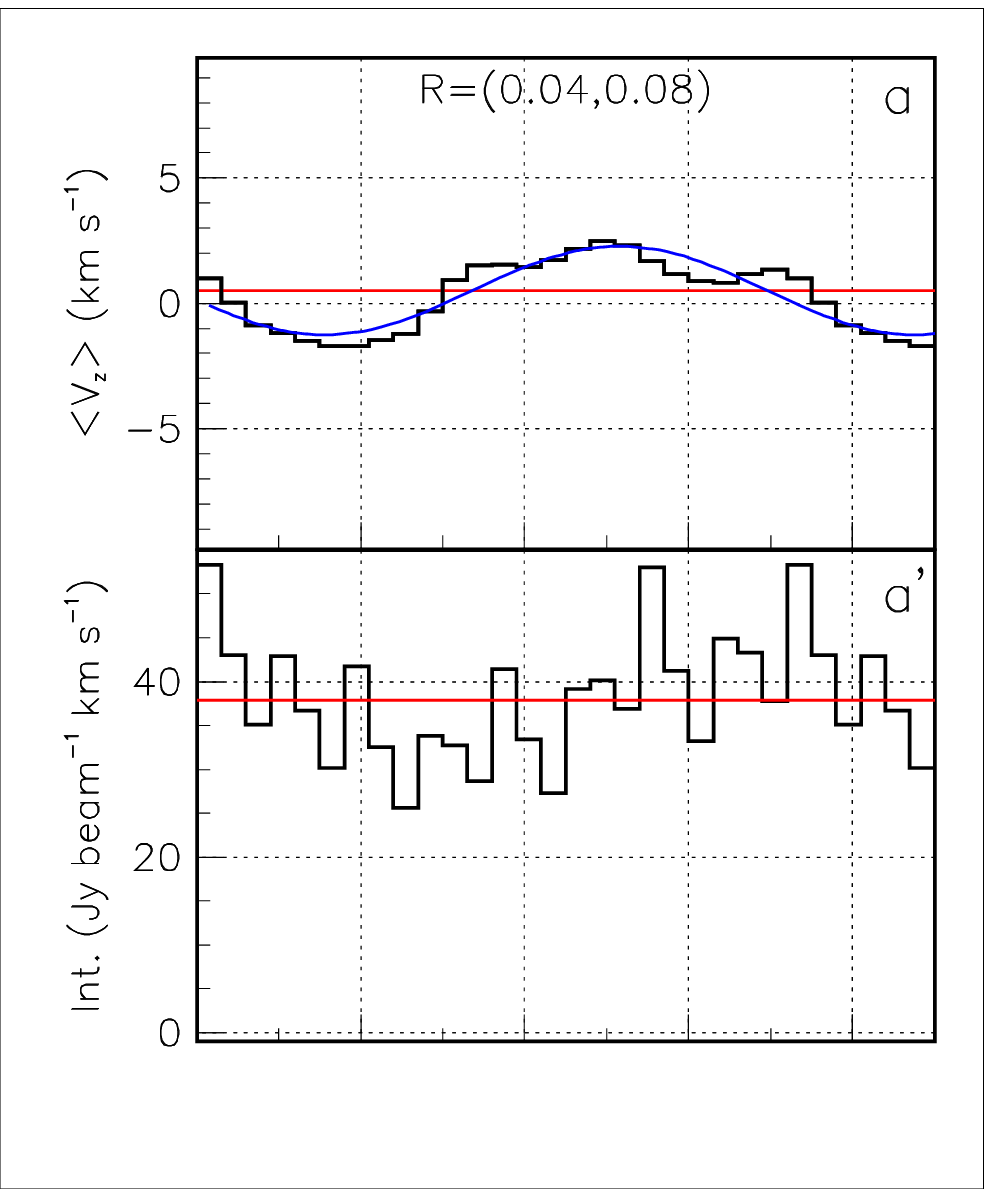}
   \includegraphics[height=5.4cm,trim=1.9cm 1.4cm 0.4cm 0.3cm,clip]{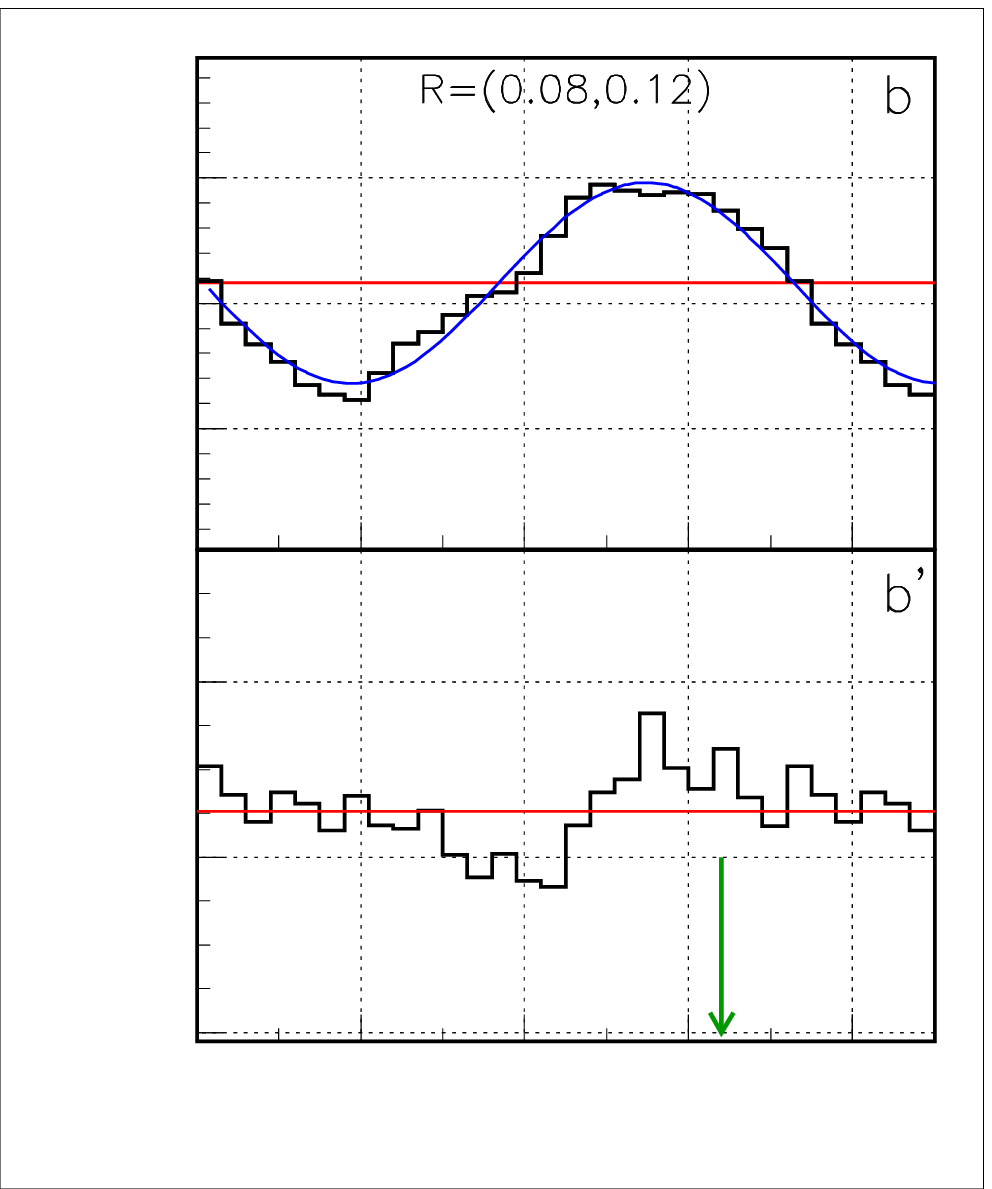}
   \includegraphics[height=5.4cm,trim=1.9cm 1.4cm 0.4cm 0.3cm,clip]{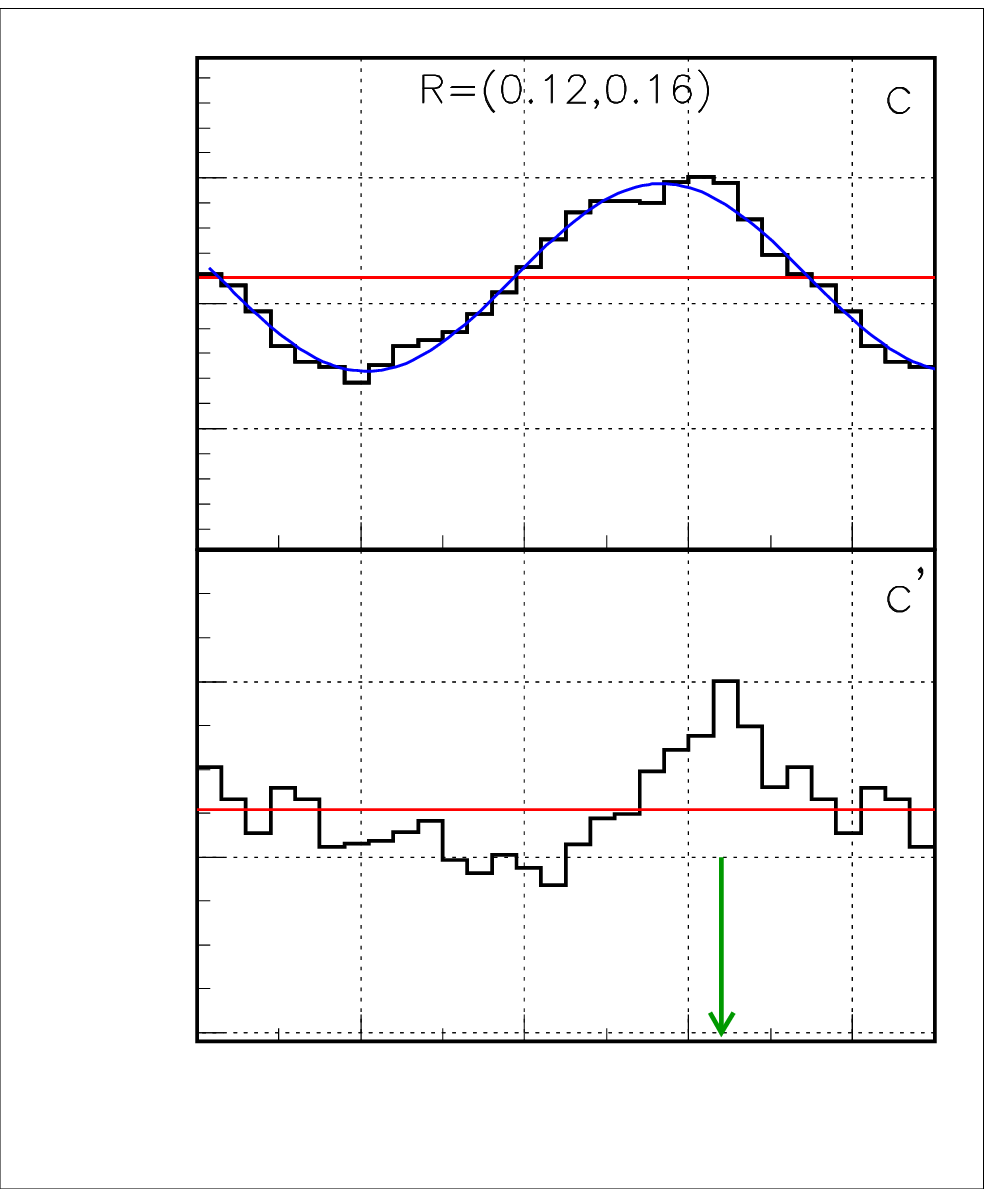}
   \includegraphics[height=5.4cm,trim=1.9cm 1.4cm 0.4cm 0.3cm,clip]{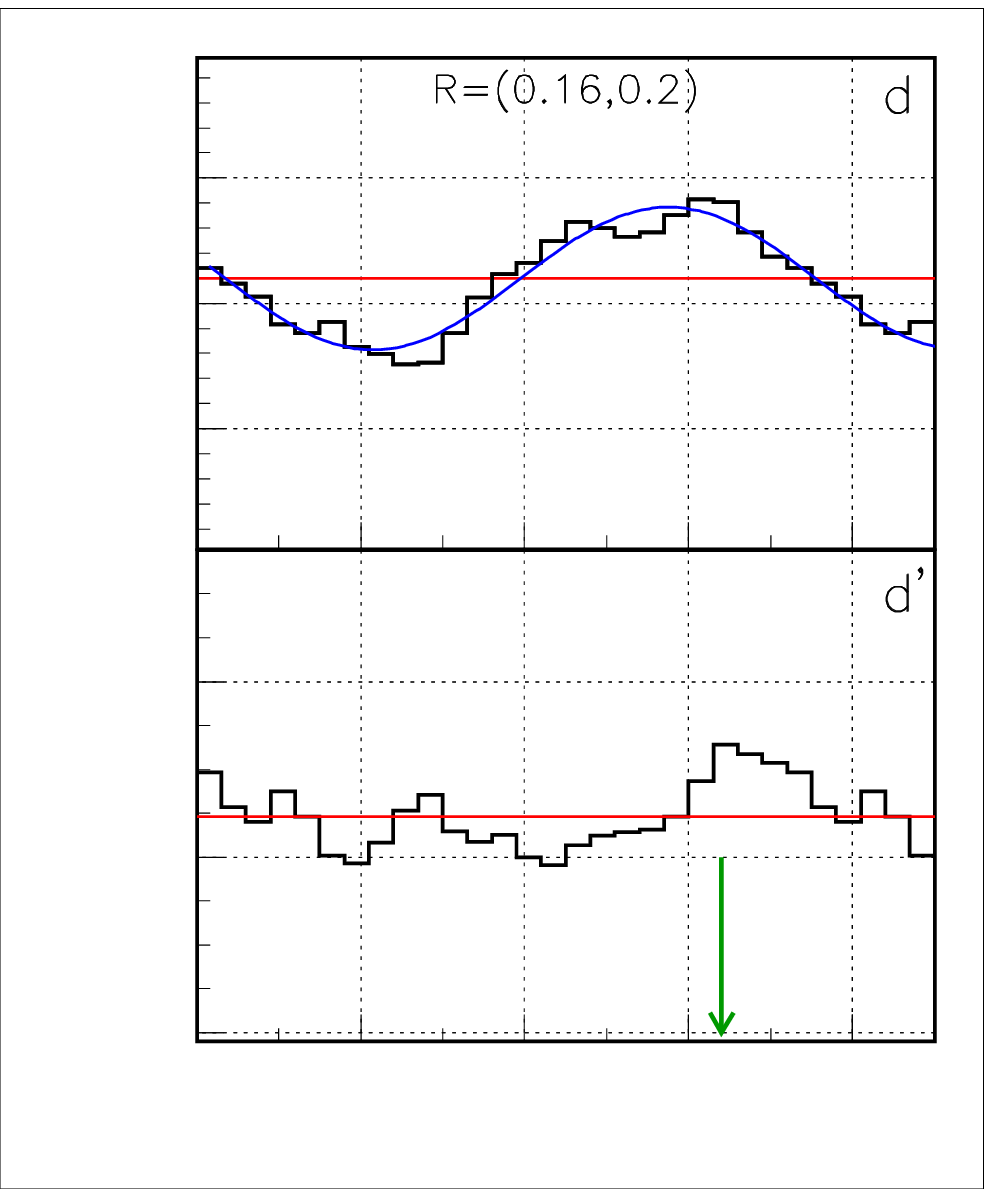}
   \includegraphics[height=5.4cm,trim=.2cm 1.4cm 0.4cm 0.3cm,clip]{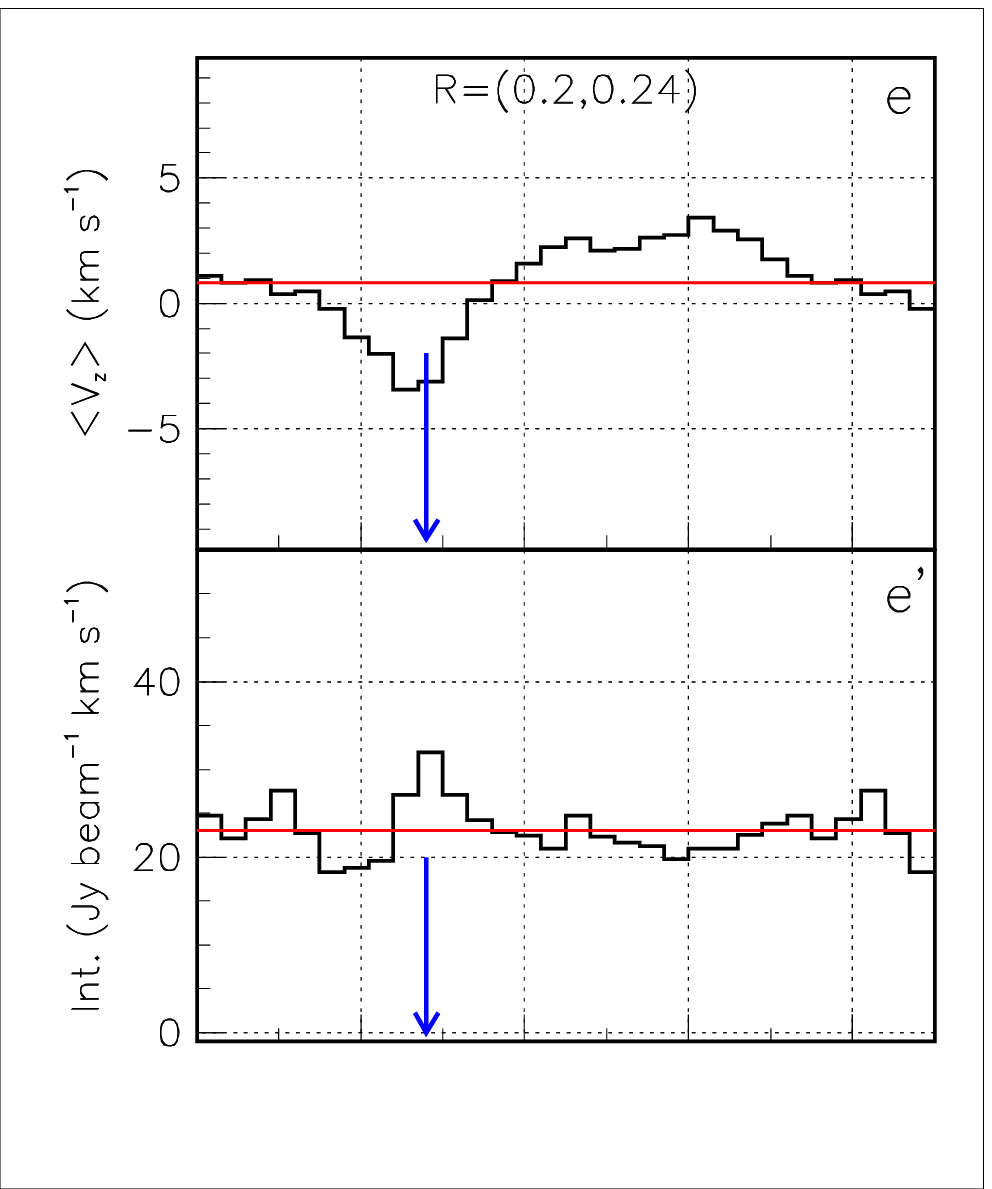}
   \includegraphics[height=5.4cm,trim=1.9cm 1.4cm 0.4cm 0.3cm,clip]{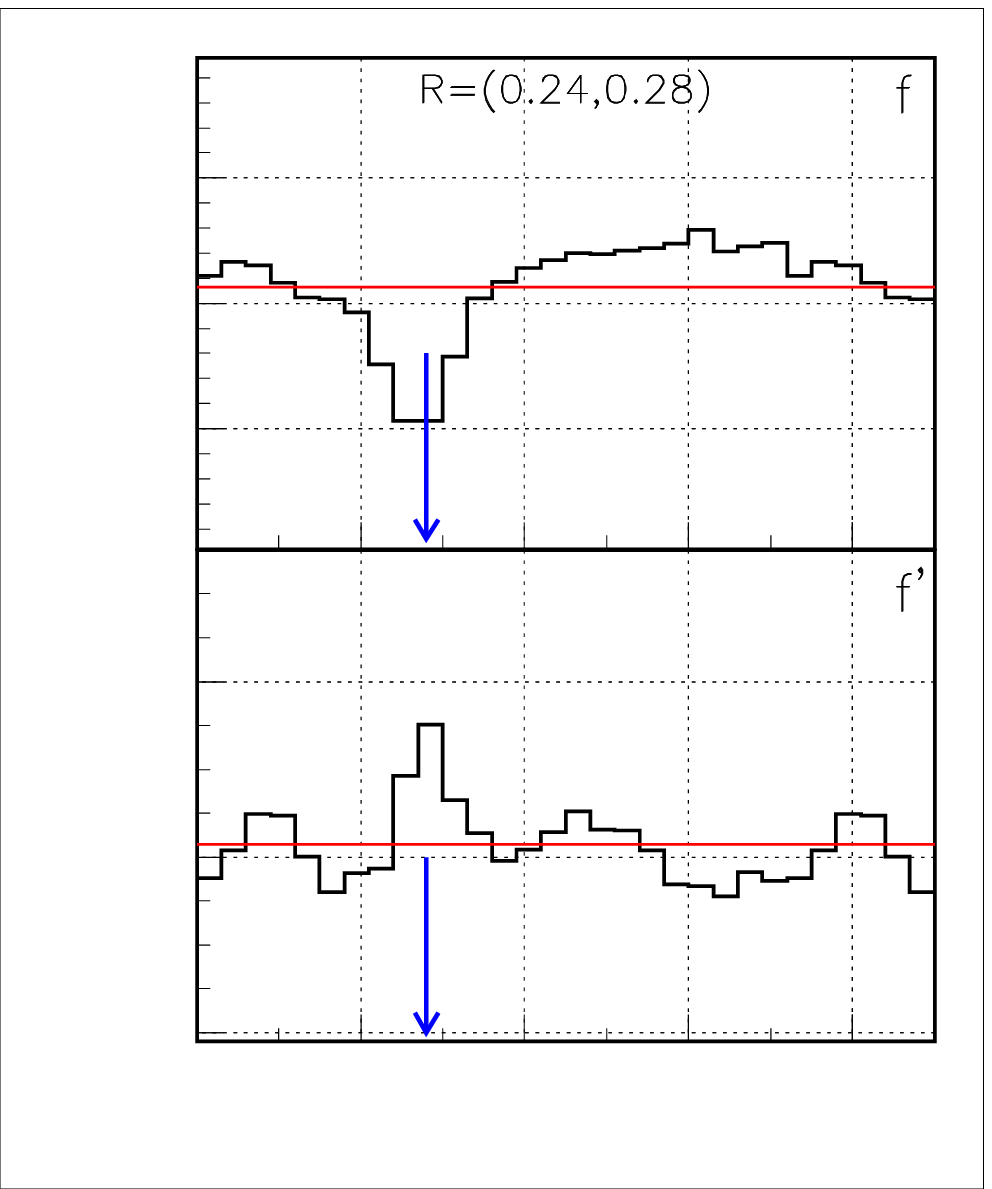}
   \includegraphics[height=5.4cm,trim=1.9cm 1.4cm 0.4cm 0.3cm,clip]{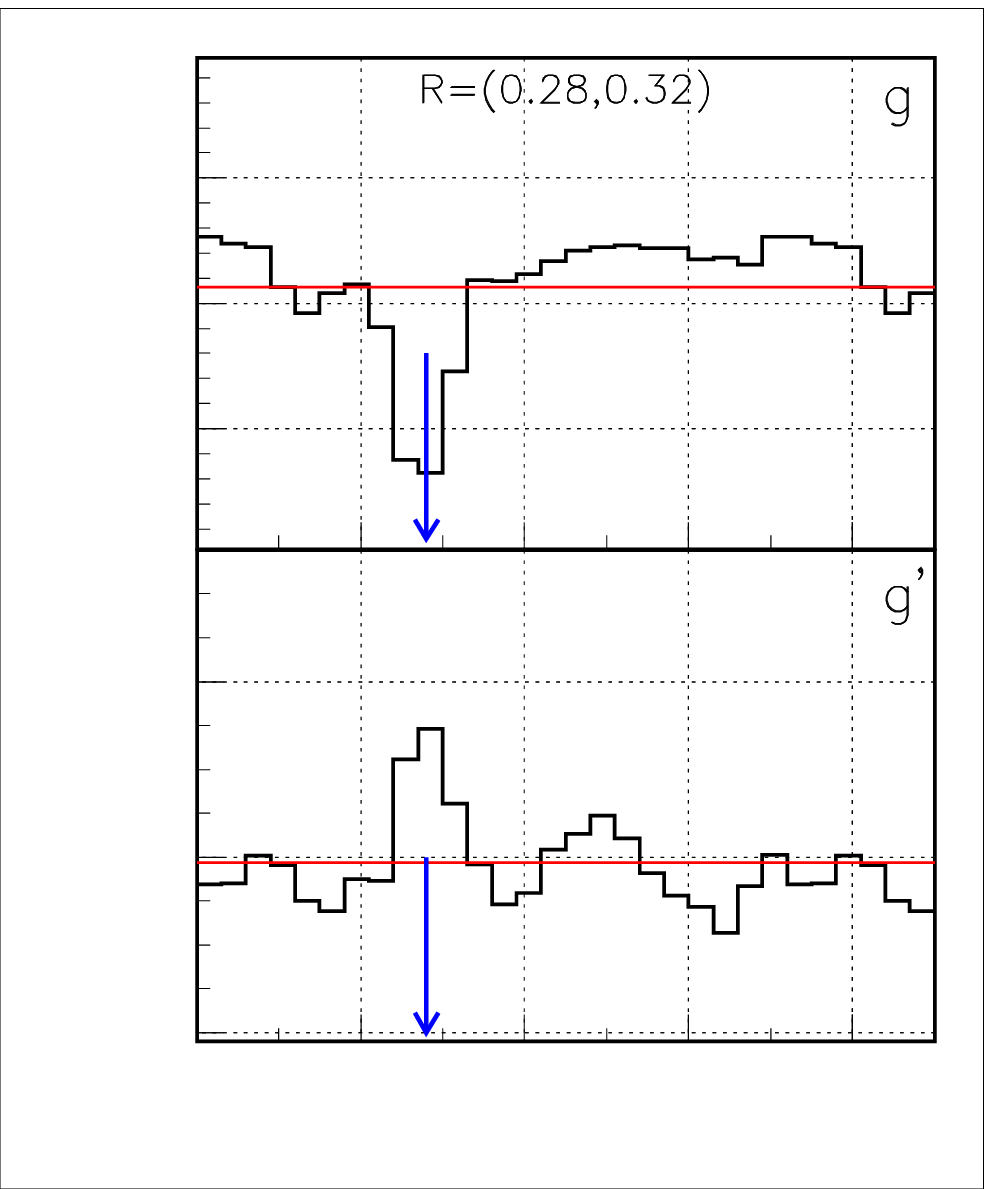}
   \includegraphics[height=5.4cm,trim=1.9cm 1.4cm 0.4cm 0.3cm,clip]{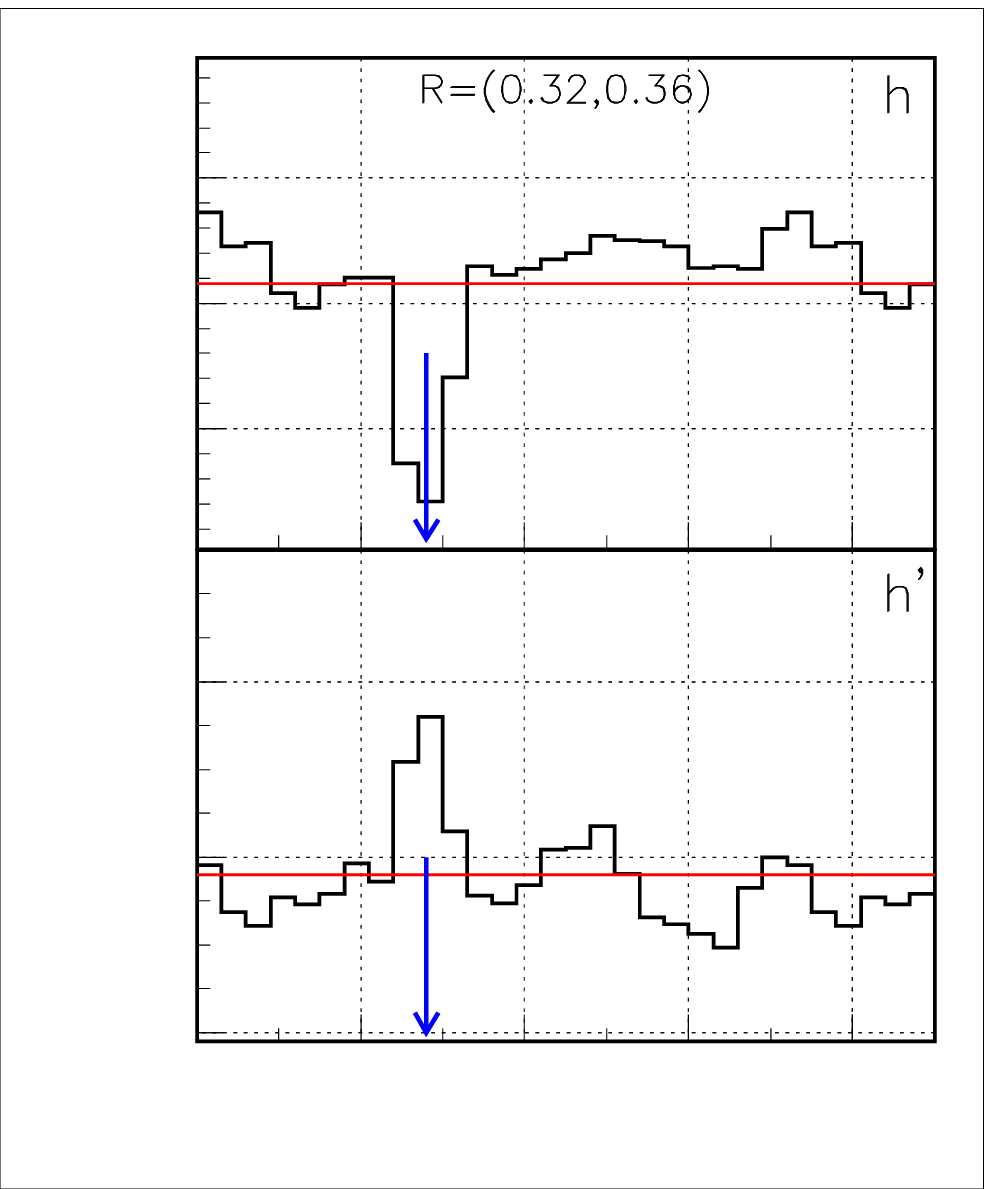}
   
   \includegraphics[height=5.611cm,trim=.2cm 1.cm 0.4cm 0.3cm,clip]{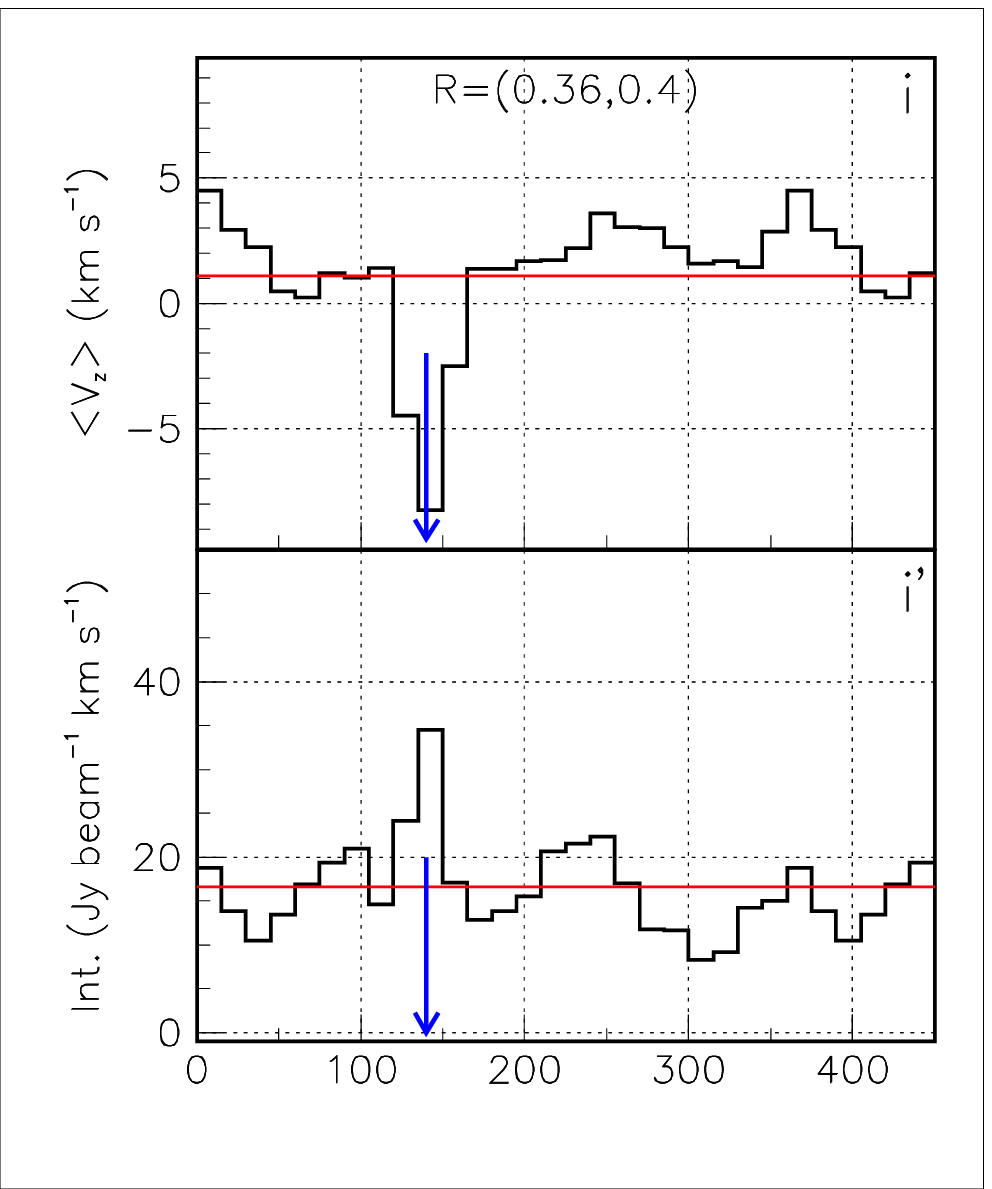}
   \includegraphics[height=5.611cm,trim=1.9cm 1.cm 0.4cm 0.3cm,clip]{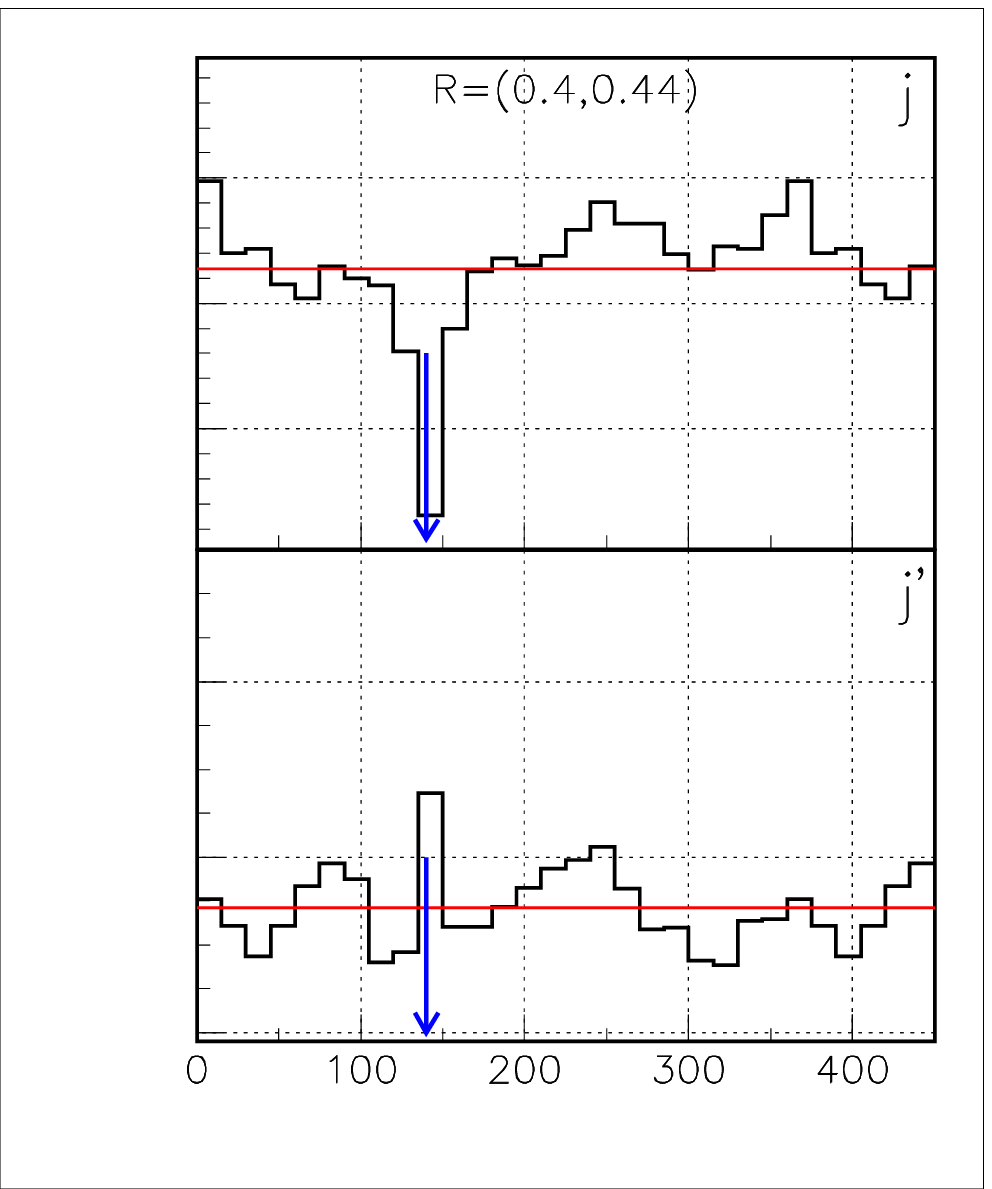}
   \includegraphics[height=5.611cm,trim=1.9cm 1.cm 0.4cm 0.3cm,clip]{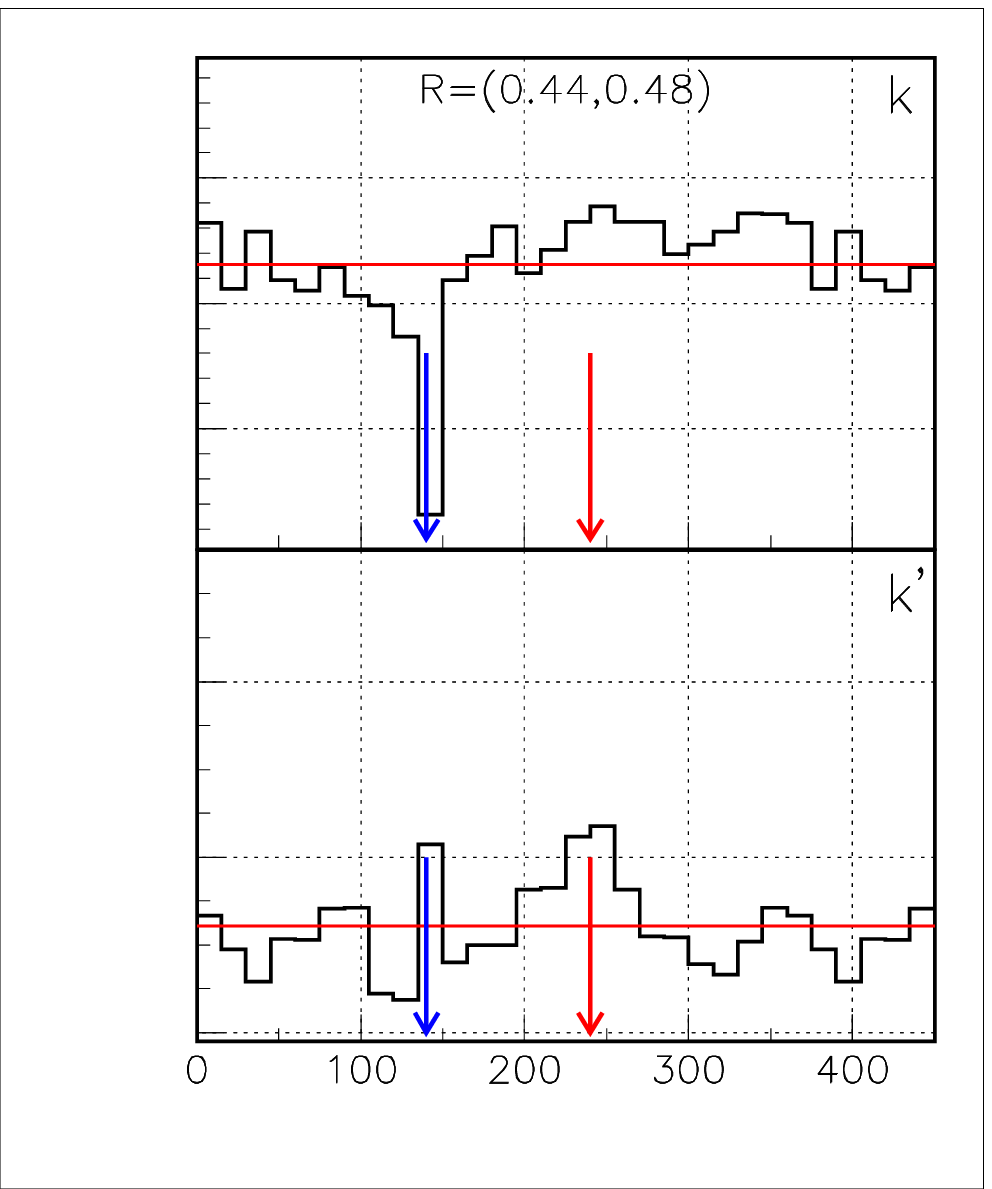}
   \includegraphics[height=5.611cm,trim=1.9cm 1.cm 0.4cm 0.3cm,clip]{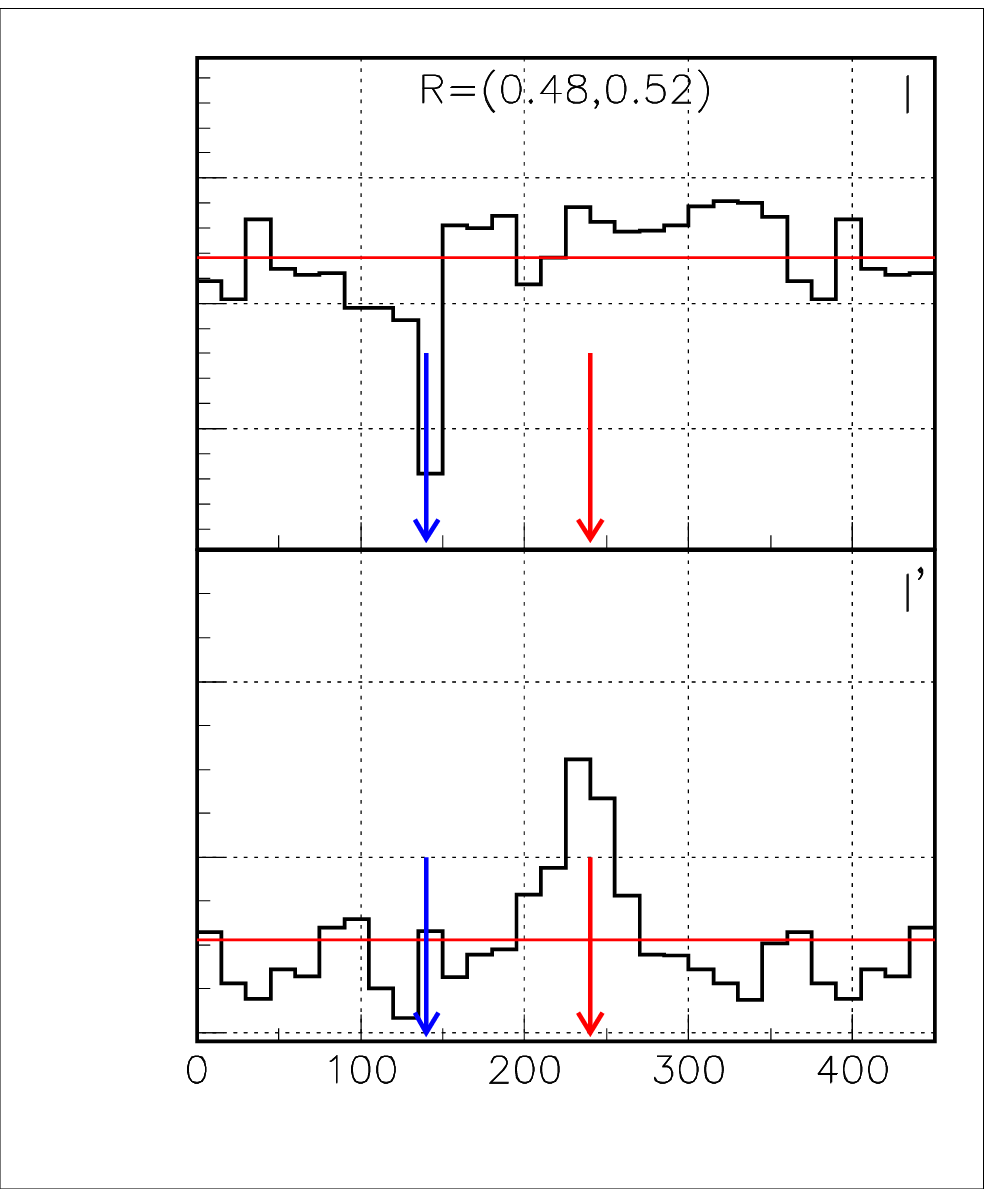}
   
   \includegraphics[height=6.13cm,trim=.2cm .1cm 0.4cm 0.3cm,clip]{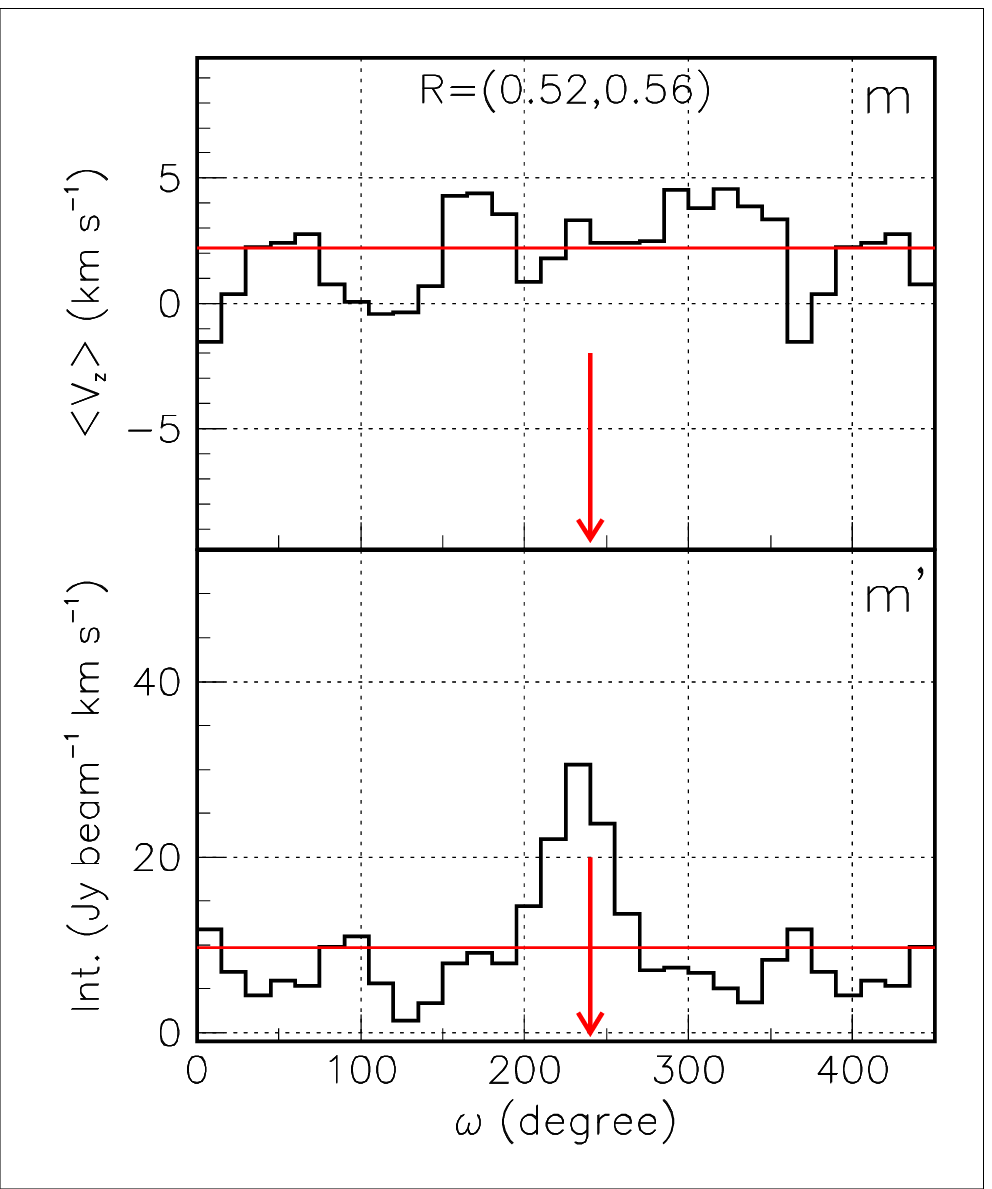}
   \includegraphics[height=6.13cm,trim=1.9cm .1cm 0.4cm 0.3cm,clip]{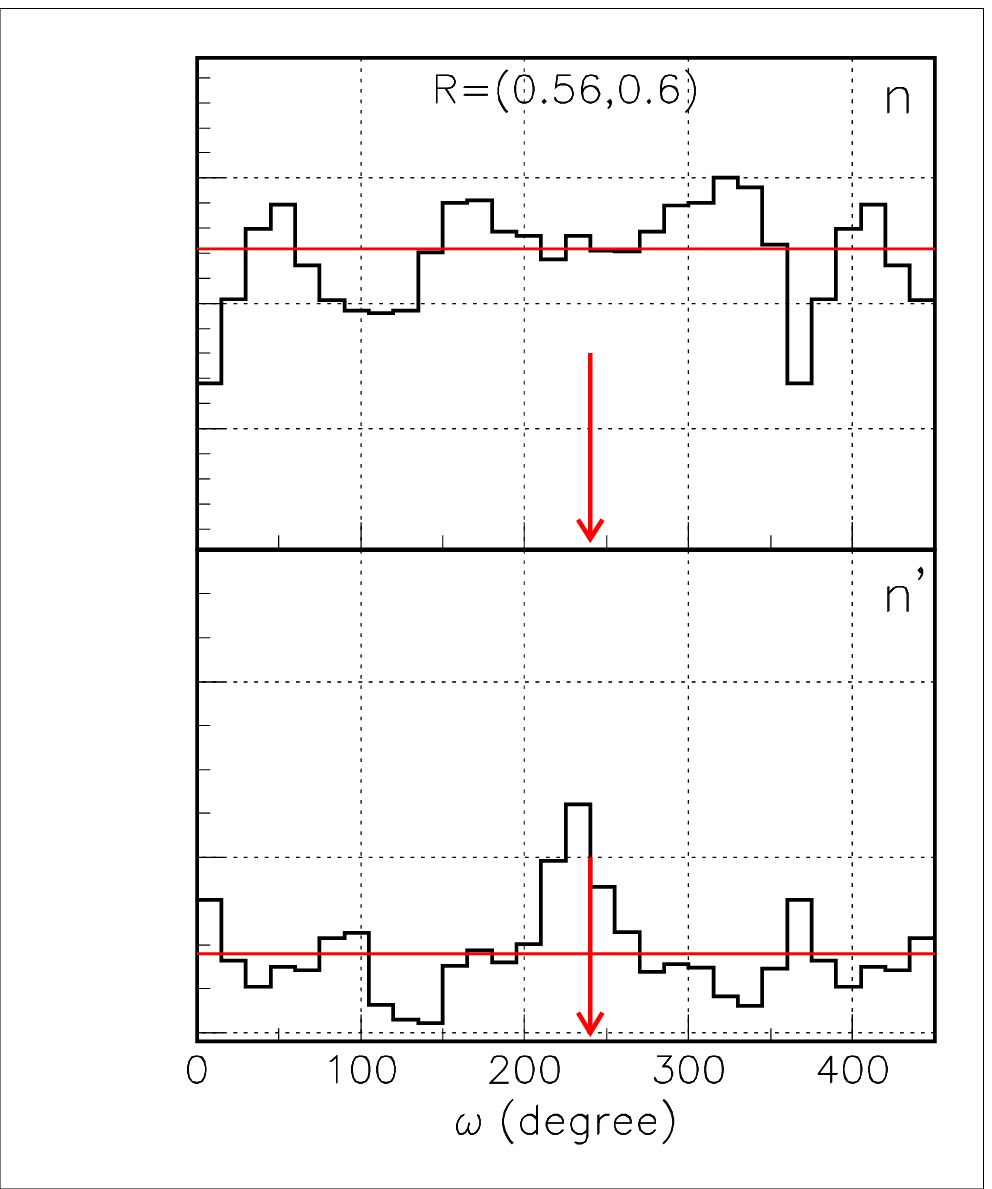}
   \caption{$^{29}$SiO emission. Dependence on position angle of the mean Doppler velocity (a to n) and of the intensity (a' to n') in successive rings (as labelled above each velocity panel). The red line shows the mean value, the blue arrow is at 140\dego, the green arrow at 320\dego\ and the red arrow at 240\dego. The blue lines are sine wave fits.}
 \label{fig9}
\end{figure*}
 
Figure \ref{fig9} displays the dependence on position angle of the $^{29}$SiO intensity and mean Doppler velocity in successive rings, each 0.04 arcsec (2.4 au) wide. It complements the channel maps displayed in the Appendix and covers not only the inner layer, with the confinement of rotation to projected distances from the star not exceeding 15 au, but also larger distances studied in the following sections, with the appearance of the blue stream at $R$$\sim$0.2 arcsec (12 au) and of the south-western outflow identified earlier in Paper I at $R$$\sim$0.45 arcsec (27 au). 

\subsection{Absorption over the stellar disc}

\begin{figure*}
  \centering
  \includegraphics[width=4.4cm,trim=0.cm 0.5cm 1.9cm 0.5cm,clip]{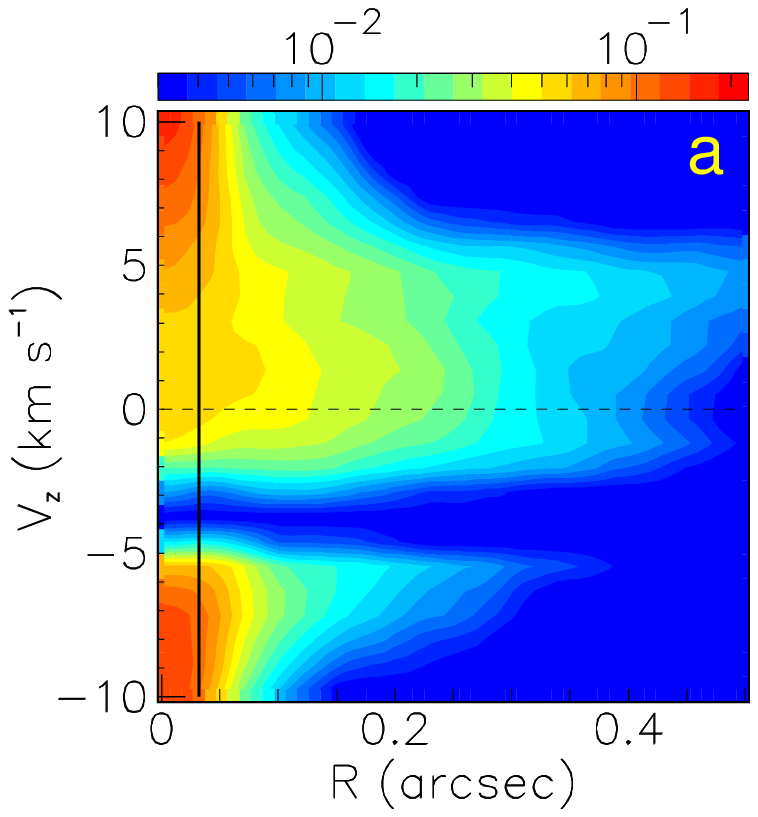}
  \includegraphics[width=4.4cm,trim=0.cm 0.5cm 1.9cm 0.5cm,clip]{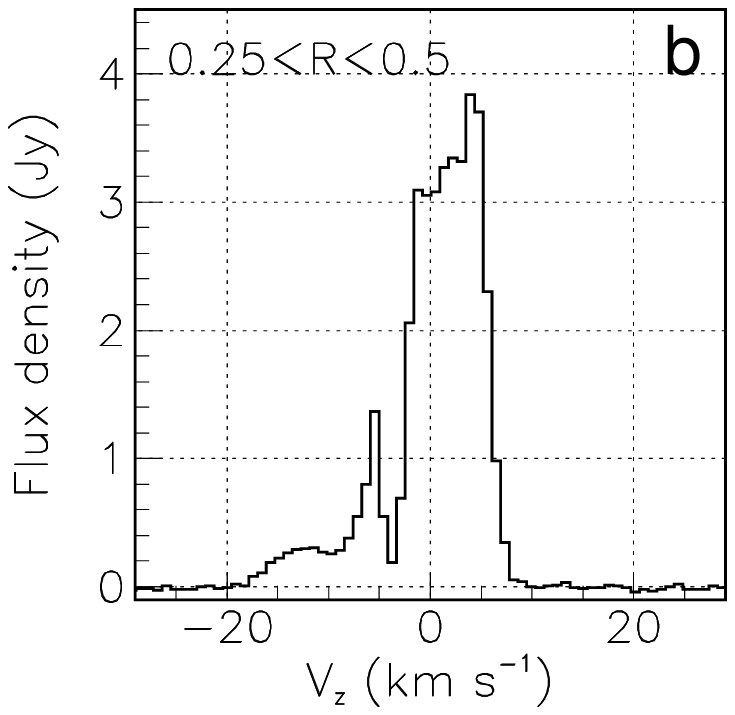}
  \includegraphics[width=4.4cm,trim=0.cm 0.5cm 1.9cm 0.5cm,clip]{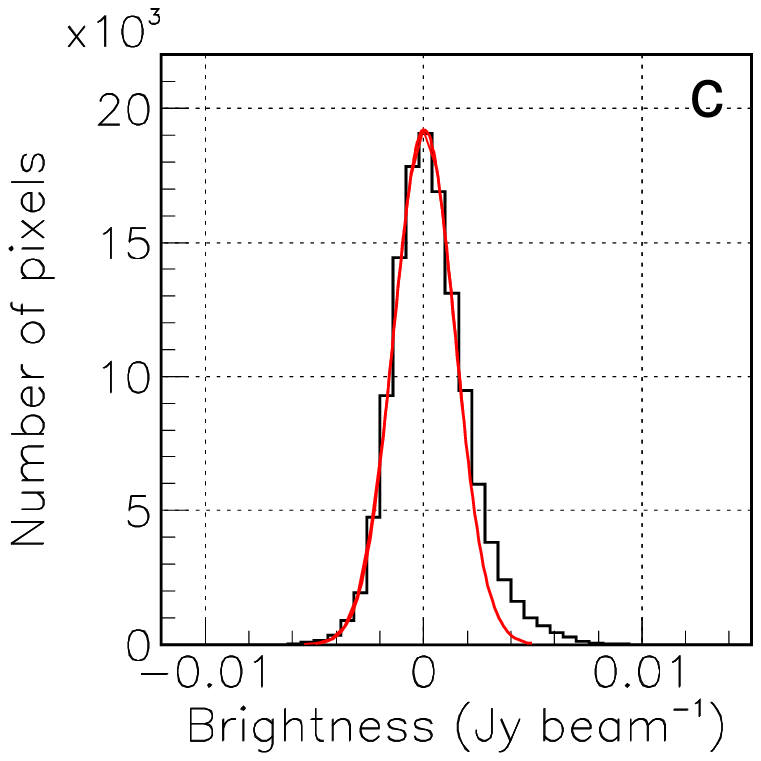}
  \includegraphics[width=4.4cm,trim=0.cm 0.5cm 1.9cm 0.5cm,clip]{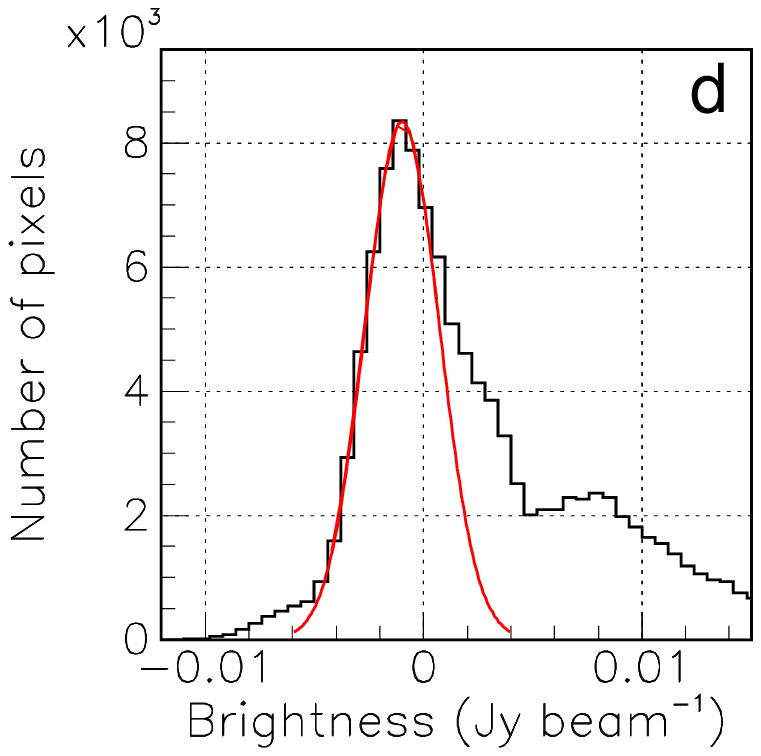}
   \caption{Absorption of the $^{29}$SiO emission. (a): PV map showing the  brightness (in Jy beam$^{-1}$) averaged over position angle as a function of Doppler velocity, $V_z$, vs projected distance from the star, $R$. The vertical line shows the size of the stellar disc. (b): Doppler velocity spectrum integrated in the ring 0.25<$R$<0.5 arcsec. (c) and (d): brightness distributions integrated over $R$<1 arcsec in the $V_z$=$-$4 \kms\ (c) and $V_z$=+4 \kms\ (d) frequency channels.}
 \label{fig10}
\end{figure*}

In the preceding section (Figure \ref{fig5} left), we noted the presence of a nearly total absorption peak at $\sim$ $-$4 \kms\ in the $^{29}$SiO data, extending to large values of $R$ as clearly illustrated in Figure \ref{fig10}: the absorption is seen to extend well beyond $R$$\sim$15 au, far from the stellar disc (panels a and b). This is also clearly apparent on the channel maps (Figure \ref{figA1}). Comparison between the brightness distributions measured in the $V_{z}$=$-$4 \kms\ and $V_{z}$=$+$4 \kms\ Doppler velocity channels shows that indeed the former detects only very little signal above noise (panels c and d). Both features, the nearly total absorption over the stellar disc and its extension beyond it, are surprising. We discuss the former below but postpone a discussion of the latter to Section 6 because it requires analyses of line emissions that explore larger distances from the star, up to 100 au or so.

Quantitatively, we can safely place a 3-$\sigma$ upper limit of 4.2 mJy\,beam$^{-1}$ (10 mJy over the stellar disc) on the brightness measured at maximal absorption, $V_{z}$$\sim$$-$4 \kms, over the stellar disc ($R$<0.035 arcsec). While local thermal equilibrium (LTE) is not expected to provide a reliable description of the CSE, in particular in the close neighbourhood of the star, its prediction provides a useful reference with which to compare observations. For a given velocity $V_{z}$ and a brightness $I_0$ [Jy arcsec$^{-2}$] on the entrance side of a slab of thickness d$z$ [arcsec], density $n$ [molecules\,cm$^{-3}$] and temperature $T$ [K], the brightness $I$ [Jy arcsec$^{-2}$] at the exit side is simply the sum $I=I_0+\varepsilon$ , where $\varepsilon$ is the emission of the slab, when absorption is neglected. The effect of absorption, described by the optical depth $\tau$, is to have instead
\begin{equation}
  I=I_0e^{-\tau}+\varepsilon(1-e^{-\tau})/\tau=\varepsilon/\tau+(I_0-\varepsilon/\tau)e^{-\tau}
\end{equation}
where $\varepsilon$ is given by Relation 2 below:
\begin{equation}
  \begin{split}
    \varepsilon&=hf/(4\pi \Delta{f})n A_{\rm{ji}}f_{\rm{pop}}{\rm{d}}z \\
    &=0.33\,10^6 N A_{\rm{ji}} f_{\rm{pop}}/\Delta{V},\,f_{\rm{pop}}=(2J + 1)e^{-E_{\rm{u}}/T}/T
  \end{split}
\end{equation}
 with $J$ and $E_u$ being the angular momentum quantum number and energy of the upper level, $N$ the column density in molecules\,cm$^{-3}$\,arcsec, $A_{\rm{ji}}$ the Einstein coefficient, $\Delta{V}$ the FWHM of the profile measured in \kms\\
and where $\tau$ is given by Relation 3 below:
 \begin{equation}
   \begin{split}
     \tau&=c^2/(8\pi f^2\Delta{f})(e^{\Delta{E}/T}-1)n A_{\rm{ji}}f_{\rm{pop}}{\rm{d}}z\\
     &=2.9\,10^7\varepsilon(e^{\Delta{E}/T}-1)/f^3 (f=343\,\mbox{GHz})
     \end{split}
  \end{equation}

  When the column density $N$=$n\times$d$z$ increases, $I$ varies from $I_0$ to $\varepsilon/\tau$. In case of absorption, $I_0$>$\varepsilon/\tau$, $I$ is always larger than $\varepsilon/\tau$, a lower limit to the emission of an optically thick shell. In the present case, where the energy of the transition, $\Delta{E}$=16.4 K, can be expected to be much smaller than the temperature $T$, we can approximate $\varepsilon/\tau$ [Jy\,arcsec$^{-2}$]$=$$T$ [K]/11.8. Integrating over the beam area, $\sim$1.6\,10$^{-3}$ arcsec$^2$, we obtain a lower limit for the brightness of 1.6\,$T$/11.8 mJy\,beam$^{-1}$ to be compared with the 3-$\sigma$ limit of 4.2 mJy\,beam$^{-1}$, meaning an upper limit of 11.8$\times$4.2/1.6$\sim$30 K on $T$. Such a low value seems unlikelyat such short distance from the star. The assumptions made are very crude but the fact remains that however large absorption may be, the outer part of the layer cannot be prevented from shining. This result is independent from the value of $\varepsilon$, namely of the column density. We note that in the case of Mira A \citep{Wong2016} the brightness measured at maximal absorption exceeds 20\% of the continuum level, compared with <3\% in the present case.

The above argument applies in particular to the outer layer, which is expected to host less turbulence, shocks and condensing dust than the inner layer does and where Doppler velocities can be assumed to stay within the interval $-$10<$V_{z}$<0 \kms. Such assumption implies that the absorption observed outside this interval is confined to the inner layer. We accordingly define three values of the flux density over the stellar disc at $V_{z}$$\sim$$-$4 \kms: inflowing, $I_{\rm{in}}$$\sim$0.34 Jy, intermediate, $I_{\rm{mid}}$$\sim$0.17 Jy measured as the mean between the values of the brightness measured at $V_{z}$=$-$10 \kms\ and $V_{z}$=0 (see Figure \ref{fig5}-left) and outflowing, $I_{\rm{out}}$<0.01 Jy. At $V_{z}$$\sim$ $-$4 \kms\ the flux density over the stellar disc decreases therefore from 0.34 Jy near the star to 0.17 Jy outside the inner layer and <0.01 Jy outside the outer layer.

We first calculate $\varepsilon$ and $\tau$ at a temperature $T$$\sim$270 K using Relations 2 and 3 for $^{29}$SiO. We use a column density $N$ [molecules cm$^{-3}$ arcsec]=0.7 obtained by integration from $r$=0.8 to 1.0 arcsec for a $^{28}$SiO/H$_2$ abundance of 5.5 10$^{-5}$ \citep{VandeSande2018}, an isotopic ratio $^{29}$Si/$^{28}$Si=8\% \citep{DeBeck2018} and an H$_2$ density of the form 1.6\,10$^{7}$ $(r/0.2)^{-2}$ [molecules\,cm$^{-3}$], meaning a mass-loss rate of $\sim$1.6\,10$^{-7}$ \msun\,yr$^{-1}$. Writing $J$=8, $E_{\rm{u}}$=74 K, $A_{\rm{ji}}$=2.1\,10$^{-3}$ s$^{-1}$, we obtain $\varepsilon$=8.2\,10$^3e^{-74/T}/(T\Delta{V})$ and $\tau$=9.7\,10$^4e^{-74/T}/(T^2\Delta{V})$. Namely, at $T$=270 K, $\varepsilon$=8.2 Jy arcsec$^{-2}$ and $\tau$=0.36 for $\Delta{V}$=2.8 \kms: the self-absorbed emission is $\sim$27 mJy over the stellar disc, nearly 3 times the upper limit placed on the observed emission. Repeating the same calculation for 0.2<$r$<0.4 arcsec gives $\varepsilon$=46 Jy arcsec$^{-2}$ and $\tau$=1.0, respectively, the self-absorbed emission being $\sim$117 mJy over the stellar disc. For the whole range, 0.2<$r$<1 arcsec, and $I_{\rm{mid}}$=0.17 Jy we find $I_{\rm{out}}$=152 mJy over the stellar disc, over an order of magnitude above the observed upper limit of 10 mJy.
  
  We can repeat the above argument in the case of SO$_2$ emission. The abundance relative to H$_2$ is now of the form 5\,10$^{-6}\exp[-(r/1.78)^2]$ \citep{Danilovich2016, Danilovich2020}, $E_{\rm{u}}$=581.92 K and $A_{\rm{ji}}$=3.45\,10$^{-4}$ s$^{-1}$, but the relation $\tau$$\sim$11.8$\varepsilon/T$ still applies. We estimate $I_{\rm{mid}}$=0.325 Jy and $I_{\rm{out}}$=0.27 Jy. For the whole range, 0.2<$r$<1 arcsec, we find $I_{\rm{out}}$=0.32 Jy. 

In summary, for $^{29}$SiO, below $\sim$12 au and at $V_z$$\sim$$-$4 \kms, the inner layer absorbs $\sim$40\% of the continuum emission ($\sim$60\% at $V_z$=0) and is expected to host shocks and turbulence. We study it in Section 4. For SO$_2$, at $V_z$$\sim$$-$4 \kms, the inner layer absorbs only 2\% of the continuum emission and the outer layer absorbs 17\% of the inner layer emission.  Beyond $\sim$12 au and at $V_z$$\sim$$-$4 \kms, we detect no emission for $^{29}$SiO, implying that the outer layer not only absorbs completely the emission of the inner layer but also its own emission, which we discuss in Section 6. We have shown that a simple LTE model cannot produce such a result if the temperature takes values as expected in the literature and predicts an emission an order of magnitude larger than observed. This result suggests that such absorption is more likely to happen at large radii, say above 40 au, where the temperature is expected to be closer to 200 K \citep{VandeSande2018} and where an enhancement of emission has been observed (Paper I). This is further discussed in Section 6.

\subsection{The blue stream}

\begin{figure*}
  \centering
  \includegraphics[height=4.8cm,trim=0.cm 0.5cm 0.cm 1.8cm,clip]{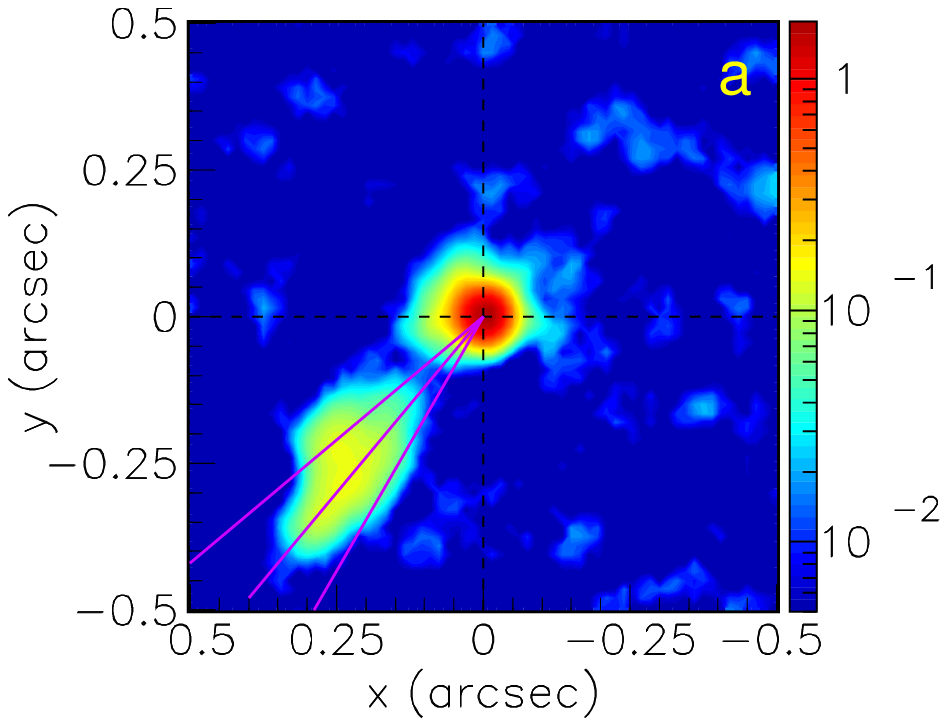}
  \includegraphics[height=4.8cm,trim=0.cm 0.5cm 0.cm 1.8cm,clip]{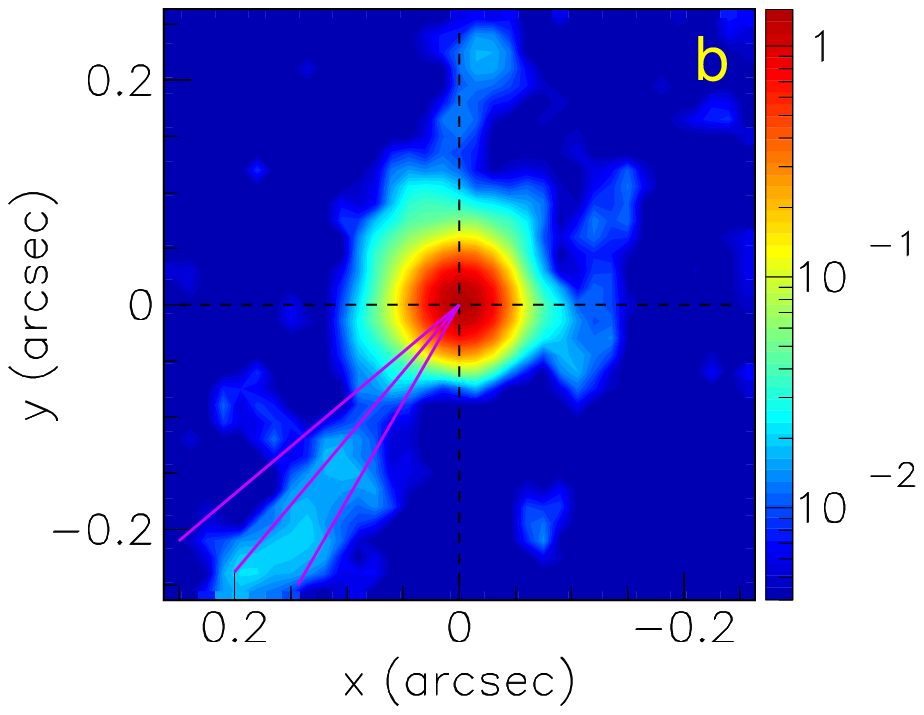}
  \includegraphics[height=4.8cm,trim=0.cm 0.2cm 1.8cm 2.6cm,clip]{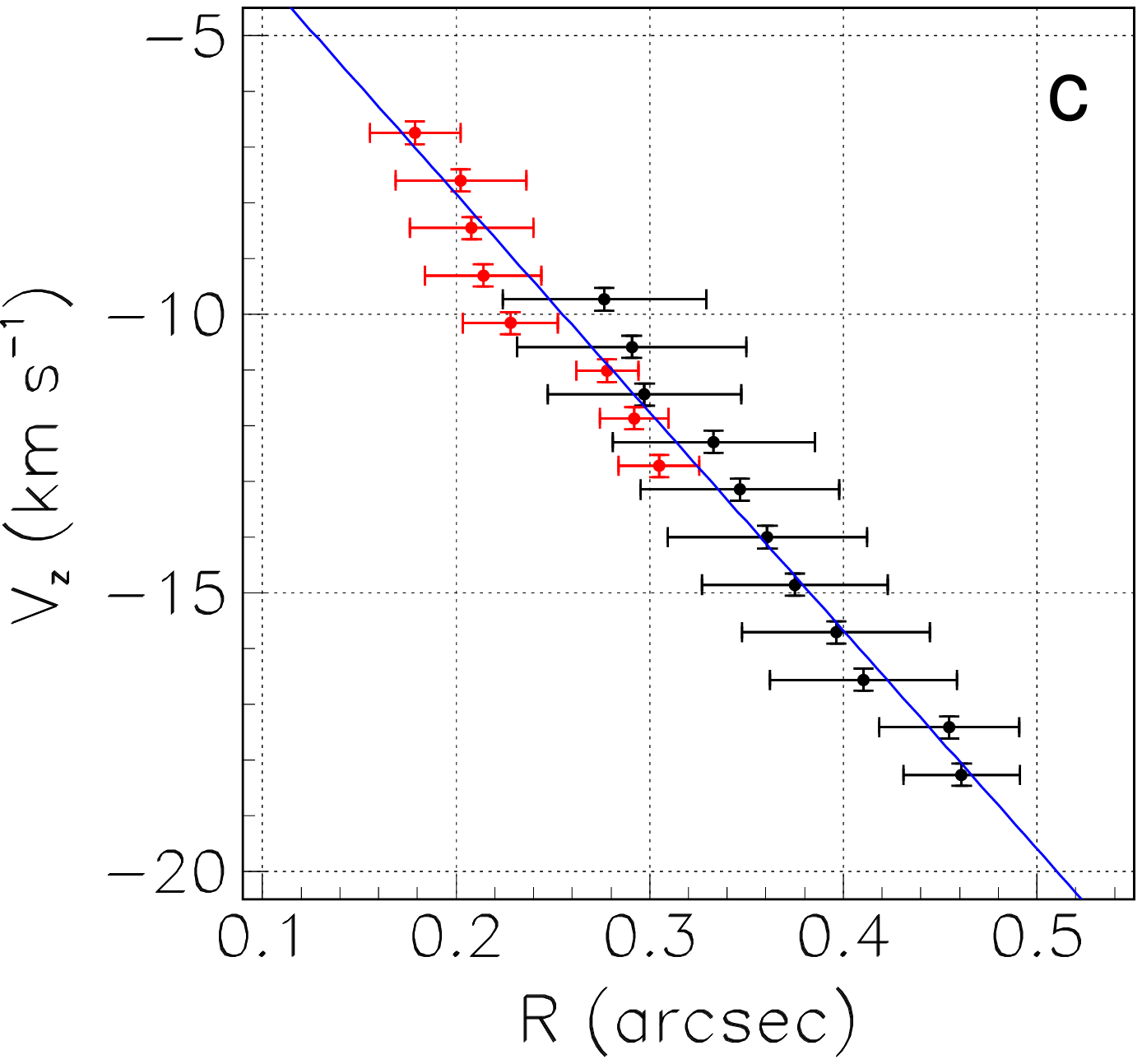}
  \caption{ Blue stream. (a): $^{29}$SiO intensity map integrated over $-$18.7<$V_z$<$-$9.3 \kms. (b): SO$_2$ intensity map integrated over $-$13.1<$V_z$<$-$6.3 \kms. On both panels lines mark position angles of 130\dego, 140\dego\ and 150\dego. The colour scales are in units of Jy beam$^{-1}$\kms. Note the different spatial scales. (c): Dependence of $V_z$ on $R$ for $^{29}$SiO (black) and SO$_2$ (red) emissions. The line shows the best fit to a form $V_z$=$kR$ with $k$=$-$39.2 \kms\, arcsec$^{-1}$. The dots and error bars show the mean and rms values of $R$ calculated over the blobs seen in the channel maps (Figures \ref{figA1} and \ref{figA2}).}
 \label{fig11}
\end{figure*}

The blue stream is seen clearly on the channel maps of both $^{29}$SiO and SO$_2$ emissions. It covers projected distances from the star reaching $\sim$30 au for the former (Figure \ref{figA1}) and $\sim$20 au for the latter (Figure \ref{figA2}) at a mean position angle of $\sim$140\dego. In order to define it precisely we measure for each line its position in each frequency interval separately and fit straight lines through the points. Figure \ref{fig11} displays intensity maps integrated over relevant Doppler velocity intervals. Panel c shows the dependence of the stream Doppler velocity on $R$ defined as described above, for each frequency interval separately. To a good approximation it displays a constant projected acceleration of 39.2$\pm$0.5 \kms\,arcsec$^{-1}$.  A priori, there is no reason for such projected linearity: it may be destroyed by de-projection and, even if it were not, it would require that the force driving the wind is approximately the sum of a constant term and an $r^{-2}$ term, the latter compensating gravity. The angle made by the stream with the line of sight is unknown. The SO$_2$ channel maps show that the stream seems to maintain its identity down to Doppler velocities of $\sim$$-$3 \kms, at which point $R$$\sim$9 au, but the constant acceleration law seems to be no longer obeyed. A natural explanation is that the stream is born within the slow wind at a space distance from the star of some 12 au, in a region where rotation has essentially faded away. A mechanism such as what \citet{Takigawa2017} call ``local acceleration'', caused by locally enhanced density of aluminium-rich dust, would then be at stake. In such a picture, the stream would make a significant but not too large angle with the line of sight, say between 30\dego\ and 45\dego. Qualitatively, this is the most natural and simple interpretation that comes to mind; however, we cannot exclude a configuration where the stream would fly at high velocity near the plane of the sky, not to mention configurations for which the linearity observed in projection is no longer obeyed once de-projected in space.

\begin{figure*}
  \centering
  \includegraphics[width=5.5cm,trim=0.cm 0.5cm 1.cm 0.5cm,clip]{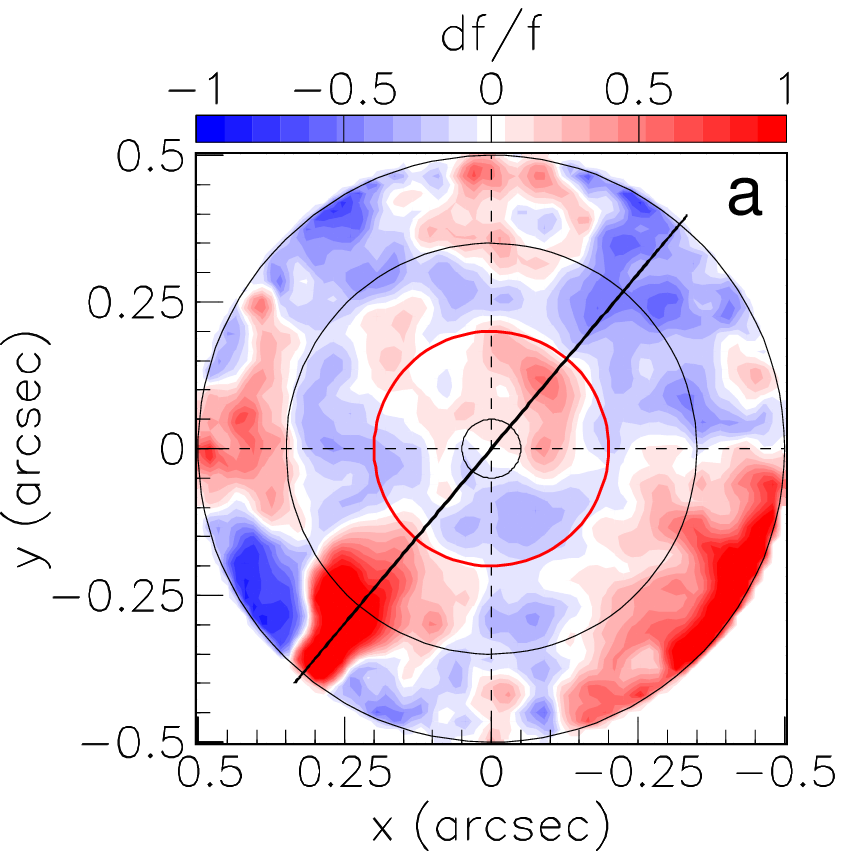}
  \includegraphics[width=5.5cm,trim=0.cm 0.5cm 1.cm 0.5cm,clip]{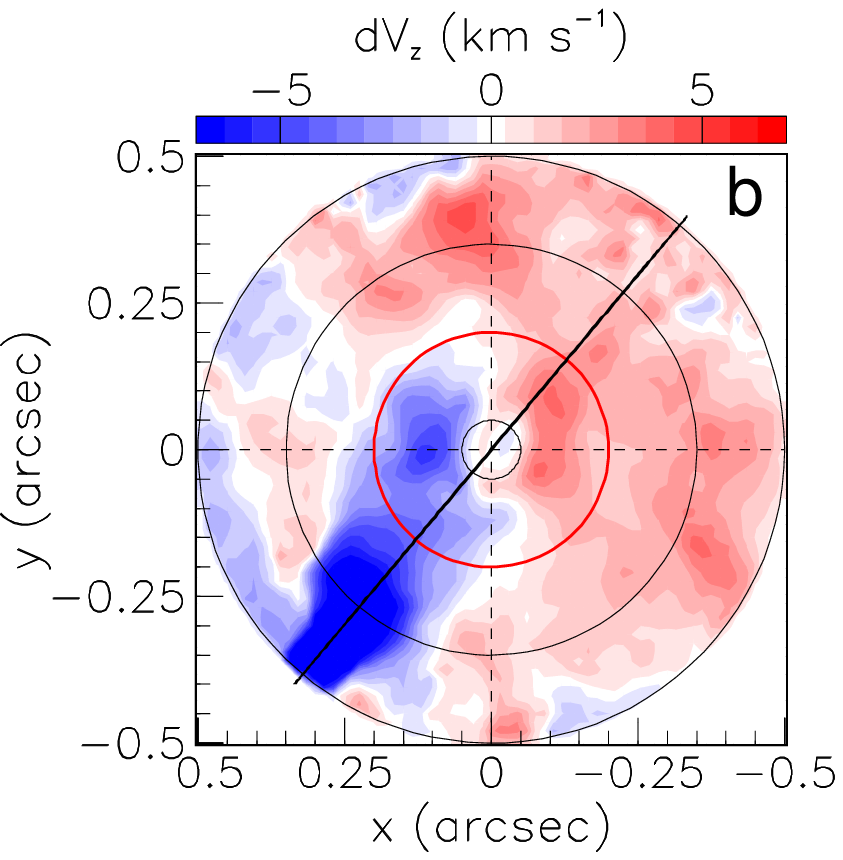}
  \includegraphics[width=5.5cm,trim=0.cm 0.5cm 1.cm 0.5cm,clip]{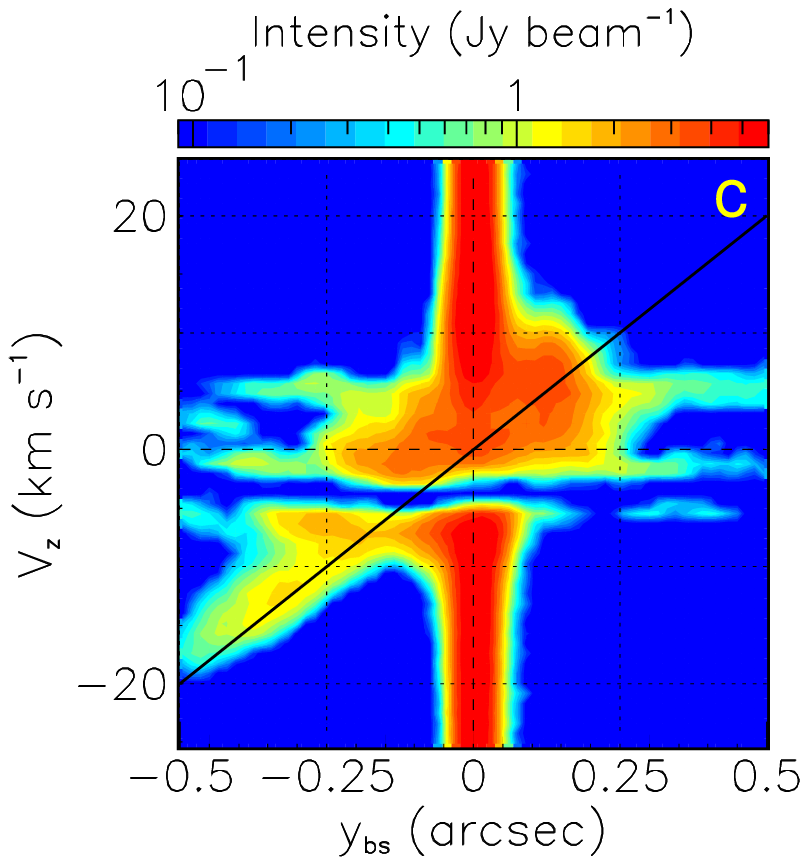}
  \caption{Blue stream in $^{29}$SiO emission. (a) relative difference d$f/f$ between the $^{29}$SiO intensity map and its average over position angle; intensity is integrated over |$V_z$|<20 \kms. (b) difference d$V_z$ between the $^{29}$SiO mean velocity <$V_z$> and its average over position angle; the mean velocity  is evaluated over |$V_z$|<20 \kms. In panels a and b the red circle has a radius of 12 au (0.2 arcsec). (c) PV diagram along the blue stream (position angle of 140\dego, slit width of $\pm$3 au).}
 \label{fig12}
\end{figure*}

Figures \ref{fig11} and \ref{fig12} show that the stream nearly points to the star; more precisely it points 1.5$\pm$1.5 au south-west of the star. Panel c of Figure \ref{fig12} displays a Position-Velocity (PV) diagram along a $\pm$3 au wide slit at position angle 140\dego. An enhancement of intensity is seen back to back to the stream, but at much smaller Doppler velocities. It is shown by a green arrow at position angle 320\dego\ in the panels of Figure \ref{fig9} covering values of $R$ between 0.1 and 0.2 arcsec.

\begin{figure*}
  \centering
  \includegraphics[width=4.4cm,trim=0.5cm 0.7cm 1.7cm 1.8cm,clip]{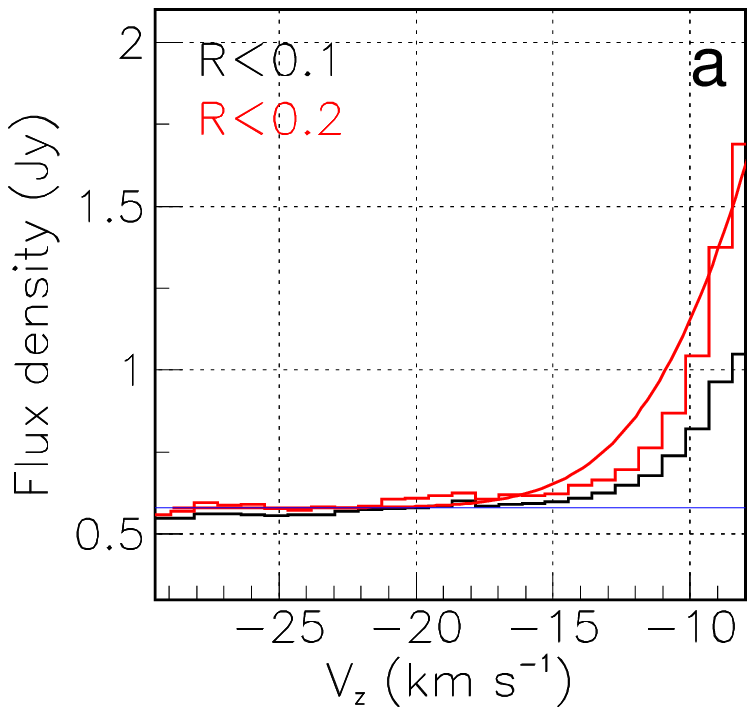}
  \includegraphics[width=4.4cm,trim=0.5cm 0.7cm 1.7cm 1.8cm,clip]{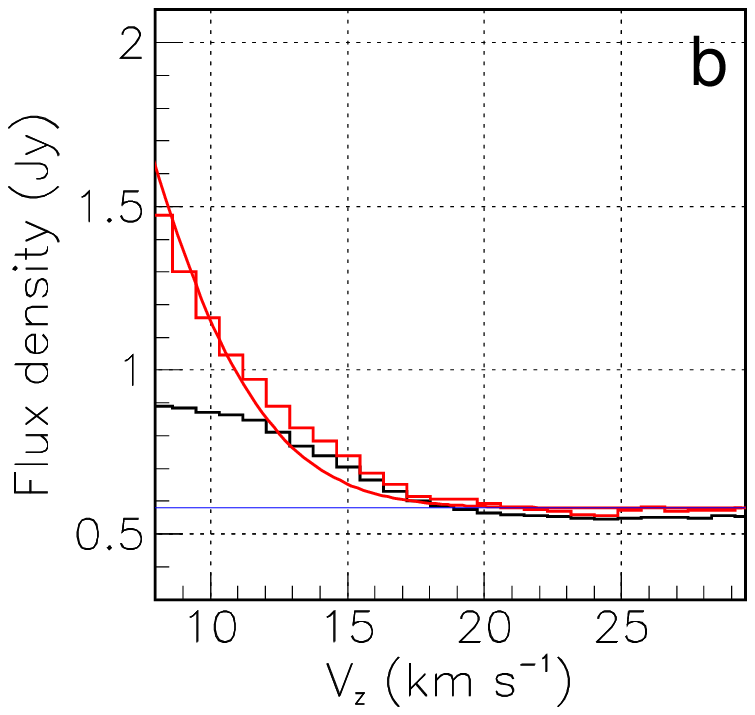}
  \includegraphics[width=4.4cm,trim=0.5cm 0.7cm 1.7cm 1.8cm,clip]{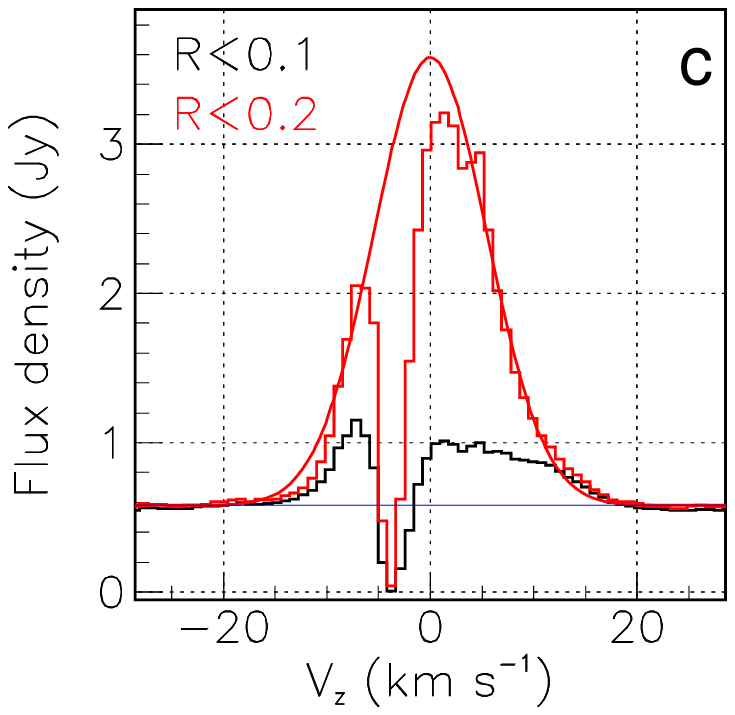}
  \includegraphics[width=4.4cm,trim=0.5cm 0.7cm 1.7cm 1.8cm,clip]{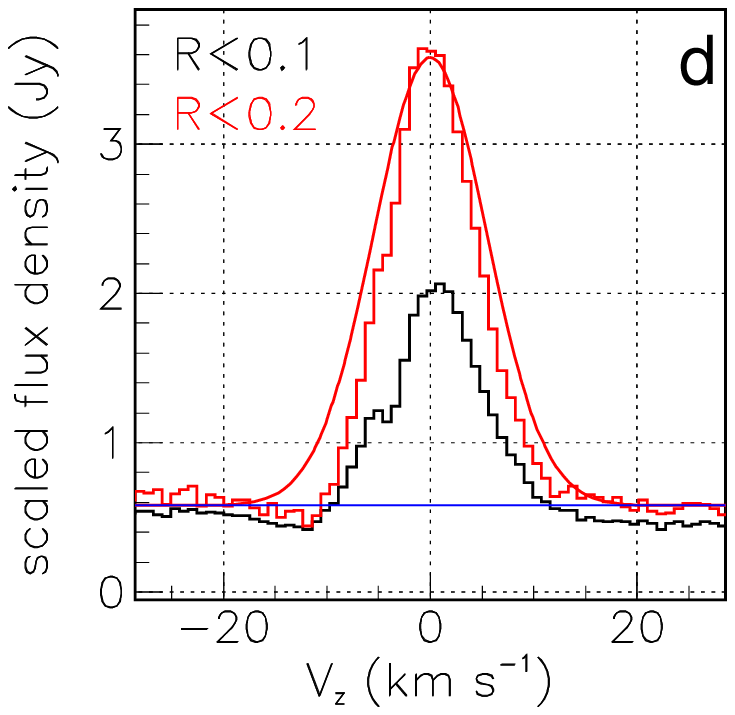}
  \caption{(a) and (b): high velocity wings of the $^{29}$SiO Doppler velocity spectrum integrated over small circles around the centre of the star ($R$<0.1 and <0.2 arcsec). (c) and (d): total Doppler velocity spectra in the same intervals of $R$ for $^{29}$SiO (c) and SO$_2$ (d). The SO$_2$ data have been scaled up by a factor 6 and shifted down by 3 Jy for convenience. The Gaussian has an amplitude of 3 Jy above continuum at zero Doppler velocity and a $\sigma$ of 5.5 \kms.}
 \label{fig13}
\end{figure*}

\begin{figure*}
  \centering
  \includegraphics[height=5.2cm,trim=0.cm 1.cm 0cm 1.5cm,clip]{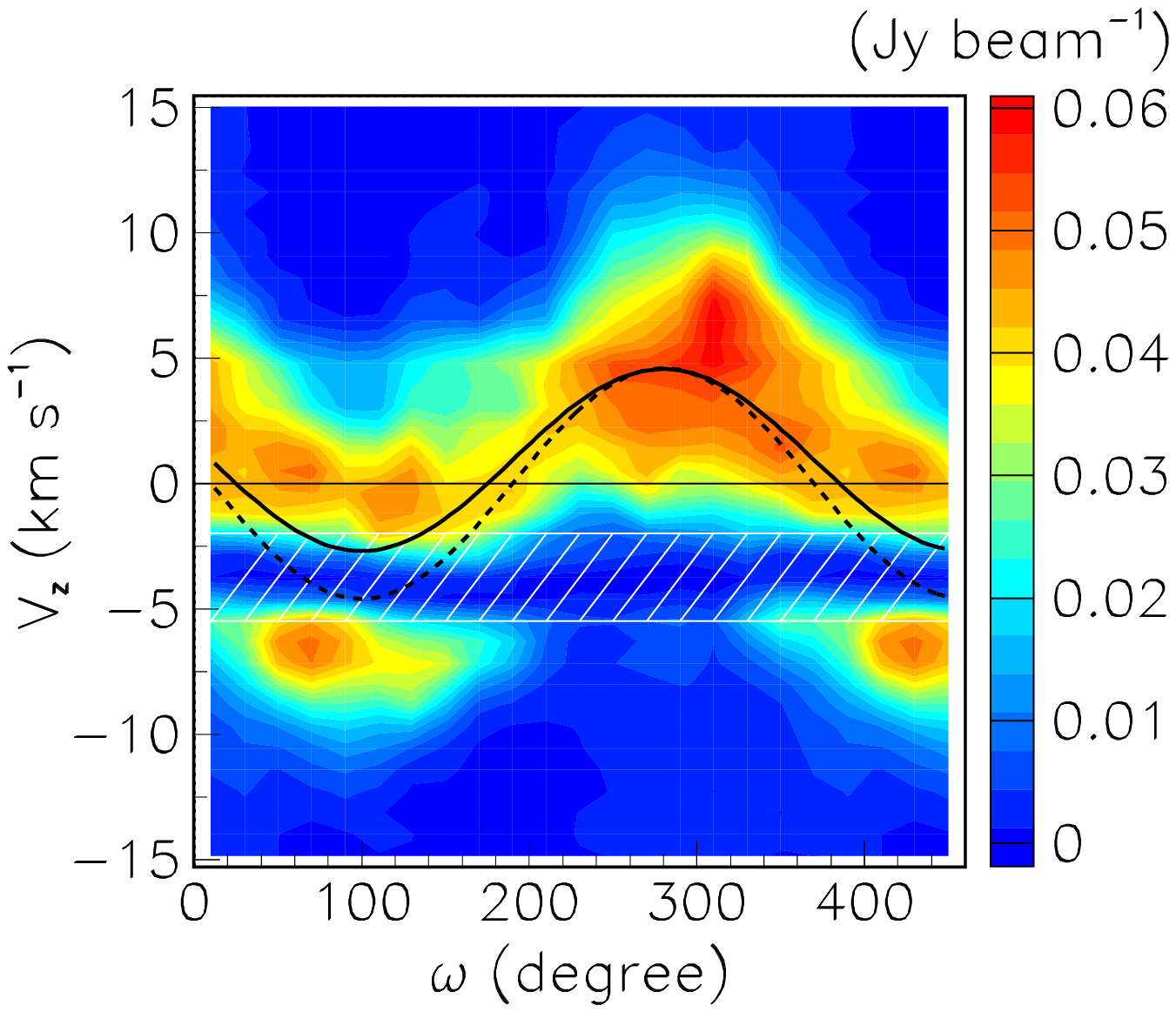}
  \includegraphics[height=5.2cm,trim=0.cm 1.cm 1.8cm 1.5cm,clip]{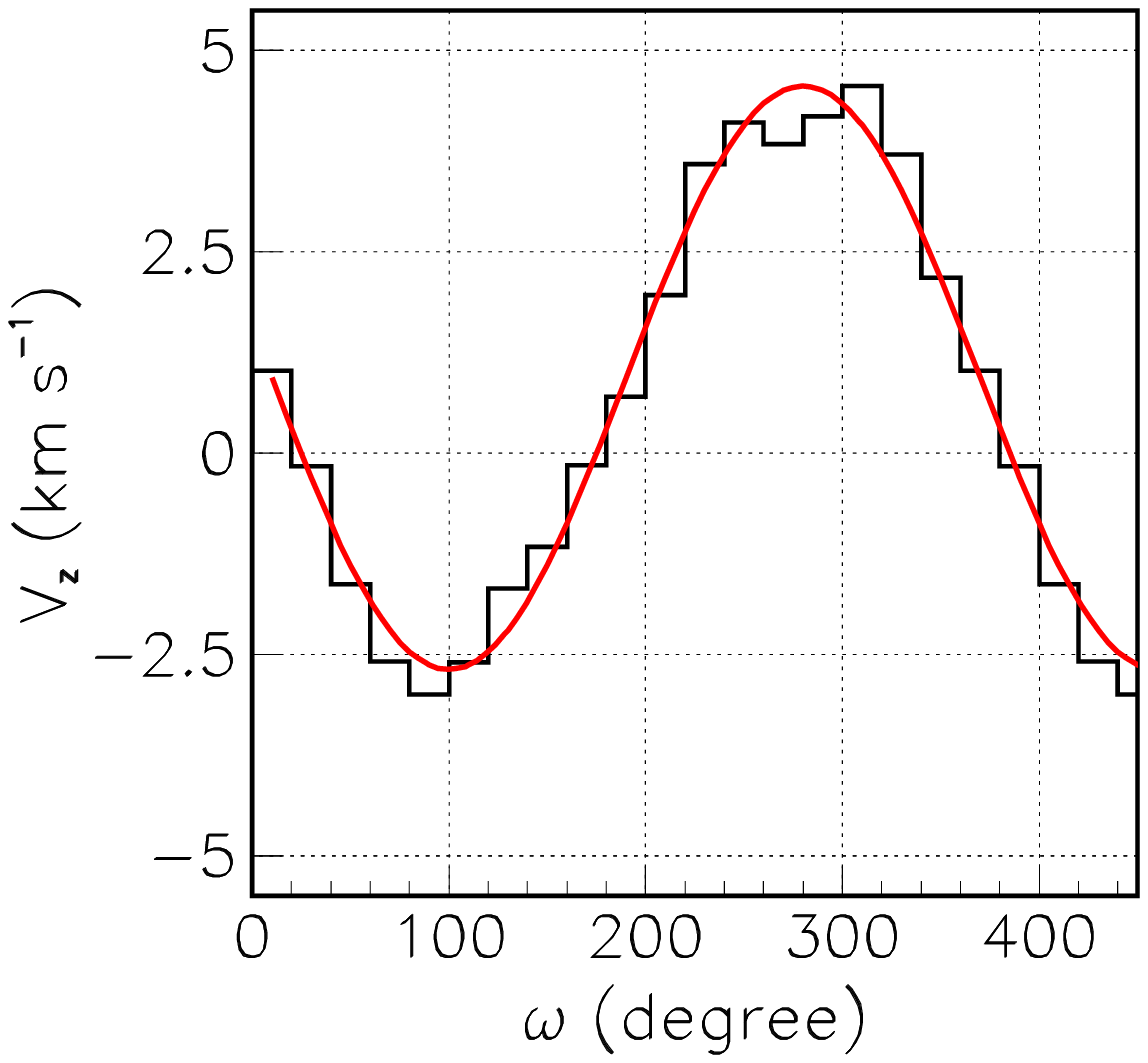}
  \caption{Left: PV map ($V_z$ vs $\omega$) of $^{29}$SiO emission in the ring 0.075<$R$<0.175 arcsec. Right: Dependence on position angle of the mean Doppler velocity averaged over |$V_z$|<12 \kms. Sine wave fits of <$V_z$> are shown in  both panels (full lines). The brightness has been linearly interpolated between $V_z$=$-$5.5 and $V_z$=$-$2 \kms (white striped region in the left panel). Sine wave fit of <$V_z$> with interpolation in the absorption interval is shown in the left panel (dashed line).}
 \label{fig14}
\end{figure*}

Evidence for important high $V_z$ components in the vicinity of the line of sight crossing the centre of the stellar disc is shown in Figure \ref{fig13}. It displays, for both $^{29}$SiO and SO$_2$ emissions, the high velocity wings of the Doppler velocity spectrum integrated over small circles around the centre of the star ($R$<0.1 and <0.2 arcsec), small enough to receive only a small contribution from the blue stream. A Gaussian line centred at the origin and having $\sigma$$\sim$5.5 \kms\ is shown as reference in the figure. The SO$_2$ line width is $\sim$15\% smaller than the $^{29}$SiO line width, probably the result of the steeper $r$ dependence of the flux density. The clear separation observed here between the blue stream and the high Doppler velocity wings underlines their different nature, the latter being the simple result of line broadening in the inner layer.

\section{The inner layer, rotation and line width}

As discussed in Section 3.2, the $^{29}$SiO emission of the star and of the inner CSE layer, $R$<$\sim$12 au, seems to be seen across a self-absorbing layer moving toward us with a velocity of $\sim$4 \kms, causing nearly total absorption over its range of Doppler velocities. In order to study the emission of the inner layer, we simply ignore this Doppler velocity interval. Figure \ref{fig14} shows a PV map, $V_z$ vs $\omega$, in the ring 0.075<$R$<0.175 arcsec. In each $\omega$ bin, the brightness in the $V_z$ absorption interval has been interpolated linearly between its values taken just above and just below. The result is essentially the same as if one keeps the absorbed interval untouched: a fit of the form $A+B\sin(\omega-\omega_0)$ gives $B$=3.6 \kms\ and $\omega_0$=191\dego\ instead of 3.6 \kms\ and 190\dego. However, the constant term is now smaller, 0.5 instead of 0.9 \kms. This shows the robustness of the result in spite of the very strong absorption. We retain a value of 10$\pm$5\dego\ for the projection angle of the rotation axis, in agreement with the value of 7$\pm$6\dego\ obtained by \citet{Vlemmings2018}.

\begin{figure*}
  \centering
  \includegraphics[width=4.4cm,trim=0.cm .8cm 1.8cm 1.8cm,clip]{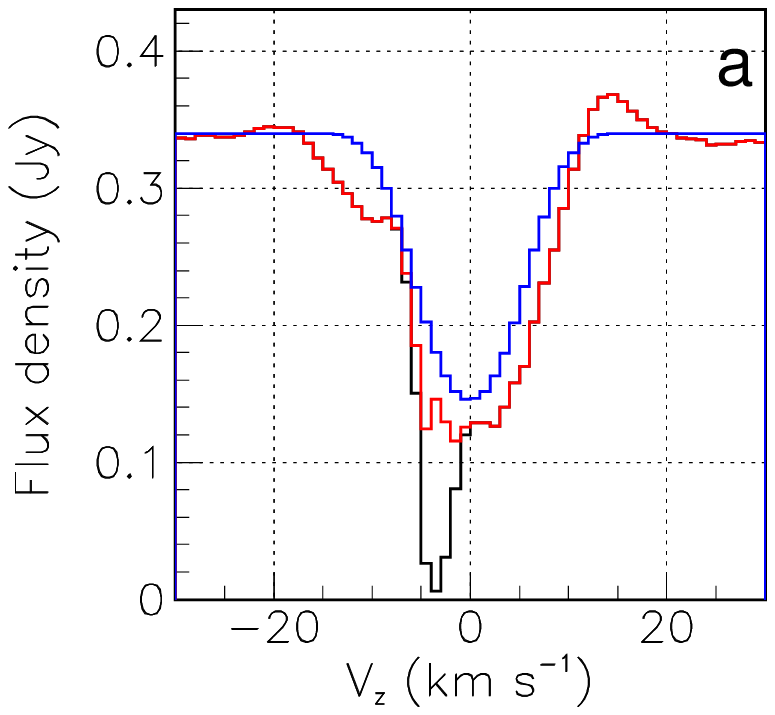}
  \includegraphics[width=4.4cm,trim=0.cm .8cm 1.8cm 1.8cm,clip]{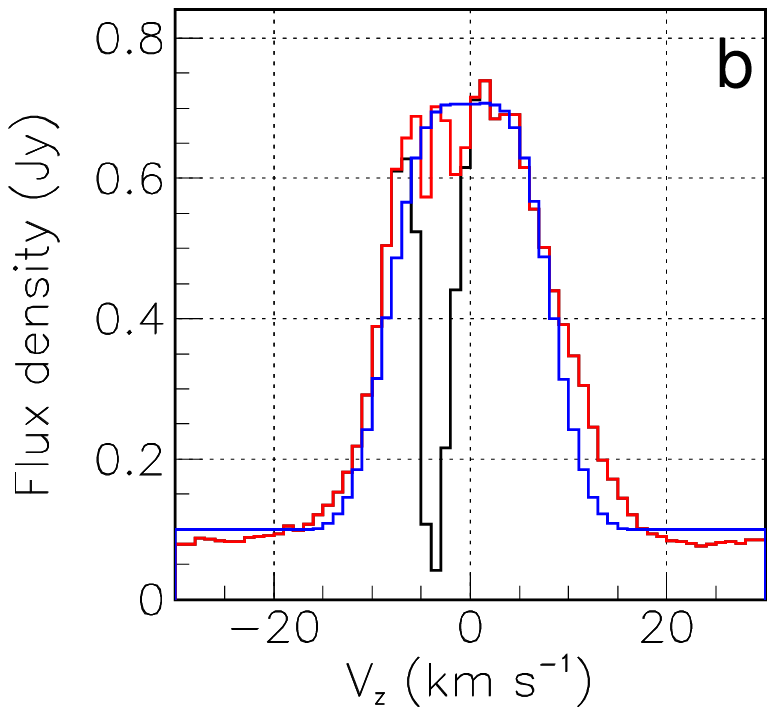}
  \includegraphics[width=4.4cm,trim=0.cm .8cm 1.8cm 1.8cm,clip]{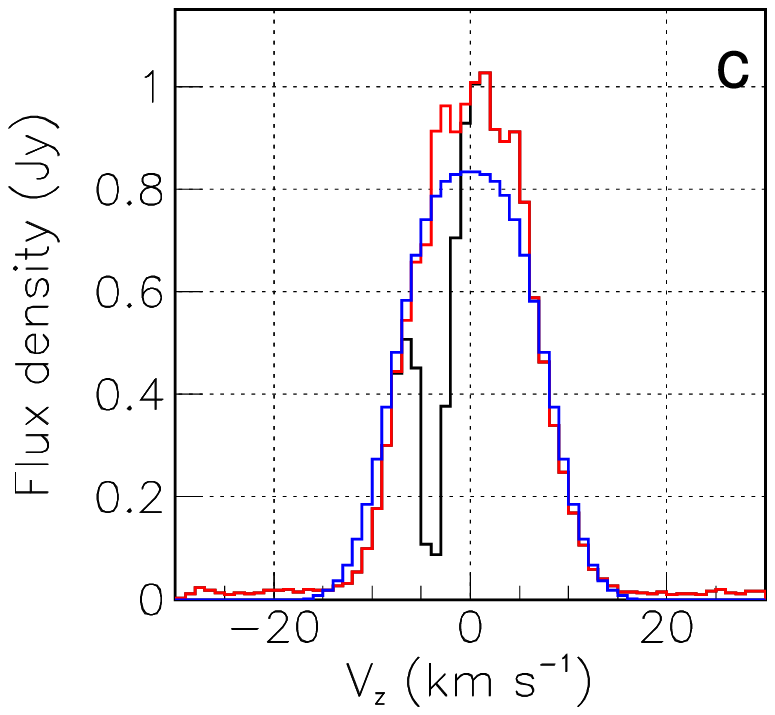}
  \includegraphics[width=4.4cm,trim=0.cm .8cm 1.8cm 1.8cm,clip]{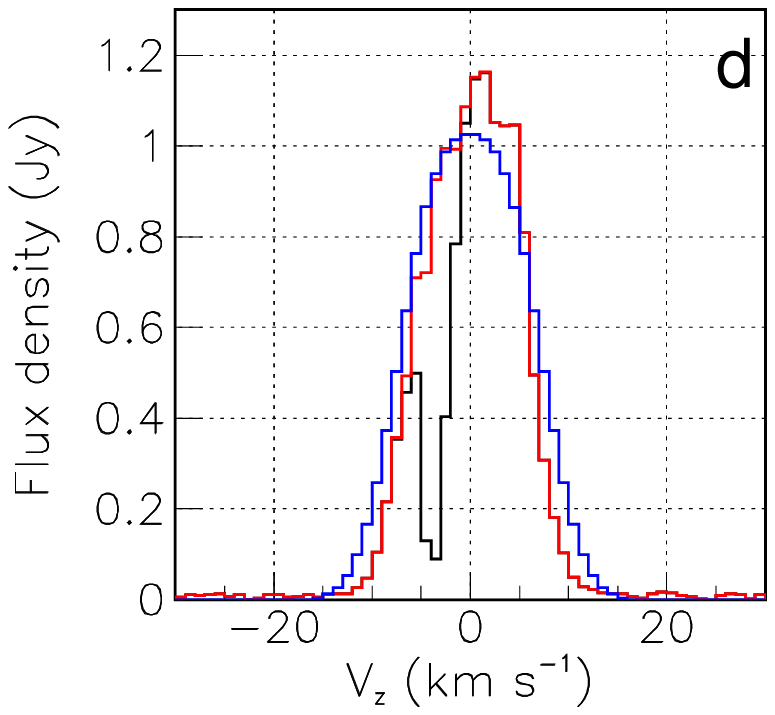}
   \caption{Doppler velocity spectra of $^{29}$SiO(8-7) emission: broad components (red) and narrow components (black) of the integrated spectra in the regions $R$<0.035 (a), 0.05<$R$<0.1 (b), 0.1<$R$<0.15 (c), 0.15<$R$<0.2 (d) arcsec. The blue histograms show the results of the best fit to the broad component. }
 \label{fig15}
\end{figure*}
 
When averaged over a ring, rotation contributes to broadening the absorption line but does not cause any shift of its mean value. The observed profile should therefore reveal by how much the wind is accelerated, if at all, in the inner layer. We study this with the help of a crude model of the morpho-kinematics of the $^{29}$SiO emission of the inner layer. Rotation is assumed to proceed about an axis projecting 10\dego\ east of north with a velocity depending on $r$ [arcsec] as $V_{\rm{rot}}$ [\kms]=64$r(1-4r)$ and $V_{\rm{rot}}$=0 beyond $r$=0.25 arcsec.The inclination of the rotation axis with respect to the plane of the sky is taken to be 20\dego. Temperature and density are assumed to depend on distance as $r^{-0.65}$ and $r^{-2}$ respectively, their values at $r$=0.1 arcsec (6 au), $T_0$  and $n_0$ respectively, being varied to optimize the fit. Line broadening is described as a Gaussian of dispersion $\sigma_{\rm{turb}}$=$V_{\rm{turb}}\exp[-\sfrac{1}{2}(r-r_{\rm{turb}})^2/\Delta{r_{\rm{turb}}}^2]$ having itself a Gaussian radial dependence. This line broadening is probably of a complex nature, including shocks from pulsations and convective cell ejections. Its modelling as Gaussian fluctuations is a grossly oversimplified approximation. The best fit, illustrated in Figure \ref{fig15}, gives $n_0=130^{+90}_{-40}$ molecules\,cm$^{-3}$ and $T_0$=$450^{+130}_{-90}$ K: the density is $(40^{+30}_{-10})$\% and the temperature $(40^{+12}_{-8})$\% of the values quoted by \citet{VandeSande2018}. The best fit does not allow for significant expansion below $r$$\sim$10 au and requires a turbulence confined to the $r$<12 au region with $V_{\rm{turb}}=4.3\pm1.1$ \kms. The assumptions made in the model are qualitatively reasonable but quantitatively somewhat arbitrary and we cannot expect the result to give more than a reference with which to compare the observations. Yet, we can reliably retain that most of the line broadening occurs in the inner layer while most of the radial expansion occurs beyond it. The value obtained for $T_0$, about half the expected value, is probably too low to be blamed uniquely on the crudeness of the model and is likely to have some reality, suggesting a steeper radial dependence than assumed in the model.

\section{The blue stream: a possible companion?}

Several authors \citep{Homan2018, Decin2018, Vlemmings2018} have entertained the idea that the blue stream might be associated with a low mass companion, probably an evaporating planet. An argument is that the observed star rotation is much too rapid to be a remnant of the rotation of the Main Sequence star and is likely to have received its angular momentum from interaction with a companion. However, such a companion may very well have already been engulfed by the expanding star and have no relation with the observed blue stream, significantly weakening the argument. \citet{Vlemmings2018} locate the companion within 8 au and \citet{Homan2018} at $\sim$12 au from the star. The present data show that, in projection on the plane of the sky, gas flows radially from $R_1$$\sim$10 au to $R_2$$\sim$30 au with mean Doppler velocity increasing from |$V_{z1}$|$\sim$7 \kms\ to |$V_{z2}$|$\sim$18 \kms. Calling $\theta_{\rm{bs}}$ the angle between the stream axis and the line of sight, this corresponds in space to a distance $(R_2-R_1)/\sin\theta_{\rm{bs}}$ covered at mean velocity $\sfrac{1}{2}(|V_{z1}|+|V_{z2}|)/\cos\theta_{\rm{bs}}$ in a time $t$=$2(R_2-R_1)/[(|V_{z1}|+|V_{z2}|)\tan\theta_{\rm{bs}}]$$\sim$$6.8/\tan\theta_{\rm{bs}}$ yr. During this time, the companion, assumed to be on a circular orbit of radius $R_1/\sin\theta_{\rm{bs}}$, covers a distance $d$=$tV_{\rm{orb}}$ where $V_{\rm{orb}}$[\mbox{au yr$^{-1}$}]=$2\pi\sqrt{(M/\mbox{\msun})\sin\theta_{\rm{bs}}/R_1[\mbox{au}]}$ is the orbital velocity; for a star mass $M$=1.4 \msun\ \citep{Danilovich2017, DeBeck2018}, $d$=$14.6\cos\theta_{\rm{bs}}/\sqrt{\sin\theta_{\rm{bs}}}$ au. As the stream points to the star within $\sim$1.5 au, this implies that $\cos\theta_{\rm{bs}}/\sqrt{\sin\theta_{\rm{bs}}}$ must not exceed $\sim$0.1 or $\theta_{\rm{bs}}$>84\dego. If it were to exceed it the space velocity would exceed 100 \kms at $R_2$. 

These simple considerations illustrate the difficulty to describe the blue stream as gas trailing behind a planetary companion: if the stream flows close to the plane of the sky, one needs a mechanism that boosts the gas velocity radially at a level well above what can be expected from standard acceleration processes; if the stream flows close to the line of sight, the time it takes to cover the distance over which it is detected allows for the companion to cover a distance much larger than acceptable from the good pointing of the stream to the star. This conclusion rests on making a large number of simplifying assumptions. As underlined by \citet{Homan2018}, such assumptions may not be obeyed in reality. However, to invalidate the conclusion requires major changes to the picture, not simple numerical adjustments within reasonable limits.

\begin{figure*}
  \centering
  \includegraphics[height=5.5cm,trim=-1.cm -1.cm 0.cm 0.cm,clip]{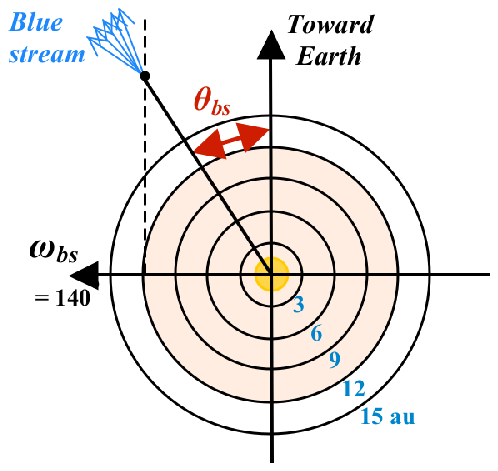}
  \includegraphics[height=4.95cm,trim=0.cm -.76cm 0.cm 0.2cm,clip]{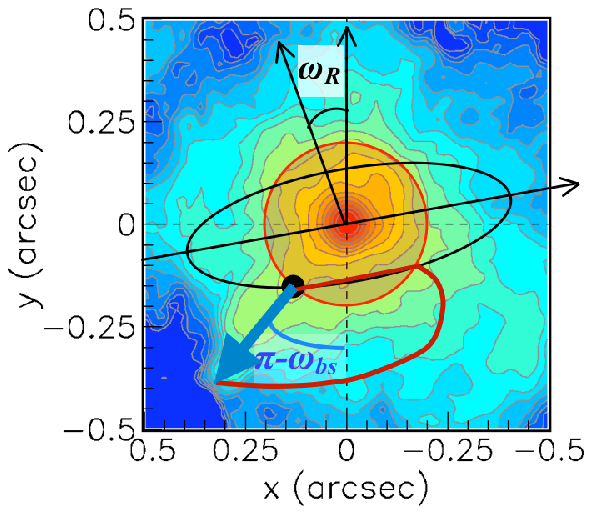}
  \includegraphics[height=4.95cm,trim=0.cm .7cm 0.cm 2.5cm,clip]{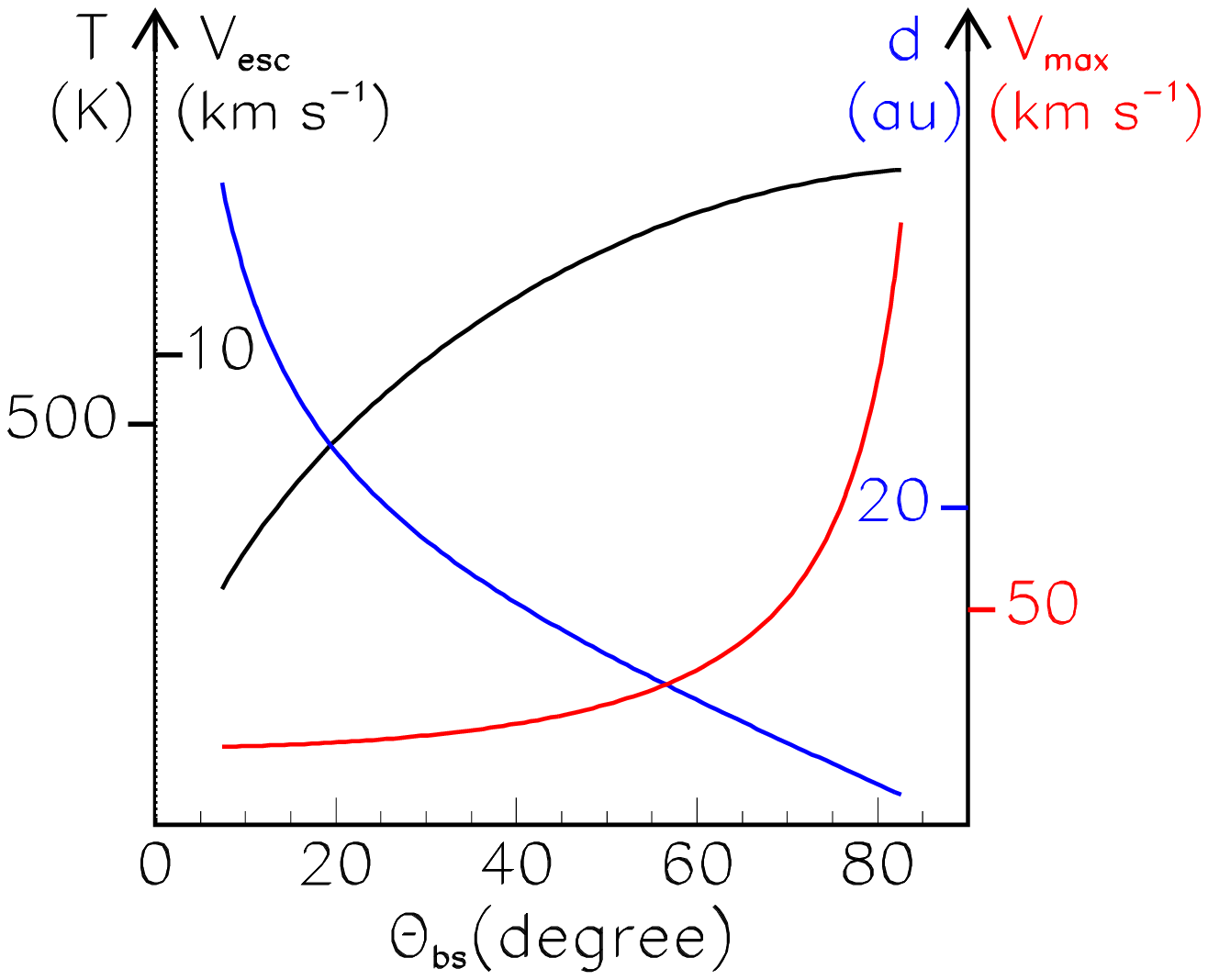}
  \caption{Schematic example of a geometry of the blue stream interpreted as gas trailing beyond an orbiting planet. Left: configuration in a plane containing the line of sight and the companion at the start of the blue stream ($\theta_{\rm{bs}}$=30\dego, $\omega_{\rm{bs}}$=140\dego, $R$=12 au, $r$=24 au). The rotation area, contained within 12 au from the star, is coloured in orange. Rings of $R$ multiples of 3 au are shown. Centre: sky map showing the companion orbit (black, $\omega_R$=10\dego, $\theta_R$=20\dego) and in red the distance covered by the companion in 12 years. The distance covered by the blue stream during the same period is shown as a blue arrow. The red spiral shows the expected density enhancement produced by the gas trailing behind the companion. Right: dependence on $\theta_{\rm{bs}}$ of the temperature (black), the escape velocity (black), the distance $d$ covered by the companion during the time where the blue stream is detected (blue) and the space velocity $V_{\rm{max}}$ reached by the blue stream gas at $R_2$=30 au (red). Note that escape velocity and temperature have both a $\sqrt{\sin\theta_{\rm{bs}}}$ dependence and the scales have been chosen for their values to overlap.}
 \label{fig16}
\end{figure*}

Figure \ref{fig16} illustrates the geometry. The left panel is drawn in the plane of position angle $\omega_{\rm{bs}}$=140\dego\ containing the line of sight and the axis of the blue stream. It is meant to illustrate the dependence of the geometry on the unknown value of $\theta_{\rm{bs}}$. It is drawn in a configuration where $\theta_{\rm{bs}}$ is small and the blue stream flies close to the line of sight, implying that it starts beyond the inner layer. If instead $\theta_{\rm{bs}}$ were large, the blue stream would fly close to the plane of the sky and would start at the outer edge of the inner layer. As the companion projects at a position angle $\omega_{\rm{bs}}$=140\dego, the inclination $\theta_R$ of the orbital axis with respect to the plane of the sky is related to $\theta_{\rm{bs}}$ by the relation  $\sin\theta_R$=$\cos(\omega_{\rm{bs}}-\omega_R)\sin\theta_{\rm{bs}}/\sqrt{1-\sin^2\theta_{\rm{bs}}\sin^2(\omega_{\rm{bs}}-\omega_R)}$, where $\omega_R$ is the position angle of the projection of the orbital axis on the plane of the sky. Assuming that the orbital axis is at the same time the rotation axis of the volume of gas we retain a value $\omega_R$=10\dego$\pm$5\dego\ east of north, from the results listed in Table \ref{tab2} and those obtained by \citet{Vlemmings2018}. To a good approximation, the relation reduces to $\theta_{\rm{bs}}$=1.5$\theta_R$ implying $\theta_R$$\sim$20\dego\ for $\theta_{\rm{bs}}$=30\dego, the same value as inferred by \citet{Homan2018} for the rotation axis of the gas volume. Figure \ref{fig16} (right) shows the dependence on $\theta_{\rm{bs}}$, at the assumed location of the companion, of the temperature $T$[K]$\sim$820$\sqrt{\sin\theta_{\rm{bs}}}$, the escape velocity $V_{\rm{esc}}$ [\kms]=14$\sqrt{\sin\theta_{\rm{bs}}}$, the distance $d$[au]=$14.6\cos\theta_{\rm{bs}}/\sqrt{\sin\theta_{\rm{bs}}}$ and the maximal gas velocity (at $R_2$=30 au), $V_{\rm{max}}$[\kms]=18$/\cos\theta_{\rm{bs}}$.

Figure \ref{fig16} invites another comment concerning the radial evolution of the Doppler velocity along the blue stream. When looking only at the projection on the sky plane apparent continuity is observed (Figure \ref{fig12}b) between the central rotation and the stream, suggesting that the acceleration of the stream might be somewhat boosted by the rotation. However, the channel maps do not display such apparent continuity. Indeed, when looking in space, one cannot make sense of such a statement: if the stream flows close to the line of sight, it is born well above the plane of the sky: the rotation velocity is parallel to the line of sight in the plane of the sky but is negligibly small where the stream is born; if it flows close to the plane of the sky, it needs a boost perpendicular to the line of sight and not along it as rotation provides.

\section{The CSE as probed at larger distances from the star}

The study of the absorption of $^{29}$SiO emission over the stellar disc presented in Section 3.2 has suggested that R Dor is seen across a self-absorbing optically thick layer moving away from the star toward us with a velocity of some 4 \kms\ and covering a region of the plane of the sky much larger than the stellar disc. While a similar effect seems to be present in the CSE of W Hya \citep{Takigawa2017}, it has not been explicitly discussed in the literature. The present section addresses this issue in the broader context of a description of the morpho-kinematics of the CSE at distances from the star covering up to some 100 au, where such absorption is expected to take place. Section 6.1 presents a critical discussion of possible instrumental effects, in particular related to the short spacing problem \citep{Gueth1997}, which might mimic the extension beyond the stellar disc of the absorption of the  $^{29}$SiO emission. The following sections probe the CSE using emissions of the SO($J_K$=$6_5-5_4$) line, together with three other molecular lines: $^{28}$SiO, CO and HCN. The emission of the SO line was studied in Paper I; Section 6.2 recalls the main results and presents a new and detailed description of the space morphology. The emission of the other lines was studied in much less detail in Paper I and Section 6.3  presents a qualitative picture of the morpho-kinematics as obtained from these multiline observations; Section 6.4 addresses the specific issue of absorption.

\subsection{Absorption of the $^{29}$SiO line emission beyond the stellar disc}

\begin{figure*}
  \centering
  \includegraphics[width=7cm,trim=2.cm 1.9cm 3cm 3cm,clip]{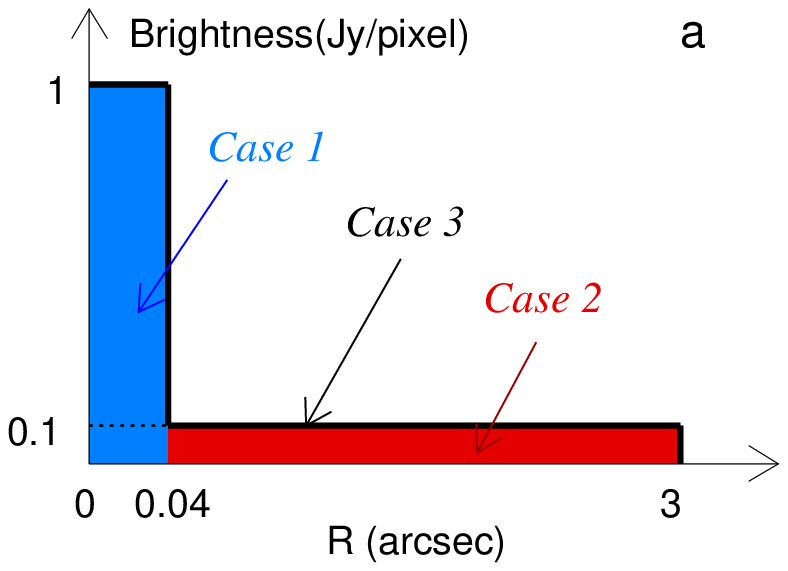}
  \includegraphics[width=4.9cm,trim=0.cm 0.5cm 1.8cm 1.8cm,clip]{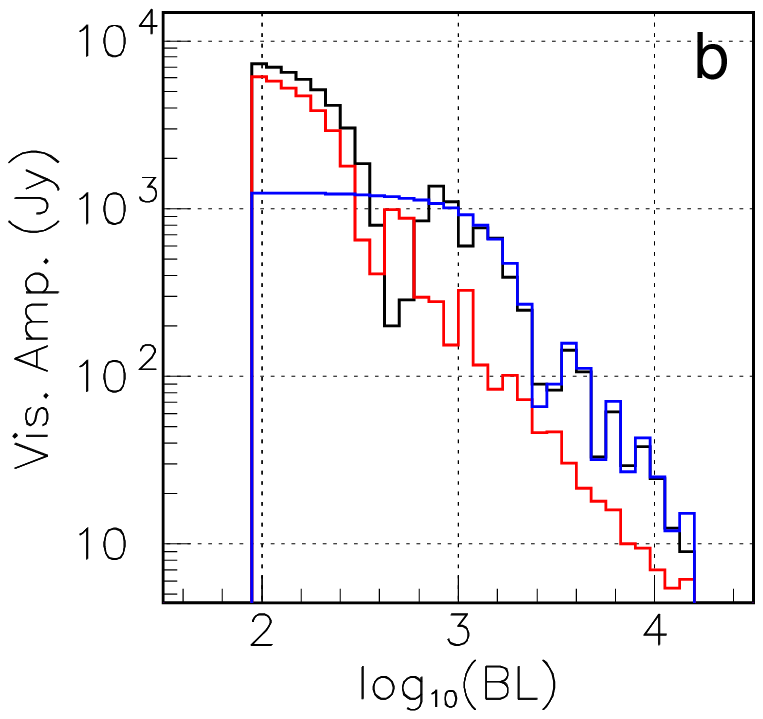}
  \includegraphics[width=4.9cm,trim=0.cm 0.5cm 1.8cm 1.8cm,clip]{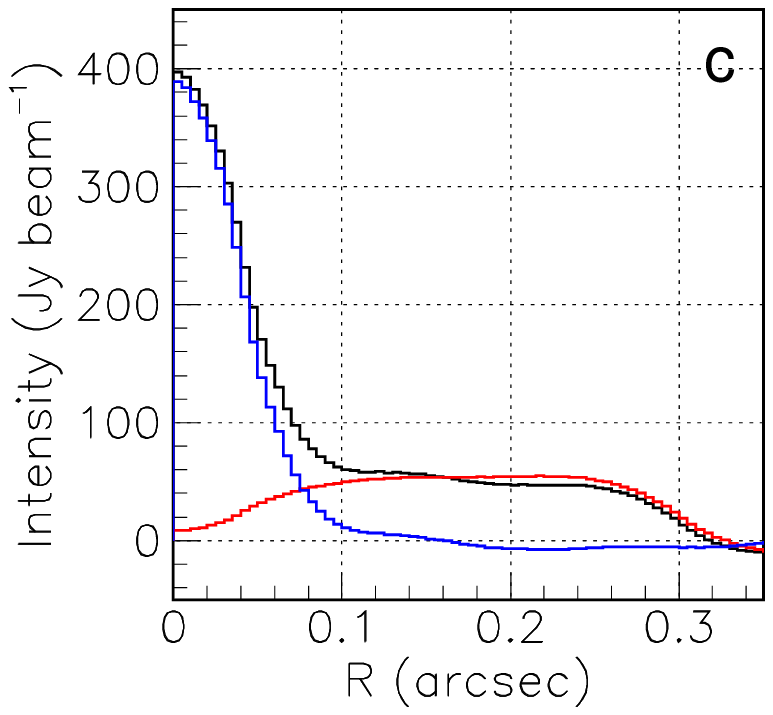}
    \caption{(a) Radial dependence of each of the three model brightness used in the evaluation of the corresponding visibilities (see text). (b) Dependence of the amplitude of the visibility on the decimal logarithm of the baseline lengths (meters) for each of the three cases 1 (blue), 2 (red) and 3 (black). (c) $R$ distributions of the flux density as evaluated from the dirty maps obtained from the model visibilities.}
 \label{fig17}
\end{figure*}

\begin{figure}
  \centering
  \includegraphics[width=4.2cm,trim=0.4cm 0.5cm 1.9cm 1.8cm,clip]{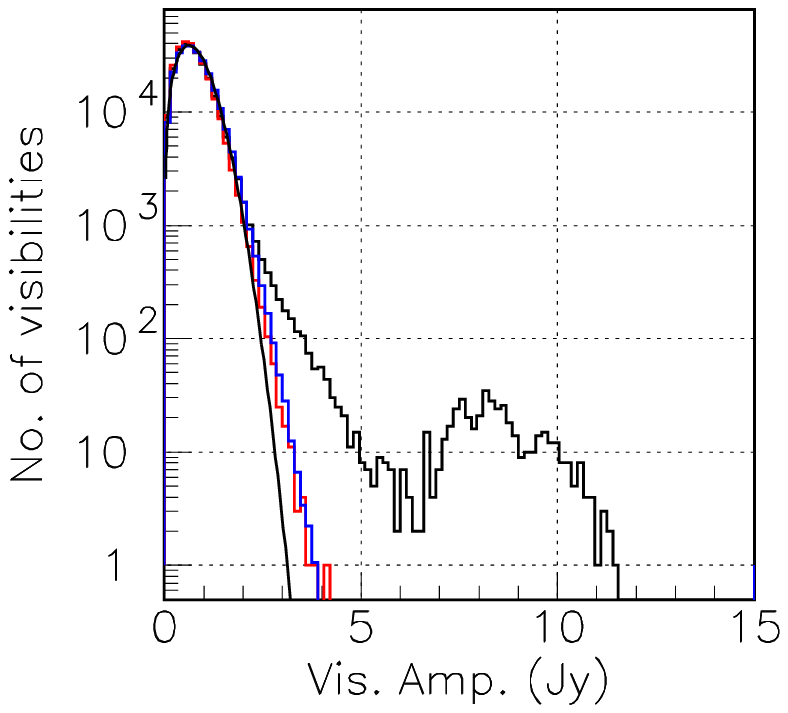}
  \includegraphics[width=4.2cm,trim=0.4cm 0.5cm 1.9cm 1.8cm,clip]{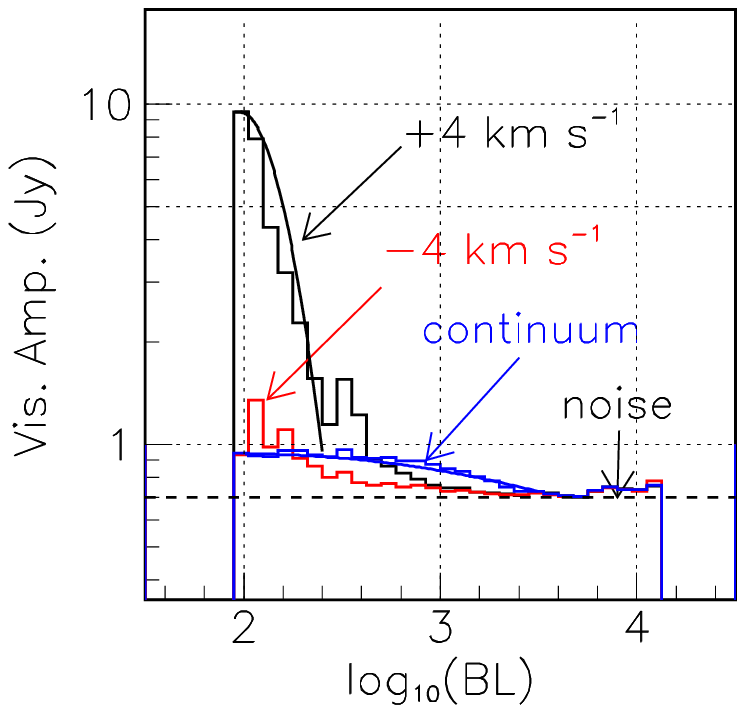}
  
   \caption{Visibilities measured from the $^{29}$SiO observations. In both panels colours identify the velocity intervals defined in the text: blue for continuum (normalized to a single frequency channel), red for $-$4 \kms\ and black for $+$4 \kms. Left: distribution of the amplitude of the observed visibility $U$ averaged over each of the frequency intervals described in the text; the fits correspond to Gaussian distributions for each of the real and imaginary parts of $U$. Right: normalised visibility distributions for each of the three frequency intervals as a function of the decimal logarithm of the baseline length (m).}
 \label{fig18}
\end{figure}

Figure \ref{fig10} has given evidence for strong absorption of the $^{29}$SiO line emission to persist well beyond the stellar disc. In order to get an idea of which absorption to expect from the antenna configuration used for the observations, we construct a very crude model that reproduces qualitatively the main characteristics of the $^{29}$SiO morpho-kinematics. It assumes the isotropy of the brightness in the plane of the sky and considers three different models, each defined by a simple radial dependence of the brightness as illustrated in Figure \ref{fig17}a. It considers three different brightness distributions on the plane of the sky. The first model, case 1, shown in blue, represents continuum emission and is described as a uniform brightness of 1 Jy/pixel over the stellar disc ($R$<0.04 arcsec). The second model, case 2, in red, shows a disc of uniform brightness, arbitrarily set to 0.1 Jy/pixel, and having a radius of 0.3 arcsec, meant to represent the emission of the CSE. Emission is completely absorbed over the stellar disc; this case is meant to mimic the real situation in the frequency channel corresponding to a Doppler velocity of $-$4 \kms. The third model, case 3, shown as a thick black line, represents instead the situation in the frequency channel corresponding to $V_z$=$+$4 \kms, expected to detect the same emission as the $-$4 \kms\ channel but without absorption over the stellar disc: emission is not absorbed and is described by a uniform brightness of 1 Jy/pixel over the stellar disc. The emission in case 3 is the sum of that in cases 1 and 2. For each of the three cases, we produce\footnote{Using SIMOBSERVE in CASA, http//casa.nrao.edu} visibilities $U$ corresponding to the actual antenna configuration illustrated in Figure \ref{fig2}. Figure \ref{fig17}b displays the baseline dependence of the visibility amplitude for each of the three cases. Case 1, meant to represent continuum, covers long baselines, up to $\sim$1000 m, with small visibilities (in spite of its larger brightness than the gas, by a factor 10, it covers a much smaller area, by a factor 130). Case 2, meant to represent $V_z$=$-$4 \kms, shows that the gas emission is indeed absorbed, but only over a small distance from the star, not exceeding $\sim$0.1 arcsec, in strong contrast with what we observe in the real case (Figure \ref{fig18}). Case 3, meant to represent $V_z$=$+$4 \kms, shows the gas emission that covers short baselines, up to $\sim$500 m, but with large visibilities. In summary, in spite of the poor $uv$ coverage at short baselines, we expect absorption to extend only slightly beyond the stellar disc, well below the maximal recoverable scale, in sharp contrast with observation.

Turning now to the actual $^{29}$SiO observations, we select the two frequency intervals covering $V_z$=$\pm$4 \kms\ and define a third interval, meant to measure continuum emission, made of two sets of 9 frequency channels each (7.7 \kms) at the beginning and end of the frequency band. The left panel of Figure \ref{fig18} displays the distributions of the measured visibility amplitude, |$U$|, averaged over these intervals. The noise is well described by a form |$U$|$\exp(-\sfrac{1}{2}|$$U$$|^2/\sigma^2)$, meaning a Gaussian distribution in each of Re($U$) and Im($U$)  with a common rms value, $\sigma$, measured to be 0.62 Jy. While the measured visibilities are large in the gas layer surrounding the star (black), they are much smaller and closer to the noise for the continuum (blue) and the maximal absorption channel (red). The right panel of Figure \ref{fig18} shows the dependence on baseline of the visibility for each frequency interval separately, again normalised to the baseline distribution of Figure \ref{fig2}c. Continuum emission (blue), corresponding to the stellar disc, covers the broader range of baselines, extending up to $\sim$2.3 km; gas emission (black), corresponding to the gas envelope around the star, covers up to baselines of $\sim$500 m. Finally, the red histogram is expected to stay at noise level down to baselines of $\sim$500 m, the disc region being nearly completely absorbed, and it does so. However, toward lower baselines, it is expected to rise near the level of the black histogram, as happens for case 2 in Figure \ref{fig17}b, but in fact it reaches much lower a level: this shows that the effect is already present at the level of the measured visibilities.

The short spacing problem is precisely the same for the $V_z$=$+$4 \kms\ and $V_z$=$-$4 \kms\ cases; the noise is the same; the calibration uses only observations made for this purpose and is also the same for the two channels. The global blank out of the $V_z$=$-$4 \kms\ observations is therefore shown to be present at raw data level already and we must conclude that it is real and not the result of some instrumental artefact.

\subsection{Overall picture of the morpho-kinematics as seen from SO emission}

Paper I has presented a detailed study of the morpho-kinematics of the CSE using ALMA observations of the emission of the SO($J_K$=6$_5$$-$5$_4$) line of which channel maps were displayed in the supplementary material. These are from project 2017.1.00824.S observed in December 2017 in band 6 with an average of 45 antennas. Distances from the star reaching up to some $\sim$120 au are probed with a circular beam of 9 au FWHM. Evidence is obtained for a slow radial wind with Doppler velocities reaching up to $\sim$6 \kms, displaying significant anisotropy and inhomogeneity.

The complexity of the morphology of the CSE makes it difficult to draw simple 2-D representations of the data cube. We find that PV maps displaying the brightness in the Doppler velocity ($V_z$) vs position angle ($\omega$) or vs projected distance ($R$) planes, are particularly convenient for this purpose. They have the advantage that a wind blowing at constant velocity $V$ in a small solid angle appears as a small clump in the former and as a stretch of uniform density in the latter. Moreover, for an isotropic wind of constant and uniform radial velocity $V$, the angle $\varphi$ of the flow with the line of sight is simply obtained from the relation $V_z$=$V\cos\varphi$, providing an evaluation of $z$=$R/\tan\varphi$ and therefore of the morpho-kinematics in space. More generally, space reconstruction can be done as long as the dependence of $V$ on $r$ is known; in particular for an acceleration law of the form $V$=$V_{\rm{term}}r/(r+r_{\sfrac{1}{2}})$, tending asymptotically to the terminal velocity $V_{\rm{term}}$ when $r$ tends to infinity and reaching the value $\sfrac{1}{2}V_{\rm{term}}$ for $r$=$r_{\sfrac{1}{2}}$, one has $r$=$[\alpha^2r_{\sfrac{1}{2}}+\sqrt{\alpha^4r_{\sfrac{1}{2}}^2+(1-\alpha^2)(R^2+\alpha^2r_{\sfrac{1}{2}}^2)}]/(1-\alpha^2)$ with $\alpha$=$V_z/V$.

\begin{figure}
  \centering
  \includegraphics[width=7.cm,trim=0.cm -1.cm 0cm 0.cm,clip]{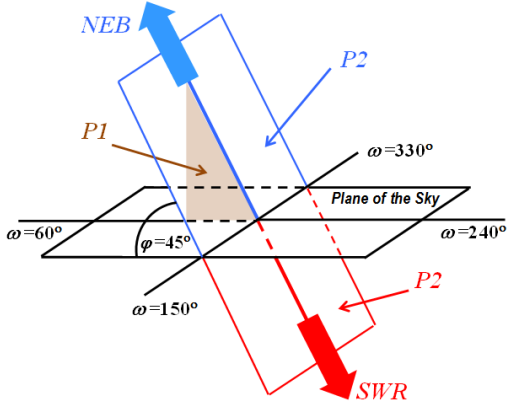}
   \caption{Outflow geometry (see text).}
 \label{fig19}
\end{figure}

\begin{figure*}
  \centering
  \includegraphics[height=0.215\textheight,trim=0.8cm 0.5cm 0.5cm 0.cm,clip]{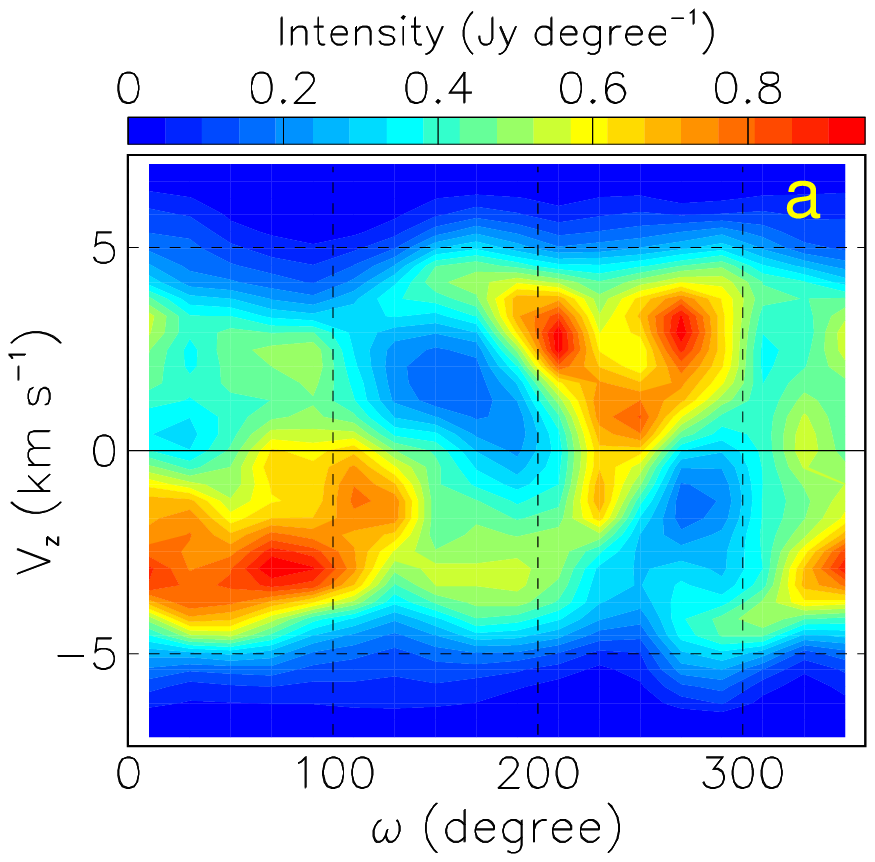}
 \includegraphics[height=0.215\textheight,trim=0.5cm 0.5cm 1.8cm 0.cm,clip]{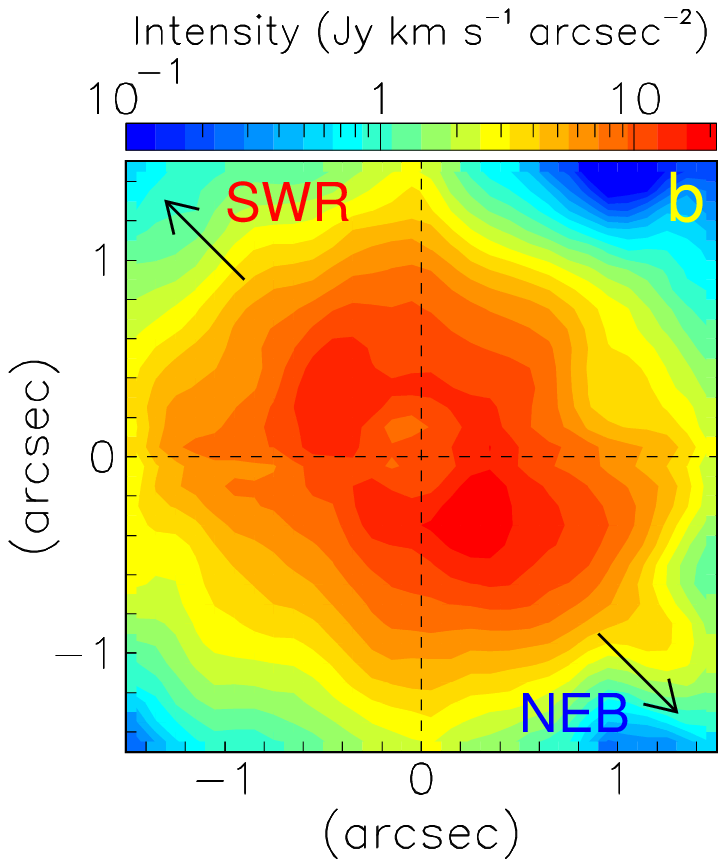}
 \includegraphics[height=0.215\textheight,trim=0.5cm 0.5cm 1.8cm 0.cm,clip]{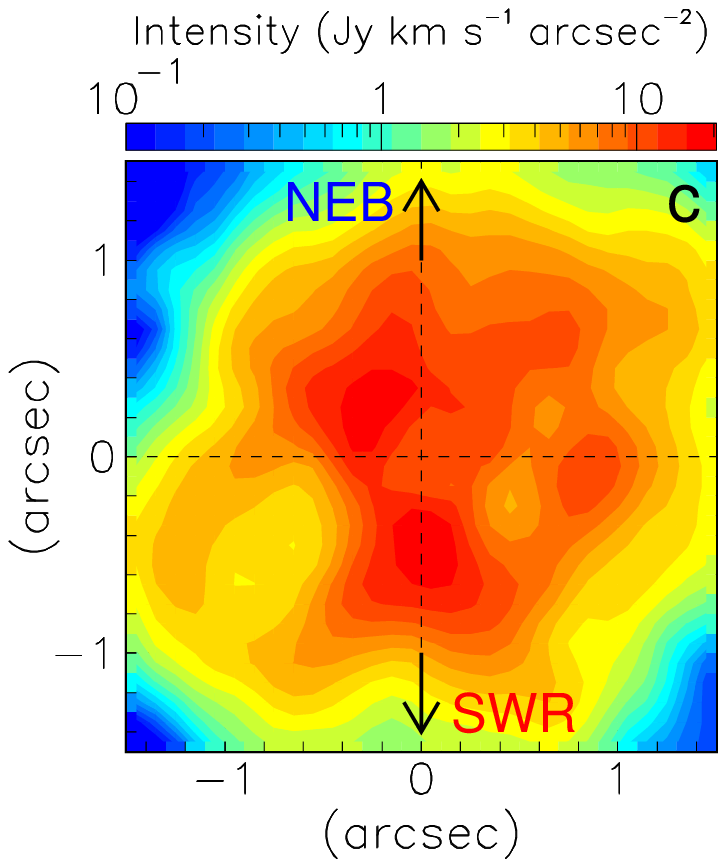}
 \includegraphics[height=0.215\textheight,trim=0.5cm 0.5cm 1.8cm 0.cm,clip]{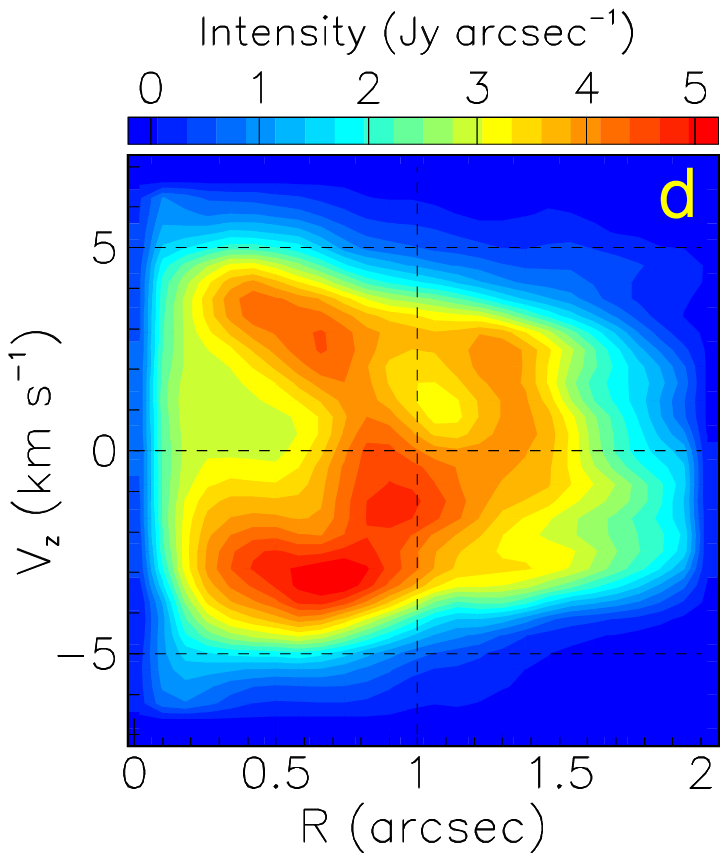}

   \caption{SO($J_K$=6$_5$$-$5$_4$) line emission. (a): intensity in the $V_z$ vs $\omega$ plane integrated over 12 au<$R$<60 au. (b): intensity projected on P1; up-left (down-right) points to SWR (NEB). (c): intensity projected on P2; up (down) points to NEB (SWR). In both panels (b) and (c) a cut $r$>0.3 arcsec (18 au) has been applied. (d): intensity projected on the $V_z$ vs $R$ plane, integrated over position angles.}
 \label{fig20}
\end{figure*}

The quality of the SO observations is excellent and the CSE is optically thin enough to suffer no significant absorption. They show two nearly back-to-back outflows, one (NEB) covering the north-eastern blue-shifted octant of the data cube and the other (SWR) the south-western red-shifted octant. Figure \ref{fig19} shows the geometry. Figure \ref{fig20}a displays the intensity in the $V_z$ vs $\omega$ plane for 12 au<$R$<60 au.  A very crude model of two back-to-back outflows is used to evaluate the orientation of the common outflow axis: it projects on the plane of the sky 60\dego$\pm$5\dego\ east of north. Both SWR and NEB have a multicore structure. We use two planes on which to project the emission of the outflows. They are perpendicular to each other and cross on the outflow axis (Figure \ref{fig19}). Plane P1 is perpendicular to the plane of the sky and projects on it at 60\dego\ position angle. Plane P2 crosses the plane of the sky at 60\dego+90\dego=150\dego\ position angle. The angle $\varphi$ of the outflow axis with the plane of the sky depends on the acceleration of the flow. Using the law introduced in the preceding paragraph with $V_{\rm{term}}$=7 \kms\ and $r_{\sfrac{1}{2}}$=30 au (0.5 arcsec) the map of the intensity projected on P1, displayed in panel b of Figure \ref{fig20}, gives $\varphi$$\sim$45\dego.  The value of $r_{\sfrac{1}{2}}$ can be chosen anywhere between 0.25 and 0.75 arcsec (15 and 45 au) without $\varphi$ changing by more than $\pm$5\dego. Panel c of the figure shows the map of the intensity projected on P2.

When looking at the radial distribution of the intensity, significant inhomogeneity is observed. Figure \ref{fig20}d displays the distribution of the intensity in the $V_z$ vs $R$ plane, integrated over all position angles. It was shown in Paper I to reveal the presence of a ring cavity surrounding the star, approximately perpendicular to the outflow axis, seen as a hole of emission at $R$$\sim$0.5 arcsec. A second hole at $R$$\sim$1.1 arcsec is also visible. This was interpreted in Paper I as evidence for episodes of enhanced mass loss. However, the ring cavity receives important contribution from the waist separating the outflows: understanding its precise nature requires additional observations.

\subsection{Overall picture of the morpho-kinematics from a comparison between SO, $^{28}$SiO, CO and HCN line emissions}

\begin{figure*}
  \centering
  \includegraphics[width=0.245\textwidth,trim=0.3cm 0.7cm 1.7cm 1.5cm,clip]{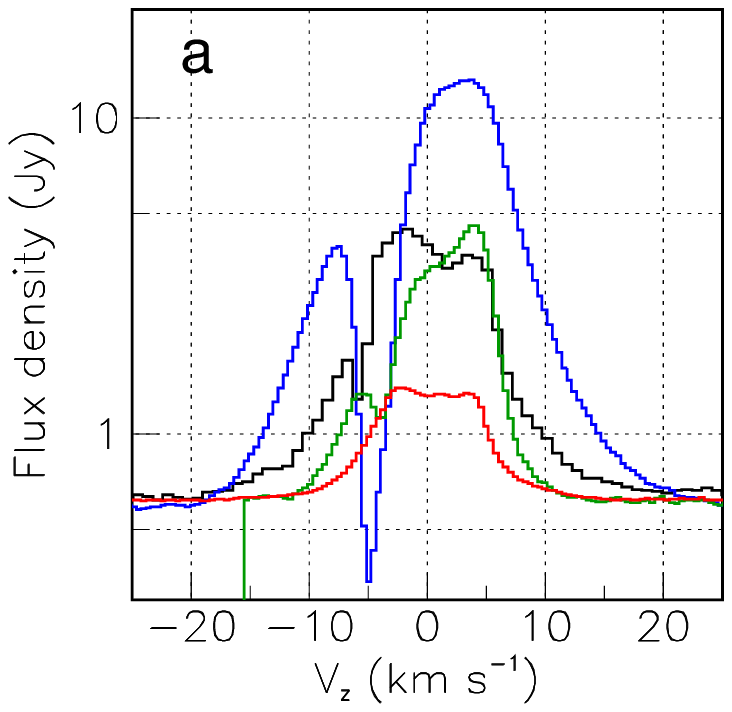}
  \includegraphics[width=0.245\textwidth,trim=0.3cm 0.7cm 1.7cm 1.5cm,clip]{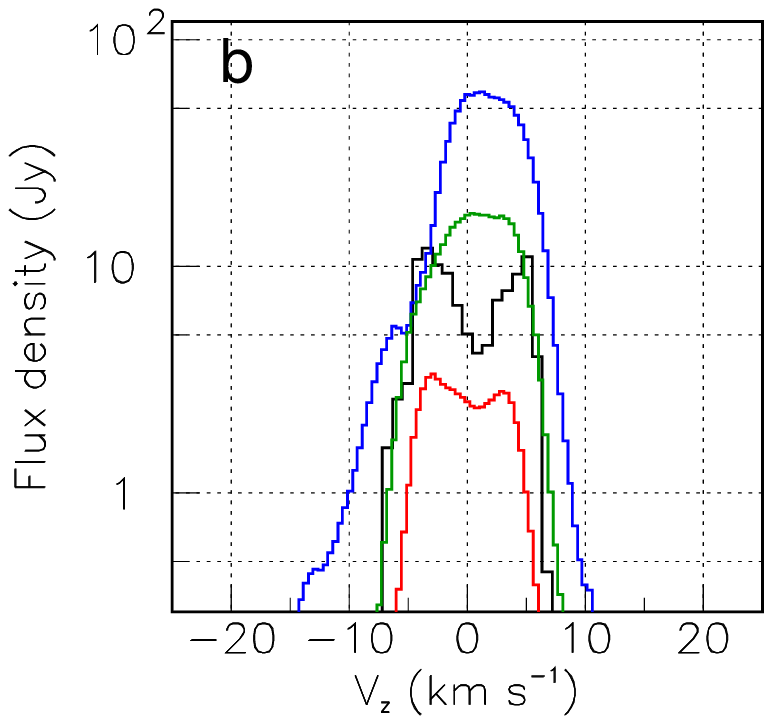}
  \includegraphics[width=0.245\textwidth,trim=0.3cm 0.7cm 1.7cm 1.5cm,clip]{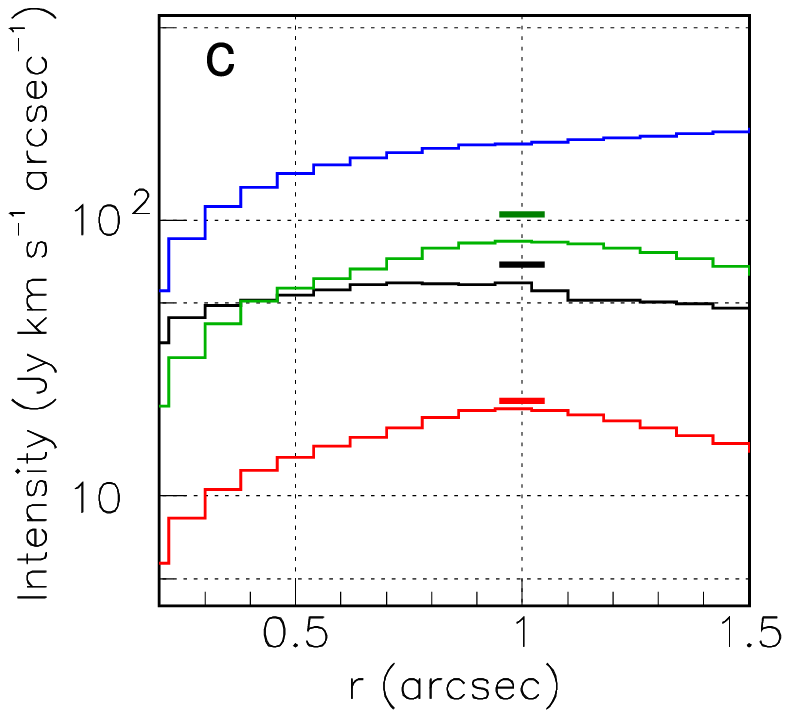}
  \includegraphics[width=0.245\textwidth,trim=0.3cm 0.7cm 1.7cm 1.5cm,clip]{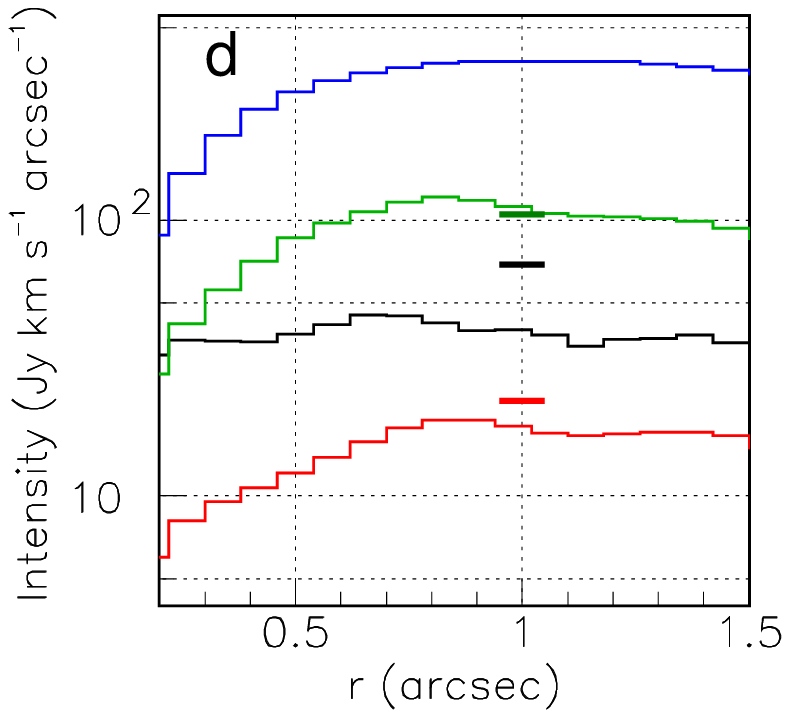}
  \caption{Doppler velocity spectra (Jy) integrated over $\omega$ and $R$<0.3 arcsec (a) or 0.3<$R$<1.0 arcsec (b). (c) and (d): $r$ distributions evaluated using the velocity law described in Section 6.2 for $V_z$<0 (c) and $V_z$>0 (d). The coloured horizontal lines show the values of $\sfrac{1}{2}F_r$ (accounting for the split in two hemispheres) calculated for a temperature of 150 K in the optically thin approximation. The values predicted for $^{28}$SiO in this approximation are considerably larger than observed and off-scale in the figure. In all panels SO is red, $^{28}$SiO blue, CO black and HCN green.}
 \label{fig22}
\end{figure*}

\begin{figure*}
  \centering
  \includegraphics[height=0.2\textheight,trim=0.5cm 0.5cm 0.5cm 0.cm,clip]{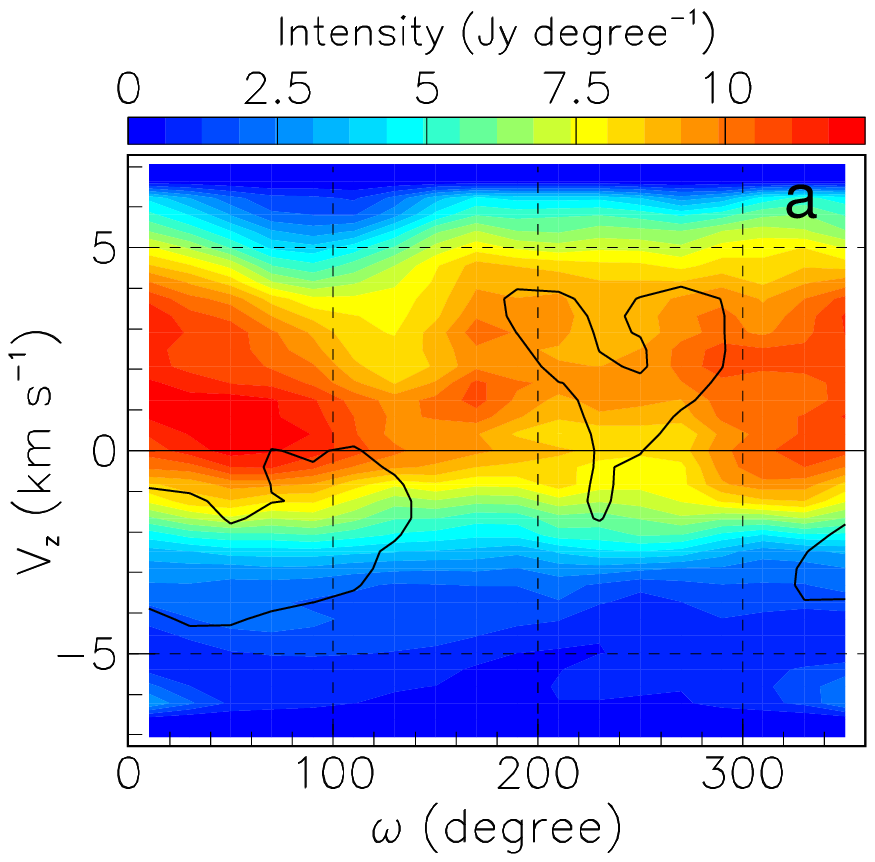}
 \includegraphics[height=0.2\textheight,trim=0.5cm 0.5cm 1.8cm 0.cm,clip]{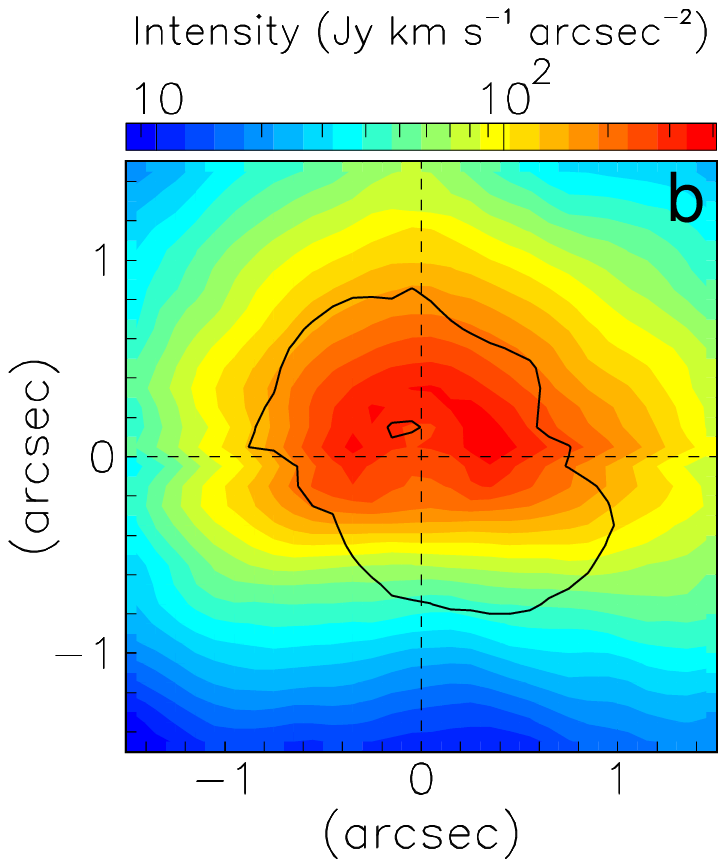}
 \includegraphics[height=0.2\textheight,trim=0.5cm 0.5cm 1.8cm 0.cm,clip]{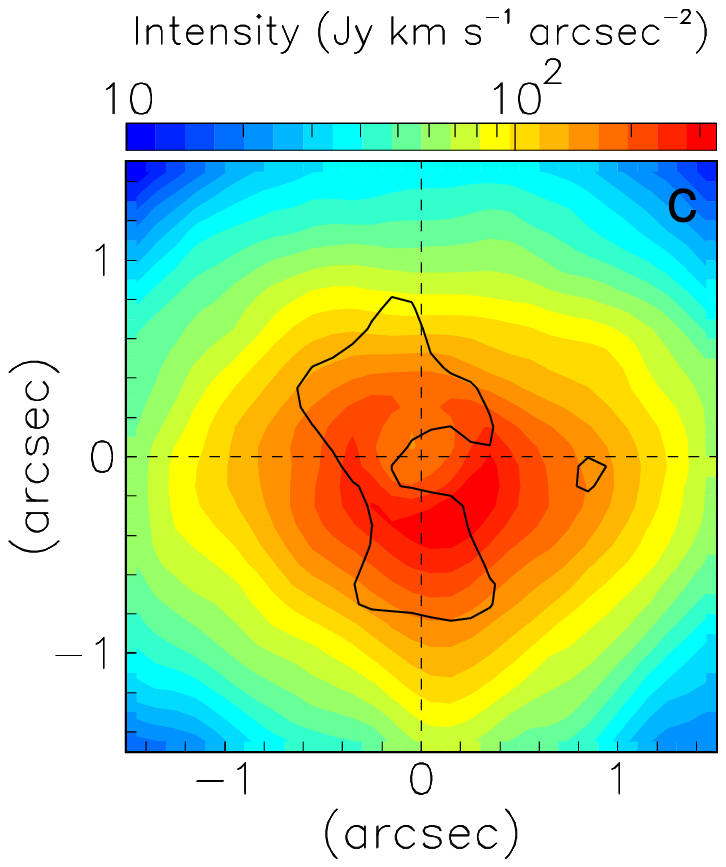}
 \includegraphics[height=0.2\textheight,trim=0.5cm 0.5cm 1.8cm 0.cm,clip]{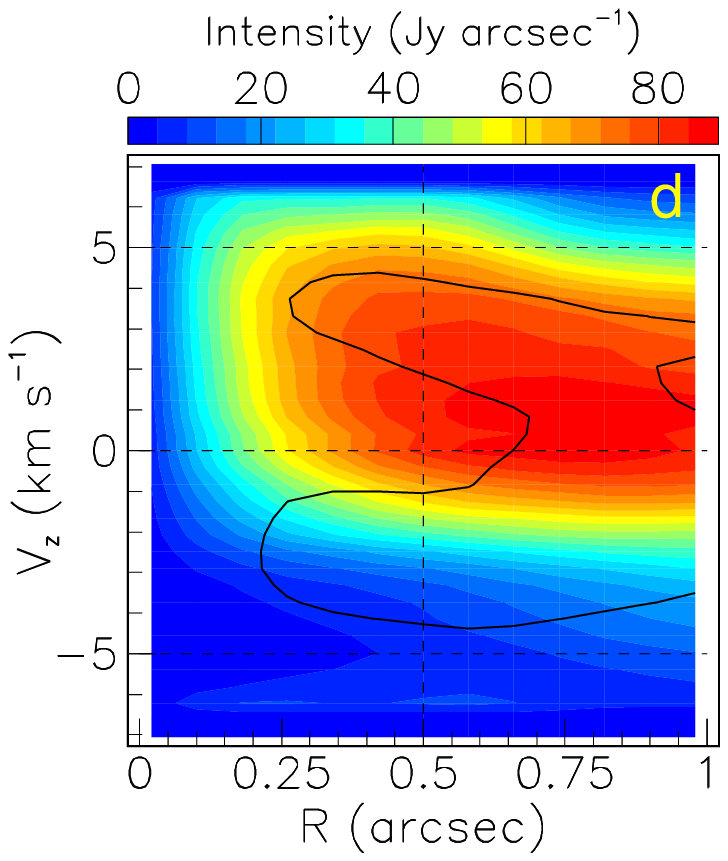}

 \includegraphics[height=0.189\textheight,trim=0.5cm .5cm 0.5cm .5cm,clip]{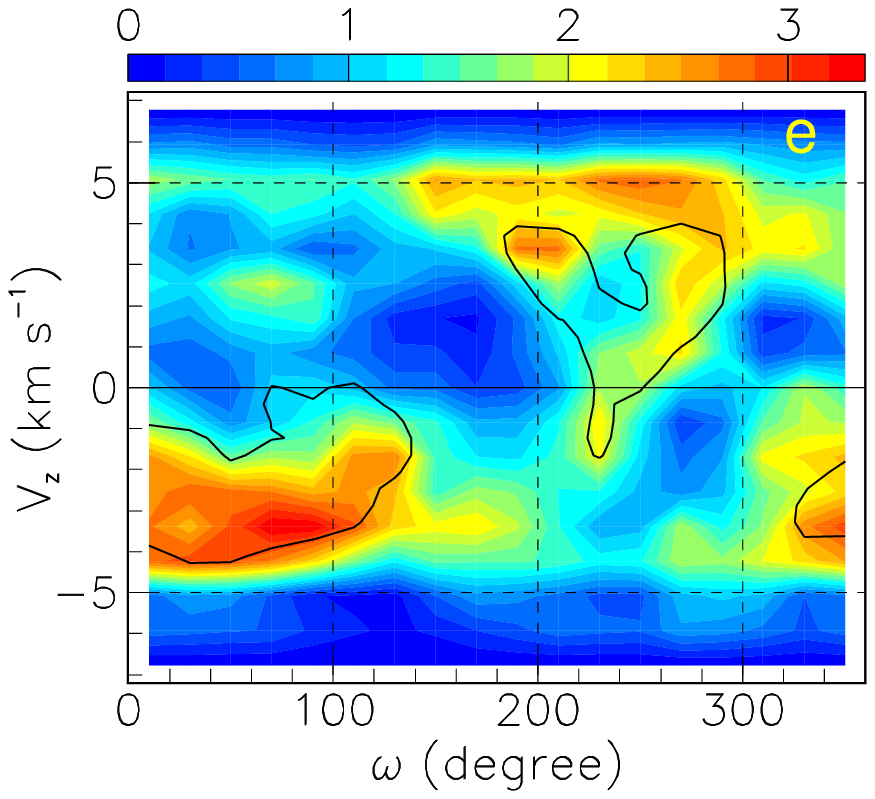}
 \includegraphics[height=0.189\textheight,trim=0.5cm .5cm 1.8cm .5cm,clip]{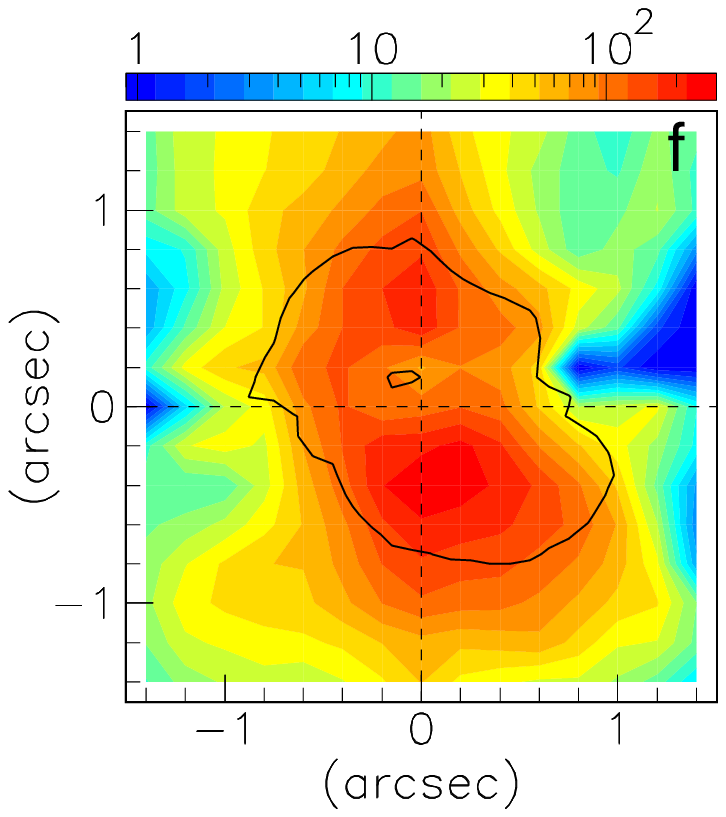}
 \includegraphics[height=0.189\textheight,trim=0.5cm .5cm 1.8cm .5cm,clip]{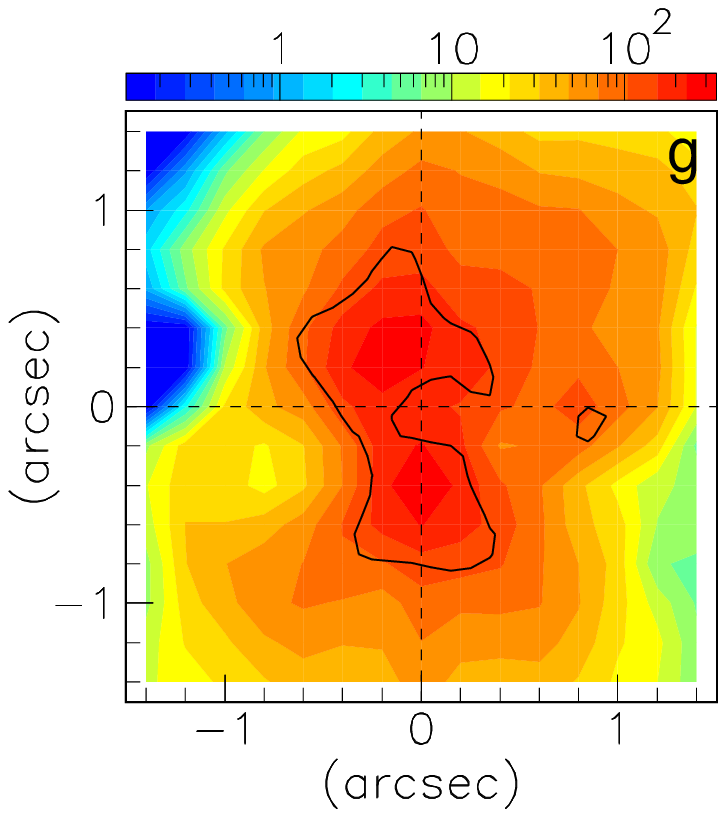}
 \includegraphics[height=0.189\textheight,trim=0.5cm .5cm 1.8cm .5cm,clip]{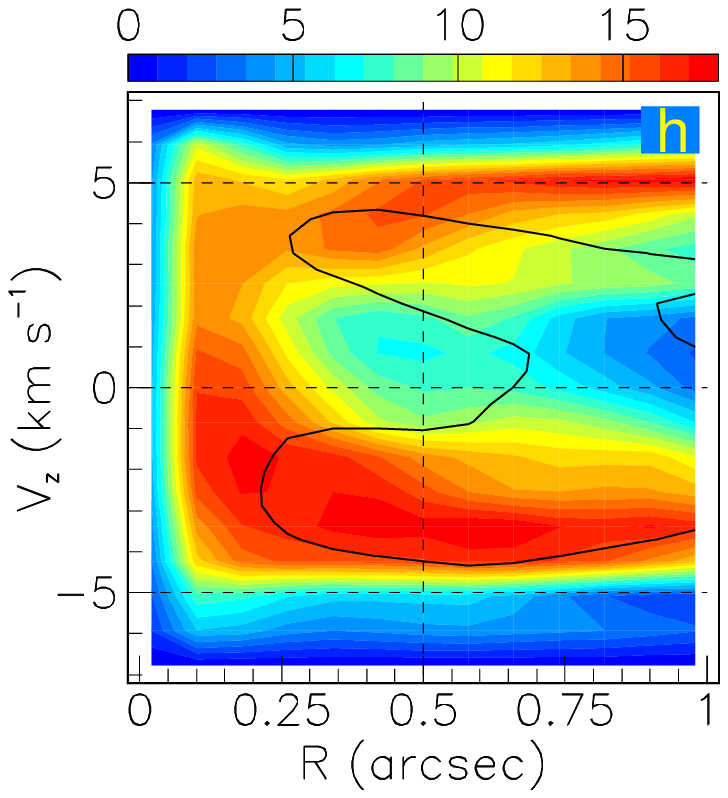}

 \includegraphics[height=0.178\textheight,trim=0.5cm 1cm 0.5cm .5cm,clip]{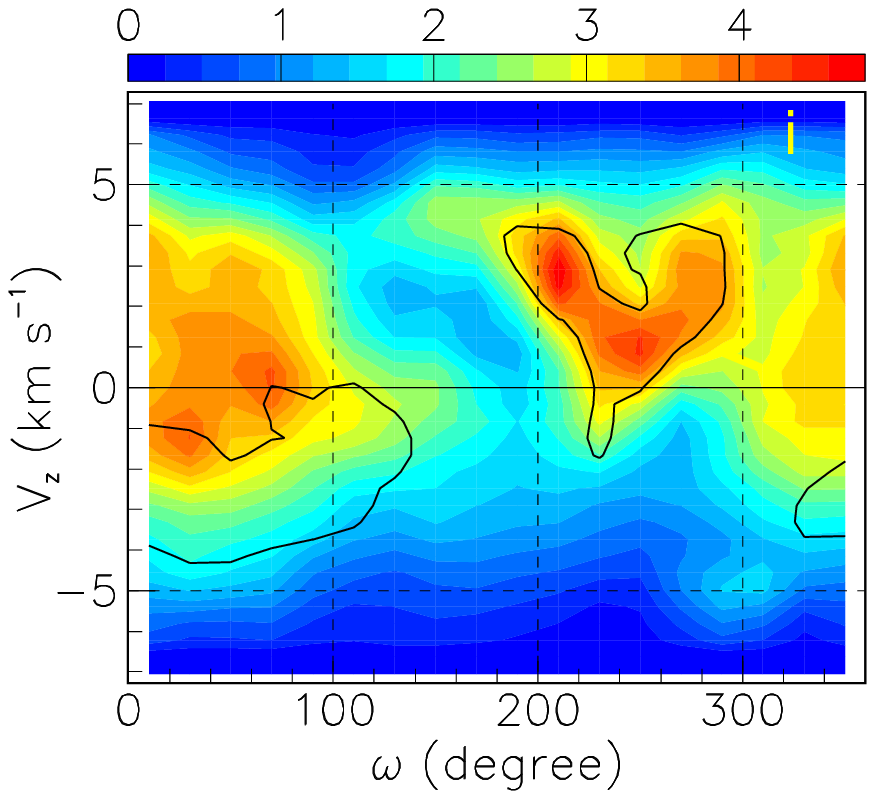}
 \includegraphics[height=0.178\textheight,trim=0.5cm 1cm 1.8cm .5cm,clip]{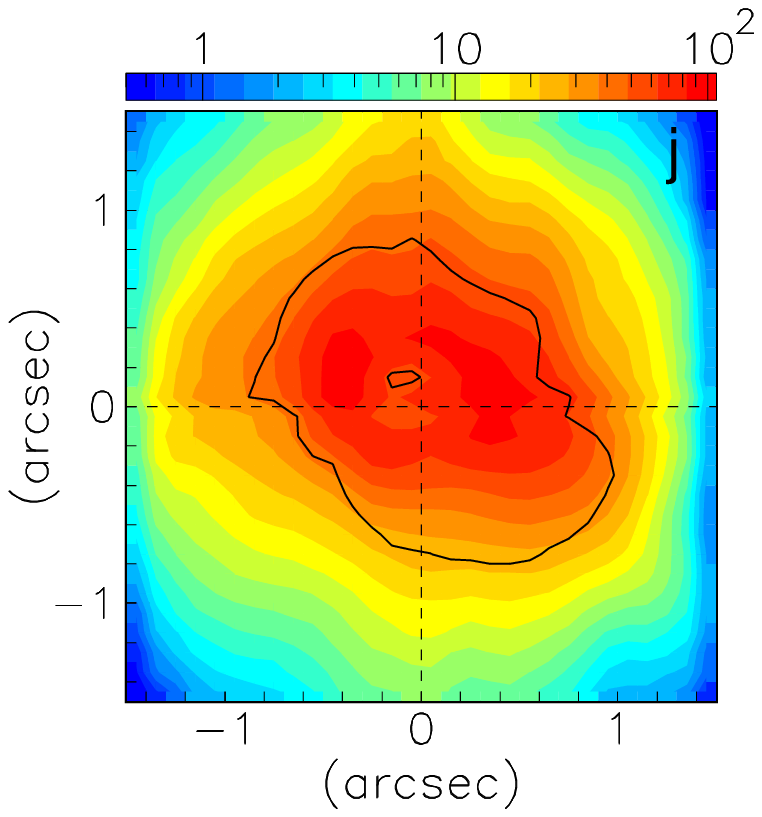}
 \includegraphics[height=0.178\textheight,trim=0.5cm 1cm 1.8cm .5cm,clip]{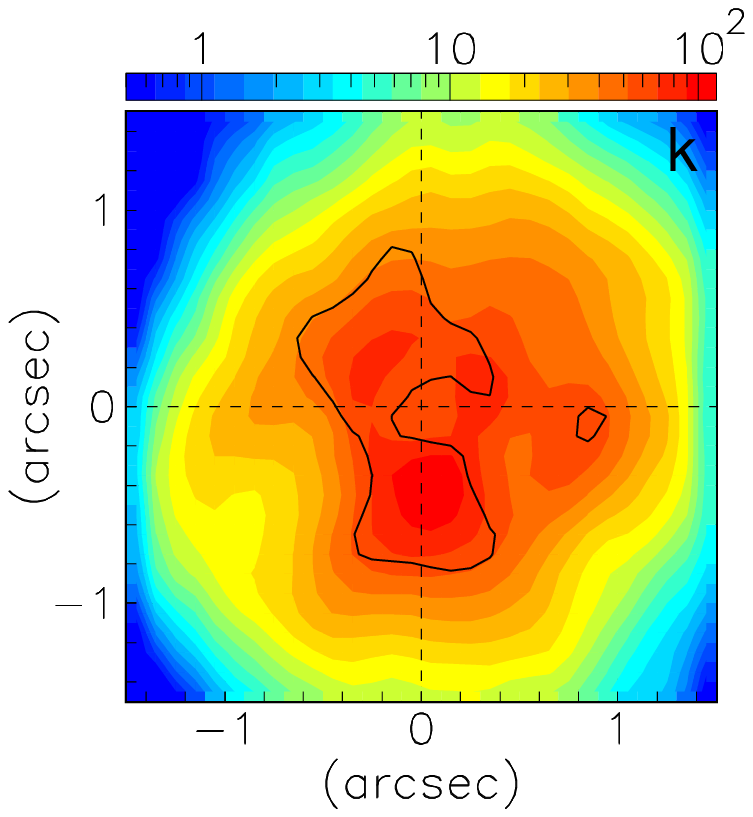}
 \includegraphics[height=0.178\textheight,trim=0.5cm 1cm 1.8cm .5cm,clip]{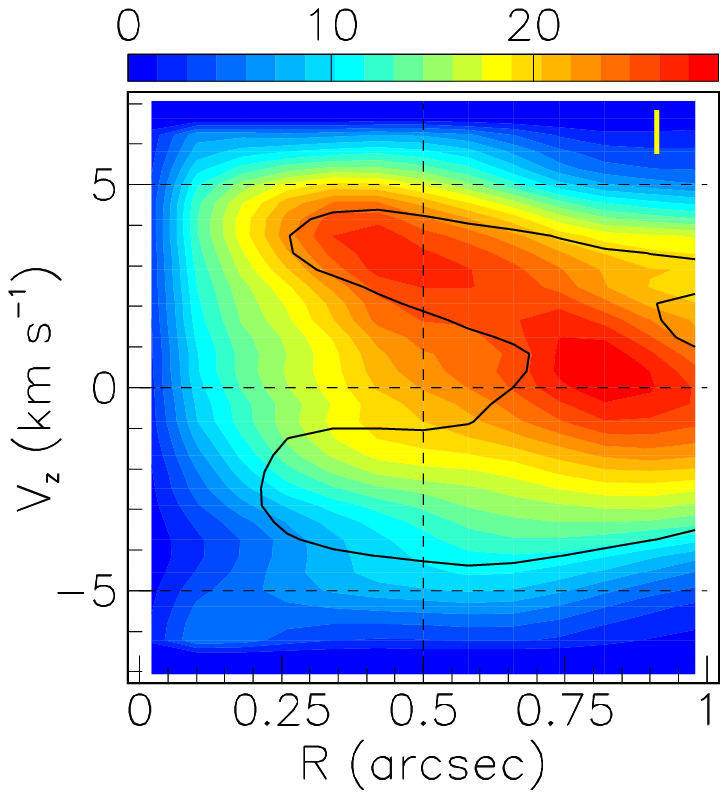}

   \caption{$^{28}$SiO (upper row), CO (middle row) and HCN (lower row) line emissions. Left: projection on the $V_z$ vs $\omega$ plane integrated over 12 au<$R$<60 au. Centre-left: projection on P1. Centre-right: projection on P2. Right: projection on the $V_z$ vs $R$ plane, integrated over position angles. Units are as in Figure \ref{fig20} from which the contour of SO emission at $\sim$60\% maximum is shown as reference in each panel.}
 \label{fig23}
\end{figure*}

In the present section we use the emission of lines $^{28}$SiO(8$-$7), CO(3$-$2) and HCN(4$-$3) to probe the CSE. They were previously considered in Paper I and have been briefly introduced in Section 2. Figures \ref{fig22} and \ref{fig23} describe the main results.
Doppler velocity spectra measured over the stellar disc (Figure \ref{fig22}a) show nearly total absorption of the $^{28}$SiO line at $V_z$$\sim$$-$4 \kms, confirming the results obtained for the $^{29}$SiO isotopologue when accounting for the much broader beam (FWHM$\sim$7 stellar radii). They also show significant but much smaller absorption of the HCN and CO lines. The CO line emission is absorbed at lower Doppler velocity than the SiO line, nearly $-$6 \kms\ instead of $-$4 \kms: this is probably the result of its much larger radial extension; it suggests that the wind acceleration from 4 \kms\ to the terminal velocity of $\sim$6 \kms\ is significant in a region where the density ratio between SiO and CO molecules is declining, namely beyond $r$$\sim$1 arcsec but the quality of the data does not allow for a more quantitative statement. Figure \ref{fig22}b displays Doppler velocity spectra outside the stellar disc where some absorption of the $^{28}$SiO line emission is seen to persist. Figure \ref{fig22}c,d displays, for the blue and red hemispheres separately, the radial distributions of the flux density integrated over Doppler velocity using the radial dependence of the velocity introduced in Section 6.2 for SO emission ($r_{\sfrac{1}{2}}$=0.5 arcsec). If the CSE were optically thin, its emission would be simply described by the $\varepsilon$ parameter. Using abundances relative to H$_2$ given by \citet{VandeSande2018} and \citet{Danilovich2016} and an H$_2$ density corresponding to a mass-loss rate of $\sim$1.6\,10$^{-7}$ \msun\,year$^{-1}$ as was done in Section 3.2, we calculate its value (Table \ref{tab4}) for a temperature of 150 K and a distance of 1 arcsec from the star and obtain the emission $F_r$ of a shell of 1 arcsec radius with which to compare the observations displayed in Figure \ref{fig22}c,d. We find that the SO and HCN observed emissions are well predicted and the CO emission some 30\% below prediction, which suggests that absorption is small for these lines (but in the blue hemisphere, HCN seems to suffer significant absorption). However, the observed $^{28}$SiO emission is considerably smaller, suggesting very strong self-absorption.

\begin{table}
  \centering
  \caption{Parameters of line emission per arcsec. $A$ is the Einstein coefficient, $J$ is the spin of the upper level, $f$ is the frequency, $\lambda$ is the wavelength, $\Delta{E}$ is the transition energy, $E_u$ is the energy of the upper level, $n$ is the number density, $r_0$ is the radius up to which the relative abundance is constant, $\sigma_0$ describes its radial decline as a Gaussian centred at $r_0$ and having a $\sigma$ equal to $\sigma_0$. $T$ is the temperature, $F_r$=$4\pi kAn(2J+1)e^{-Eu/T}/T$ is the emission in a shell of 1 arcsec radius with $k$ accounting for the different partition functions.}
  \label{tab4}
  \begin{tabular}{|c|c|c|c|c|}
  \hline
  Molecule&  $^{28}$SiO&  CO&  HCN&  SO\\
  \hline
  $A$ [10$^{-4}$]&  22.0&  0.025&  20.5&  1.93\\
  \hline
  $2J+1$&  17&  7&  9&  13\\
  \hline
  $k$ [10$^5$]&  3.3&  9.2&  6.9&  0.97\\
  \hline
  $f$ [GHz]&  347.3&  345.8&  354.5&  251.8\\
  \hline
  $\lambda$ [mm]&  0.864&  0.867&  0.846&  1.191\\
  \hline
  $\Delta{E}$[K]&  16.7&  16.7&  17.1&  12.1\\
  \hline
  $E_u$ [K]&  75&  33&  43&  51\\
  \hline
  $n/n(\rm{H}_2$) [10$^{-4}$]&  0.6&  2&  0.005&  0.07\\
  \hline
  $r_0$ [arcsec]&  1.5&  10&  0&  0\\
  \hline
  $\sigma_0$ [arcsec]&  1.1&  5&  1.6&  1.1\\
  \hline
  \makecell{$F_r$ [Jy\,\kms\,arcsec$^{-1}$]\\ @ $T$=150 K, $r$=1 arcsec}&  24080&  138&  210&  44\\
  \hline
  \end{tabular}
\end{table}

 Figure \ref{fig23} compares the $^{28}$SiO, CO and HCN emissions to SO emission by displaying the same distributions as displayed in Figure \ref{fig20} for the latter. While CO and HCN emissions show qualitatively similar distributions as SO, $^{28}$SiO seems to display a significantly different morphology. This had been noted in Paper I but the information contained in Figure \ref{fig23} offers a more detailed description of the differences at stake. The $V_z$ vs $\omega$ map (panel a) shows a slight depression rather than enhancement of emission at the position angle of the south-western outflow, $\omega$$\sim$240\dego.  When seen in the outflow planes (panels b and c) the emission shows no sign of an outflow but looks instead as isotropic, simply absorbed in the blue hemisphere. The same picture is suggested by the $V_z$ vs $R$ maps displayed in the right panels where the larger radial extension of the CO emission plays an important role. In order to explain such significant differences between the $^{28}$SiO emission and that of the other lines, we need to understand the impact of the strong self-absorption. 

\subsection{Absorption of SiO emission}

The comparison of the observed $^{28}$SiO emission with the calculated value of the unabsorbed emission parameter $\varepsilon$ performed in the preceding section suggests a strong self-absorption in both the red and blue hemispheres. In an idealised LTE picture, the emission reads $I$=$\varepsilon(1-e^{-\tau})/\tau$ and the optical thickness is $\tau$=2.9\,10$^7\varepsilon$$(e^{\Delta{E}/T}$$-$1)/$f^3$.  The ratio $\tau/\varepsilon$ is nearly the same for all lines: at 150 K, it is $\sim$0.08, 0.08, 0.08 and 0.15 for SiO, CO, HCN and SO respectively. Namely, for a 0.1 arcsec thick layer, $\tau$=15, 0.09, 0.13 and 0.05 at 150 K and $r$=1 arcsec, meaning an effective absorption by a factor $\sim$15 for SiO.

\begin{figure*}
  \centering
  \includegraphics[width=0.24\textwidth,trim=0.3cm 0.5cm 1.5cm 1.5cm,clip]{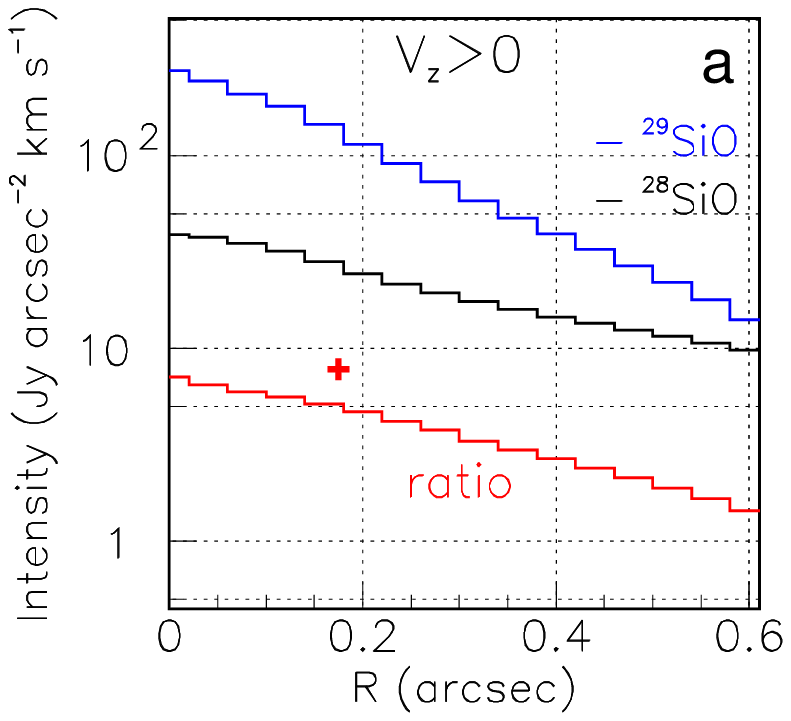}
  \includegraphics[width=0.24\textwidth,trim=0.3cm 0.5cm 1.5cm 1.5cm,clip]{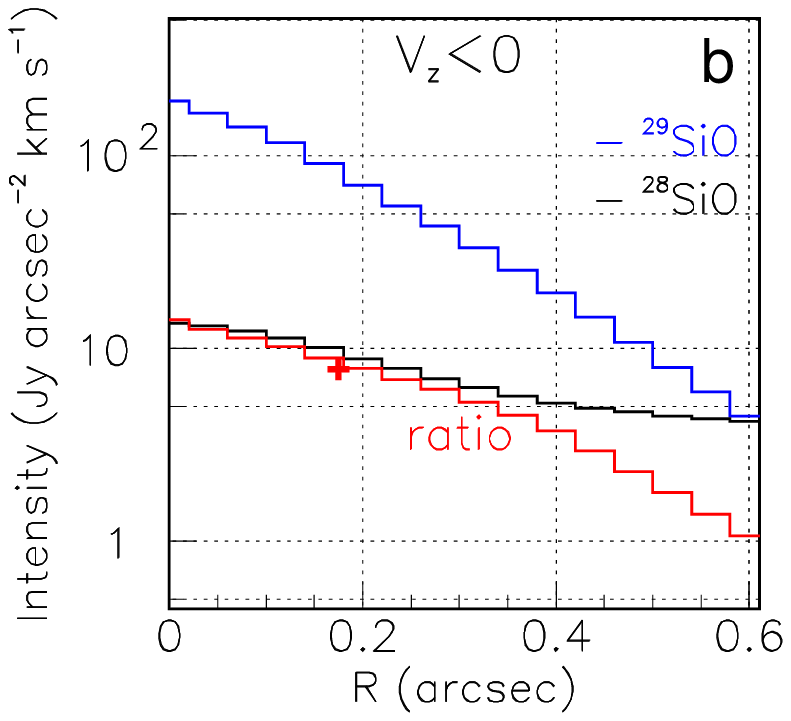}
  \includegraphics[width=0.24\textwidth,trim=0.3cm 0.5cm 1.5cm 1.5cm,clip]{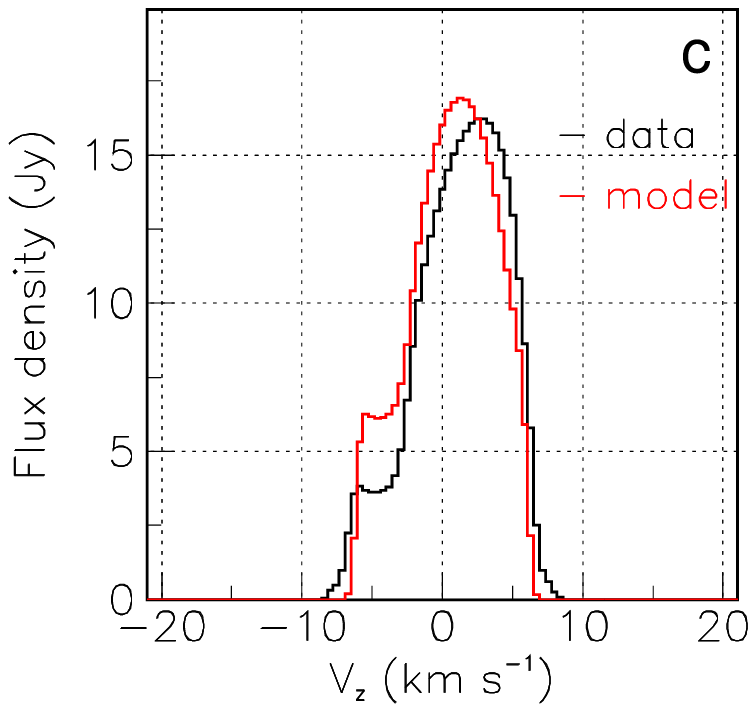}
  \includegraphics[width=0.24\textwidth,trim=0.3cm 0.5cm 1.5cm 1.5cm,clip]{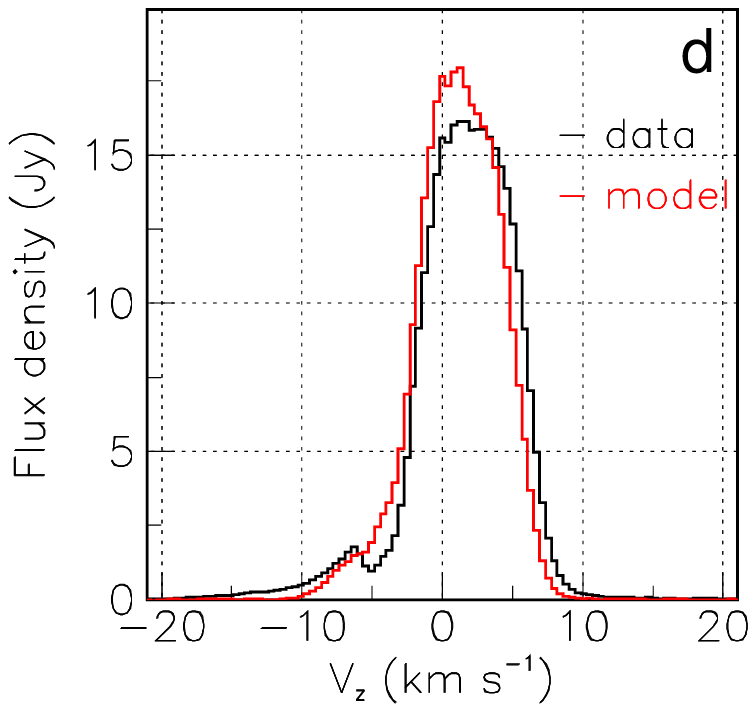}
  \caption{(a) and (b): $R$ distributions in the red (a) and blue (b) hemispheres of the $^{29}$SiO emission smeared with the $^{28}$SiO beam (blue) and the $^{28}$SiO emission multiplied by 0.08 (black); their ratio is shown in red. Continuum is not subtracted and the MRS of $^{29}$SiO is only 0.36 arcsec, that of $^{28}$SiO is 1.3 arcsec. The cross shows the increase of the absorption predicted by the model in a ring 0.15<$R$<0.20 arcsec when increasing the $^{29}$SiO density by a factor 1/0.08 (see text). (c) and (d): $^{28}$SiO Doppler velocity spectra (black) in rings 0.35<$R$<0.55 arcsec (c) and 0.65<$R$<0.85 arcsec (d) are compared with the crude model calculation (red) described in the text.}
 \label{fig24}
\end{figure*}

Another comparison of relevance is between the best fit to the $^{29}$SiO emission in a ring 0.15<$R$<0.20 arcsec obtained in Section 4 and the emission predicted using the same model for a density increased by a factor 1/0.08. We find that the ratio of absorbed to non-absorbed emissions decreases from 0.48 for the $^{29}$SiO density to 0.062 for the increased density, namely an increase of the absorption by a factor $\sim$7.7. Figure \ref{fig24} compares the $R$ distributions of the emission of both isotopologues, the $^{29}$SiO data having been smeared to account for the larger $^{28}$SiO beam size and the $^{28}$SiO data having been multiplied by 0.08 to account for the isotopic ratio. The observed increase of absorption between $^{29}$SiO and $^{28}$SiO is $\sim$5 in the red hemisphere and $\sim$10 in the blue hemisphere, well bracketing the predicted value of 7.7.

Finally, we model the $^{28}$SiO emission over two rings, 0.35<$R$<0.55 arcsec and 0.65<$R$<0.85 arcsec, far enough from the stellar disc for continuum emission to be negligible. The model includes the same form of Gaussian line broadening as described in Section 4 with the addition of a constant width $V_{\rm{turb0}}$. The radial expansion has the form described in Section 6.2. It assumes a $r^{-2}$ radial dependence of the density. The temperature is taken of the form $T$=$T_{0.5}r^{-p}$. The parameters being adjusted are the expansion parameter $r_{\sfrac{1}{2}}$, the density $N_{0.5}$ at $r$=0.5 arcsec, the line width parameters and the temperature parameters $T_{0.5}$ and $p$. The result is shown in panels c and d of Figure \ref{fig24}. The fit requires again a much steeper radial dependence of the temperature than used in earlier works: the fits displayed in Figure \ref{fig24} are for $T_{0.5}$=315 K and a 1/$r$ dependence ($p$=1) meaning $T$=158 K at $r$=1 arcsec. The model is too simple and makes too many simplifying assumptions for these numbers to be taken at face value. Yet, the trend for a steep radial dependence of the temperature, which was already suggested by the study of the inner layer presented in Section 4, is likely to be qualitatively valid. The best fit value of $r_{\sfrac{1}{2}}$ is 0.4 arcsec and the fit is not very sensitive to its precise value. The line broadening is the sum of a constant term, $V_{\rm{turb0}}$=0.7 \kms, and a Gaussian of amplitude 2 \kms\ centred at $r$=0.3 arcsec and having a $\sigma$ of 0.20 arcsec. The density at $r$=0.5 arcsec is 1625 molecules per cm$^3$, in agreement with the value found for the $^{29}$SiO fit, $\sim$40\% of that given by \citet{VandeSande2018}.  Better fits than shown in Figure \ref{fig24} can be obtained by fine-tuning the forms adopted for the radial dependence of the various parameters, but we refrain from showing these because the crudeness of the model makes such an exercise pointless.

Qualitatively, the several arguments developed in the present and preceding sections have all shown that the SiO layer covering from $\sim$0.2 arcsec to >1 arcsec distance from the star is optically thick and absorbs a large fraction of its own emission in both the red and blue hemispheres. The observed strong absorption is a result of the concurrence of a large Einstein coefficient and a large abundance: it is not expected to be specific to R Dor but should be also present in other oxygen-rich AGB stars. It is indeed the case, at least qualitatively, for Mira Ceti \citep{Wong2016} and W Hya \citep{Takigawa2017}. A quantitative reliable modelling would require observations of higher sensitivity and better $uv$ coverage than currently available.   

\section{Summary and conclusions }

The observations presented and analysed in the present article contribute important new information on five major features of the morpho-kinematics of the CSE of R Dor: rotation, absorption, presence of a south-eastern blue-shifted stream, large Doppler velocity wings and the detailed structure of the CSE up to $\sim$60 au distance from the star.

i) Rotation is seen to reach a maximal velocity of $\sim$5 to 7 \kms\ at some 8$\pm$2 au from the centre of the star, where the Keplerian orbital velocity around a 1.4 \msun\ star  is $\sim$13 \kms. Its axis projects on the plane of the sky at a position angle of 10$\pm$5\dego\ and makes an unknown angle with the plane of the sky; yet, qualitative considerations favour a low value, typically 20\dego$\pm$20\dego\ as suggested by the analysis of \citet{Homan2018}. The picture obtained from the analysis of the present observations reconciles those presented earlier by \citet{Vlemmings2018} and \citet{Homan2018}, claimed to be in conflict by the former authors. Qualitatively, it confirms an increase of the rotation velocity from $\sim$1 \kms\ at the surface of the star to its maximal value, described by \citet{Vlemmings2018} as ``solid body rotation''; and it confirms the decrease of the rotation velocity from its maximal value down to zero around $\sim$15 au, described by \citet{Homan2018} as ``sub-Keplerian''. Quantitatively, evidence is given for an important contribution of the blue stream to the kinematics for projected distances from the star exceeding $\sim$12 au, giving the appearance of a larger rotation velocity. Both $^{29}$SiO and SO$_2$ emission data give very consistent results. An interpretation suggested by \citet{Vlemmings2018} is the presence of a companion at some 6 au from the star, resulting in solid-body rotation of the close-by atmosphere extending up to a very few stellar radii and decreasing beyond. It might equally be that such a companion has already been engulfed in the recent past. The morphology of the rotating volume does not show strong anisotropy and, in particular, fails to provide evidence for a disc-like flattening about the equator as suggested by \citet{Homan2018}. It shows instead a broad enhancement covering the north-western quadrant, reaching close to 20\% for $^{29}$SiO and to 30\% for SO$_2$.

ii) Absorption of the $^{29}$SiO emission is nearly total at $V_z$$\sim$$-$4 \kms\ over the stellar disc and extends beyond it, an observation that had not been explicitly made in earlier studies. Strictly speaking, total absorption is not possible because one cannot prevent the outer part of an optically thick layer from shining. Quantitatively, using as reference the prediction of a simple LTE model, it would require the presence of a low temperature self-absorbing layer, expanding with a velocity of $\sim$4 \kms\ at a distance exceeding some 40 au from the star. A detailed study of the morpho-kinematics of the CSE at distances reaching $\sim$100 au, using ALMA observations of SO, $^{28}$SiO, CO and HCN suggests that such is indeed the case. New observations of higher sensitivity and better $uv$ coverage are needed to reliably confirm this important result. A qualitatively similar effect is expected to occur in the environment of other oxygen-rich AGB stars, as seems indeed to be the case for Mira Ceti and W Hya.

iii) The blue stream is observed at $\sim$140\dego\ position angle to extend radially from $\sim$10 au to $\sim$30 au with a mean projected acceleration of 39.2$\pm$0.9 \kms\,arcsec$^{-1}$. Interpreted as a gas stream in rectilinear motion, it points to the star to within some 1.5 au. This makes it difficult to describe it as gas trailing beyond a possible companion: unless it flows near the plane of the sky, its trajectory covers too long a time during which the companion is expected to travel too long a distance: it should have the shape of a spiral rather than of a straight line and should therefore not be seen to point precisely to the star. If instead it flows close to the plane of the sky, it must reach very large space velocities, up to 100 \kms\ or so. It is difficult to conceive a sensible scenario that would simply relate the presence of the blue stream to that of an evaporating planetary companion, as suggested earlier by \citet{Decin2018, Homan2018} and \citet{Vlemmings2018}. Qualitatively, as the blue stream is seen in continuity with the expansion of the slow wind, a more natural interpretation of its dynamics is that it is associated with a local enhancement of aluminium-rich dust density at some 12 au from the star and some 30\dego\ to 45\dego\ from the line of sight \citep{Takigawa2017}.

The absence of evidence for a planetary companion should not be interpreted as evidence for the absence of such companions. Indeed, it would be unreasonable to ignore that AGB stars are very likely to host many planets in their CSE, as do the parent Main Sequence stars. However, while the engulfment process  is expected to be extremely fast \citep[][and references therein]{Villaver2014} we must remember that the dust mass around R Dor is two orders of magnitude smaller than the Earth mass \citep{Decin2018} and many events may happen before and during engulfment that might leave a mark in the CSE.

iv) The high Doppler velocity wings studied earlier have been shown to be well separated from the blue stream and to be caused by a significant line broadening occurring within the inner layer of the CSE, below 12 au from the star. Its precise nature, including turbulence, shocks and even a possible effect of pulsations, remains to be clarified.

v) The presence of two broad outflows covering the south-western red-shifted and north-eastern blue-shifted octants respectively has been confirmed and parameters defining their morpho-kinematics have been estimated. Observation of other lines, from $^{28}$SiO, CO and HCN molecules, probing the CSE up to some 60 au from the star, has provided a qualitative confirmation of the SO result. However, the observed morpho-kinematics of $^{28}$SiO emission differs significantly from that of the other lines, probably because the very strong self-absorption hides the inner layers from observation. New observations with good $uv$ coverage, resolution and sensitivity, are very much needed in order to refine the general picture obtained from the existing SO observations.

The picture of the R Dor CSE that emerges from the present study is consistent with an early launch of the wind by aluminium-rich dust, followed by a progressive increase of the contribution of silicon-rich dust, giving support to a three-step picture of the generation and acceleration of the wind: in a first step pulsations and convective cell ejections give a first boost to the gas, with velocities reaching typically some 10 \kms, sufficient to eject part of it to a few au beyond the photosphere. There, in a second step, large transparent grains start to condense, mostly aluminium-rich, reaching grain sizes of a few tenths of a micron within some 10 au from the star; they are accelerated by photon scattering and transfer part of their momentum to the gas, providing a first pre-acceleration that brings it into a region where silicate grains start to condense. These grains are accelerated by radiation pressure and reach up to several tens au beyond the star surface. There, in a third step, the gas is progressively accelerated by collisions with the dust grains to reach a terminal velocity at the 5 to 10 \kms\ scale at some 100 au from the star, where it has finally escaped the star gravity.
  
Yet, to make sense of what we observe, we need to answer a number of open questions concerning the possible presence of planetary companion(s) and its impact, the mechanism causing the generation of significant rotation, the precise nature of the blue-shifted stream and the mechanism beyond the shaping of the pair of back-to-back outflows. In particular, new observations with good $uv$ coverage, resolution and sensitivity, of line emissions probing the CSE up to some 60 au from the star, such as $^{28}$SiO, CO and HCN molecules, are very much needed in order to refine the general picture obtained from the existing SO observations.

We are still a long way away from being able to model precisely the physics and chemistry at stake in the CSE of R Dor. Such modelling cannot afford to ignore the complex reality of the observed morpho-kinematics. While progressing in its description, we still fail to give a coherent interpretation of three apparently unrelated features: star rotation, blue stream and pair of broad back-to-back outflows.

\section*{Acknowledgements}
We express our deep gratitude to the anonymous referee for a very detailed and pertinent review that helped to significantly improve the presentation of the manuscript and we thank Professors Eva Villaver, Leen Decin, Ward Homan, Pierre Lesaffre and St\'{e}phane Guilloteau for useful discussions. This paper makes use of the following ALMA data: ADS/JAO.ALMA\#2017.1.00191.S (PI: Theo  Khouri), ADS/JAO.ALMA\#2017.1.00824.S (PI: Leen Decin) and ADS/JAO.ALMA\#2013.1.00166.S (PI: Leen Decin). ALMA is a partnership of ESO (representing its member states), NSF (USA) and NINS (Japan), together with NRC (Canada), MOST and ASIAA (Taiwan), and KASI (Republic of Korea), in cooperation with the Republic of Chile. The Joint ALMA Observatory is operated by ESO, AUI/NRAO and NAOJ. The SO data are retrieved from the JVO/NAOJ portal. We are deeply indebted to the ALMA partnership, whose open access policy means invaluable support and encouragement for Vietnamese astrophysics. In particular, we acknowledge the time and effort dedicated by the PIs of the projects used in this article for producing proposals of observations selected to best serve the AGB community. Financial support from the World Laboratory, the Odon Vallet Foundation and VNSC is gratefully acknowledged. This research is funded by the Vietnam National Foundation for Science and Technology Development (NAFOSTED) under grant number 103.99-2019.368.

\section*{Data Availability}
The raw data are available on the ALMA archive ADS/JAO.ALMA\#2017.1.00191.S, ADS/JAO.ALMA\# 2017.1.00824.S, ADS/JAO.ALMA\#2013.1.00166.S. The calibrated and imaged data underlying this article will be shared on reasonable request to the corresponding author.



\bibliographystyle{mnras}
\bibliography{rdorhires} 


\appendix

\section{}

\begin{table}
  \centering
  \caption{Parameters of the best fits to the Doppler velocity distributions displayed in Figure \ref{fig6}.}
  \label{tabA1}
  \begin{tabular}{cccccc}
    \hline
    \multicolumn{6}{c}{SiO}\\
    \hline
    ring&quadrant&$I^*$&$I_0$& $V_{z0}$&$\sigma_0$\\
    & &(mJy)&(mJy)&(\kms)&(\kms)\\
    \hline
    2&N&28
    &159
    &0.2
    &7.6\\
2
&E
&14
&266
&$-$3.1
&5.4\\

2
&S
&13
&171
&0
&6.6\\

2
&W
&19
&190
&3.5
&7.5\\
3
&N
&8
&292
&1.1
&5.1\\
3
&E
&$-$4
&453
&$-$3.0
&4.1\\
3
&S
&$-$6
&278
&$-$0.7
&4.6\\
3
&W
&9
&325
&4.4
&4.1\\
4
&N
&$-$1
&351
&1.1
&5.0\\
4
&E
&$-$4
&461
&$-$2.6
&3.8\\
4
&S
&$-$3
&318
&$-$0.5
&4.4\\
4
&W
&8
&367
&3.3
&3.1\\
2
&NE
&21
&220
&$-$2.5
&6.2\\
2
&SE
&14
&205
&$-$1.7
&5.8\\
2
&SW
&15
&172
&2.8
&7.2\\
2
&NW
&25
&164
&2.2
&7.3\\
3
&NE
&$-$2
&362
&$-$2.1
&4.7\\
3
&SE
&$-$1
&368
&$-$2.4
&4.1\\
3
&SW
&4
&259
&2.6
&4.2\\
3
&NW
&8
&328
&4.0
&4.3\\
4
&NE
&$-$7
&384
&$-$1.3
&4.5\\
4
&SE
&$-$3
&418
&$-$2.4
&3.9\\
4
&SW
&3
&339
&2.6
&2.9\\
4
&NW
&10
&362
&3.4
&3.7\\
\hline
  \end{tabular}
\end{table}

\begin{table}
  \centering
  \caption{Parameters of the best fits to the Doppler velocity distributions displayed in Figure \ref{fig7}.}
  \label{tabA2}
  \begin{tabular}{cccccc}
    \hline
    \multicolumn{6}{c}{SO$_2$}\\
    \hline
     ring&quadrant&$I^*$&$I_0$& $V_{z0}$&$\sigma_0$\\
     & &(mJy)&(mJy)&(\kms)&(\kms)\\
     \hline
2
&N
&31
&74
&0
&4.0\\
2
&E
&12
&73
&$-$2.6
&3.9\\
2
&S
&14
&67
&0.1
&3.9\\
2
&W
&21
&61
&2.3
&4.7\\
3
&N
&8
&52
&0.7
&3.9\\
3
&E
&$-$3
&44
&$-$2.3
&3.5\\
3
&S
&$-$4
&46
&$-$2.1
&3.3\\
3
&W
&8
&55
&3.1
&3.7\\
4
&N
&3
&26
&0.3
&3.9\\
4
&E
&$-$3
&35
&$-$2.0
&3.1\\
4
&S
&0
&35
&$-$2.2
&3.1\\
4
&W
&3
&39
&1.6
&3.1\\
2
&NE
&21
&76
&$-$1.8
&3.8\\
2
&SE
&13
&69
&$-$1.6
&3.8\\
2
&SW
&18
&63
&1.9
&4.3\\
2
&NW
&26
&65
&1.5
&4.4\\
3
&NE
&3
&52
&$-$1.7
&3.3\\
3
&SE
&$-$1
&48
&$-$2.7
&3.0\\
3
&SW
&2
&40
&1.0
&3.8\\
3
&NW
&6
&62
&3.0
&3.9\\
4
&NE
&$-$1
&39
&$-$1.5
&2.6\\
4
&SE
&$-$1
&33
&$-$2.9
&3.5\\
4
&SW
&2
&37
&0.3
&3.0\\
4
&NW
&5
&29
&2.3
&3.9\\
\hline
  \end{tabular}
\end{table}

\begin{figure*}
  \centering
  \includegraphics[width=0.8\textwidth,trim=0.2cm -0.2cm 0cm 0.cm,clip]{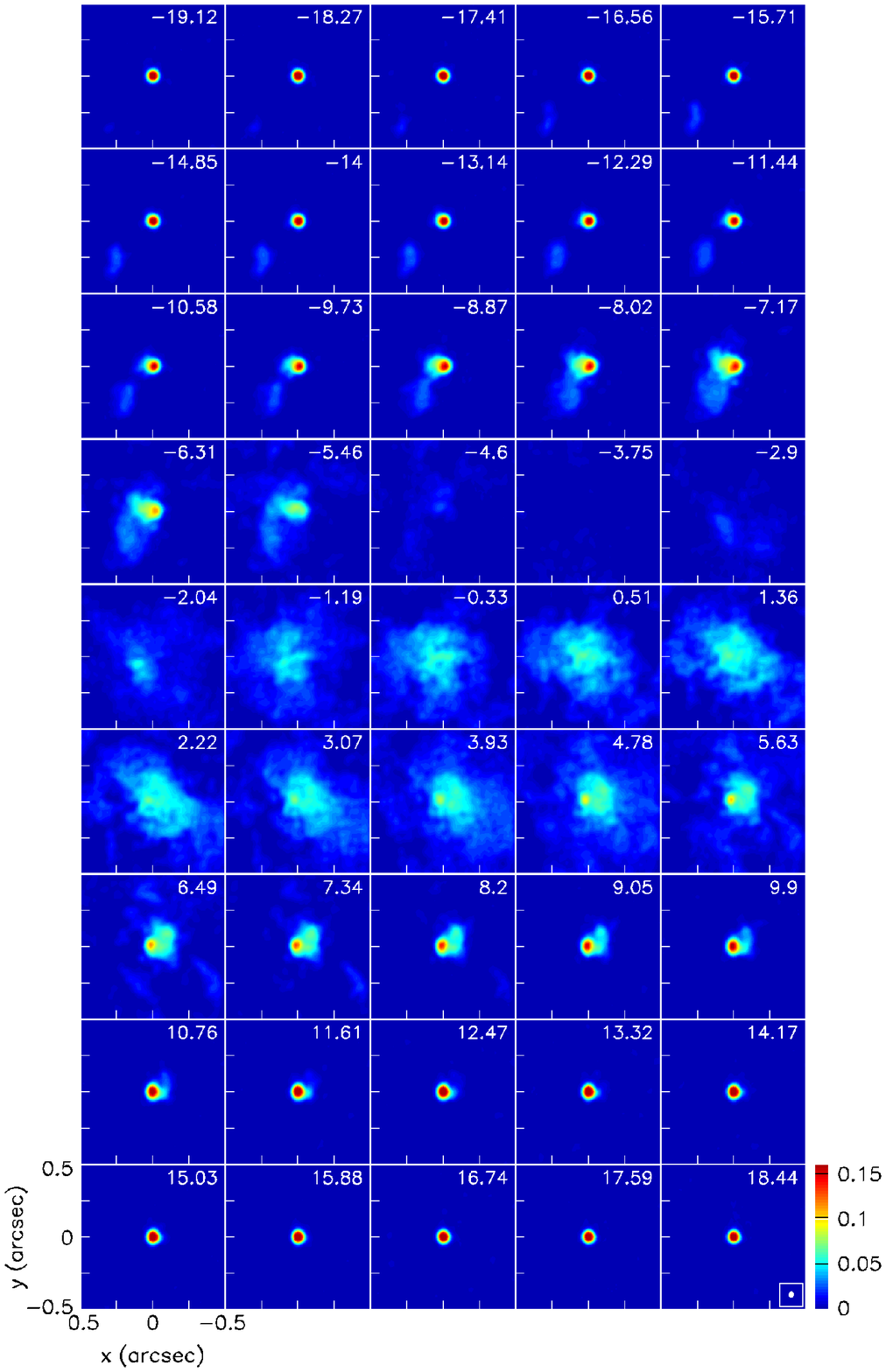} 
  \caption{Channel map of $^{29}$SiO emission (linear colour scale). The colour scale is in units of Jy beam$^{-1}$. The beam is shown in the lower-right corner of the last panel.}
 \label{figA1}
\end{figure*}

\begin{figure*}
  \centering
  \includegraphics[width=0.8\textwidth,trim=0.cm 0.cm 0cm 0.cm,clip]{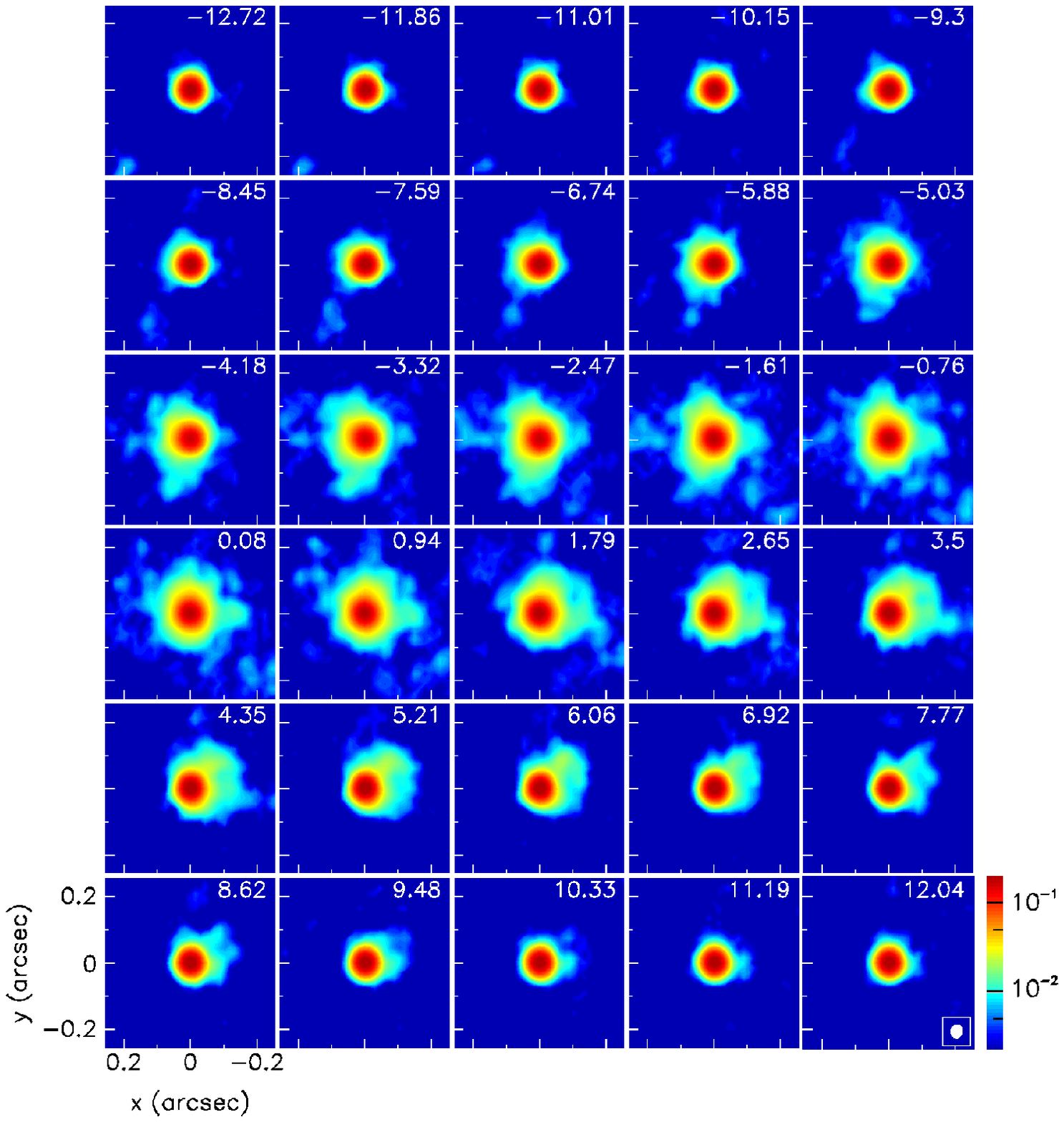}  
  \caption{Channel map of SO$_2$ emission (log colour scale). The colour scale is in units of Jy beam$^{-1}$. The beam is shown in the lower-right corner of the last panel.}
 \label{figA2}
\end{figure*}


\bsp	
\label{lastpage}
\end{document}